\documentclass[11pt]{article}

\usepackage{lineno,hyperref}
\usepackage{color}
\usepackage{here}
\usepackage{listings}
\usepackage{ifthen}
\usepackage{multirow}
\usepackage{array}
\usepackage{alterqcm}
\usepackage{pifont}
\usepackage{exscale, relsize}
\usepackage{soul}
\usepackage{makeidx}
\usepackage{graphicx}
\usepackage{graphics}
\usepackage{newlfont}
\usepackage{t1enc}
\usepackage[french2]{minitoc}
\usepackage{exscale,relsize}
\usepackage{subfig}
\usepackage{amsmath}
\usepackage{soul}
\usepackage[affil-it]{authblk}

\newtheorem{thm}{Theorem}[section]

\newtheorem{rmq}{Remark}[section]

\newcommand*{\affmark}[1][*]{\textsuperscript{#1}}
\usepackage[round]{natbib}
\usepackage{textcomp}
\newenvironment{acknowledgements}%
{\null
		\textbf{Acknowledgements}}%
{\null}
\begin{document}

\title{Effect of rainfall and fire frequency on tree--grass dynamics: Capturing the forest--savanna distributions along biogeographic gradients}



\author{A. Tchuint\'{e} Tamen\protect\affmark[1]\footnote{Corresponding author: alexis.tchuinte@yahoo.fr},
	P. Couteron\affmark[2,4], Y. Dumont\affmark[3,4,5]
}

\affil{$^{1}$Faculty of Science, University of Yaound\'e I, Cameroon\\
  $^{2}$IRD, UMR AMAP, Montpellier,  France\\
$^{3}$CIRAD, UMR AMAP,  Pretoria, South Africa\\
$^{4}$AMAP, University of Montpellier,  CIRAD, CNRS, INRA, IRD, Montpellier, France\\
$^{5}$Department of Mathematics and Applied Mathematics, University of Pretoria, South Africa}



%
%

\maketitle
\begin{abstract}

In this work, we improve a previous minimalistic  tree-grass savanna model by taking into account water availability, in addition to fire, since both factors are known to be important for   shaping savanna physiognomies along  a climatic gradient.
As in our previous models, we consider two nonlinear functions of grass and tree biomasses  to respectively take  into account grass-fire feedbacks, and the response of trees to fire of a given intensity. The novelty is that  rainfall is  taken into account in the tree and grass growth functions and in the biomass carrying capacities. Then, we derive a qualitative analysis of the ODE model, showing existence of equilibria, and studying their stability conditions.  We also construct a two dimension bifurcation diagram based on rainfall and fire frequency. This led  to summarize   different scenarios for the model including multi-stabilities that are proven possible.  Next,   to bring more realism in the model,  pulsed fire events are modelled as part of an IDE (Impulsive differential Equations) system analogous to the ODE system.  Numerical simulations are provided and we discuss some important ecological outcomes that our ODE and IDE models are  able to predict. Notably, the  expansion of forest into tree-poor physiognomies (grassland and savanna) is systematically predicted when fire return period increases, especially in mesic and humid climatic areas.
\end{abstract}



\noindent



\section{Introduction}

\label{intro}

Savannas, as broadly defined as systems where tree and grass coexist (\citealp{Scholes1997}), occupy about $20\%$ of the Earth land surface and are observed in a large range of Mean Annual Precipitation (MAP). In Africa, they particularly occur between 100 mm and 1500 mm (and sometimes more) of total mean annual precipitation (\citealp{Lehmann2011};  \citealp{Baudena2013}), that is along a precipitation gradient leading from dense tropical forest to desert. There is widespread evidence that fire and water availability are variables which can exert determinant roles in mixed tree-grass systems (\citealp{Scholes1997}; \citealp{vanLangevelde2003}; \citealp{Sankaran2005determinants}; \citealp{Staver2012integrating}; \citealp{Baudena2014forests}). Fires kill aerial parts of seedlings and shrubs while tree parts above the 'flame zone' are little affected (\citealp{Scholes2003convex}). On the other hand, water availability is considered to be the primary determinant of biomass production and indirectly of vegetation dynamics and structure in savannas (\citealp{Frost1986};   \citealp{Sankaran2005determinants}; \citealp{Abbadie2006lamto}; \citealp{Bond2008}). Moreover, empirical studies showed that vegetation properties such as biomass, leaf area, net primary production, maximal tree height and annual maximum standing crop of grasses vary along gradients of precipitation (\citealp{PenningDjiteye1982}; \citealp{Abbadie2006lamto}). It is widely accepted that water availability directly limits woody vegetation in the driest part of the rainfall gradient (\citealp{Frost1986}; \citealp{Sankaran2005determinants}; \citealp{Bond2008}). Along the rest of this gradient, rainfall is known to influence indirectly the fire regime trough what can be referred to as the grass-fire feedback (\citealp{Higgins2010stability}; \citealp{Staver2011tree}; \citealp{Baudena2014forests}): grass biomass that grow during rainfall periods is fuel for fires occurring in the dry months. Sufficiently frequent and intense fires are known to prevent or at least delay the development of woody vegetation (\citealp{vanWilgen2004response}; \citealp{Bond2005global}; \citealp{Govender2006}), thereby preventing trees and shrubs to depress grass production through shading. The grass-fire feedback is widely acknowledged in literature as a force able to counteract the asymmetric competition of trees onto grasses, at least for climatic conditions within the savanna biome that enables sufficient grass production during wet months.
\par
\begin{figure}[h!]
	\centering
	\subfloat[][]{\includegraphics[scale=0.5]{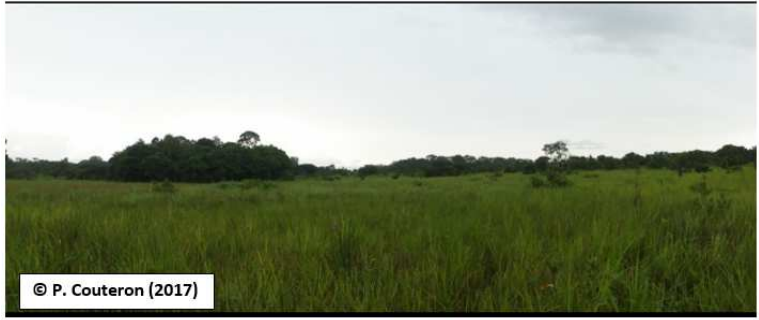}}
	\vspace{0.25cm}
	\subfloat[][]{\includegraphics[scale=0.5]{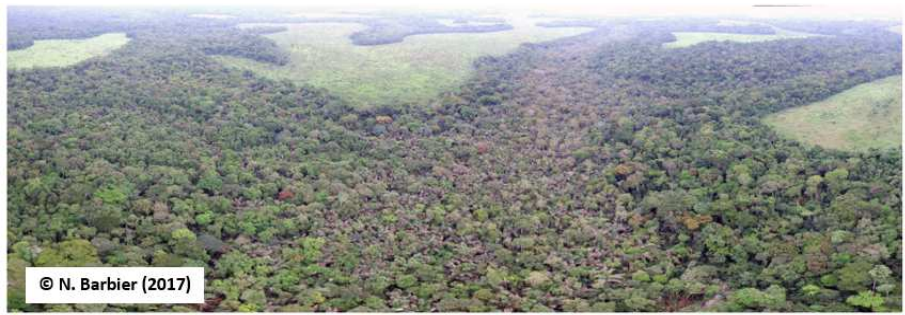}}	
	\caption{{\scriptsize (a) Photo of forest--grassland boundary in Mpem \& Djim National Park, Central Cameroon. (b) An abrupt Forest--savanna (grassland) mosaics in Ayos, Cameroon.}}
	\label{forest_savanna_mosaic}
\end{figure}

 
Dynamical processes underlying savanna vegetation have been the subject of many models. Some of them explicitly considered the influence of soil water resource on the respective productions of grass and woody vegetation components  (\citealp{Walker1981stability}; \citealp{Walker1982aspects}; \citealp{Rietkerk1997site}; \citealp{Rietkerk2002self}; \citealp{vanKoppel2000herbivore}; \citealp{vanLangevelde2003}; \citealp{DOdorico2006probabilistic}; \citealp{DeMichele2008minimal}; \citealp{Accatino2010tree}; \citealp{DeMichele2011}; \citealp{YuDOdoricco2014ecohydrological}). Most of these models also incorporated the grass-fire positive feedback, several of them distinguishing fire-sensitive small trees and shrubs from non-sensitive large trees (\citealp{Higgins2000fire}; \citealp{Beckage2009}; \citealp{Baudena2010}; \citealp{Staver2011tree}; \citealp{Yatat2014}), while the rest stuck to the simplest formalism featuring just grass and tree state variables (\citealp{vanLangevelde2003}; \citealp{DOdorico2006probabilistic}; \citealp{Higgins2010stability}; \citealp{Accatino2010tree}; \citealp{Beckage2011grass}; \citealp{YuDOdoricco2014ecohydrological}; \citealp{Tchuinte2014}). Models featuring the grass-fire feedbacks have shown that physiognomies displaying tree-grass coexistence (i.e. savannas) may be stable (\citealp{vanLangevelde2003}; \citealp{DOdorico2006probabilistic}; \citealp{Baudena2010}; \citealp{Accatino2010tree}; \citealp{Yatat2014}; \citealp{Tchuinte2014}) as well as more 'trivial' equilibria such as desert, dense forest or open grassland. Some models also predict alternative stable physiognomies under similar rainfall conditions (\citealp{Accatino2010tree}; \citealp{Staver2011tree}; \citealp{Yatat2014}; \citealp{Tchuinte2014}), a feature that is suggested as plausible by remote-sensing studies (\citealp{Favier2012abrupt}; \citealp{Hirota2011}; \citealp{Sanchez2016african}) as well as field observations of contrasted mosaics (see Fig. \ref{forest_savanna_mosaic}). However, just a subset of the published models addressed the entire rainfall gradient of the savanna biome (\citealp{Accatino2010tree}; \citealp{Higgins2010stability}; \citealp{Baudena2010}; \citealp{DeMichele2011}; \citealp{Beckage2011grass}; \citealp{YuDOdoricco2014ecohydrological}). The ability to predict all the physiognomies that are suggested by observations as possible stable or multi-stable outcomes was not established but for  \citealp{Accatino2010tree}.\par

The need to understand vegetation dynamics over very large territories for which information to calibrate models is scarce makes however desirable simple and robust models. The  \citealp{Accatino2010tree} model which built on  \citealp{DeMichele2008minimal} was pioneering in the sense that it allowed these authors to provide a "broad picture", by delimitating stability domains for a variety of possible vegetation equilibria as functions of gradients in rainfall and fire intensity. This result was especially  interesting that the considered model was sufficiently simple (two vegetation variables, i.e. grass and tree covers) to provide analytical forecasts.  However, results from  \citealp{Accatino2010tree} were questionable regarding the role of fire return time. In fact, all over the rainfall gradient their model predicted that decreasing fire frequency would lead to increase in woody cover which contradicts most of empirical knowledge on the subject. The features of the model that led to this problem were barely debated in the ensuing publications. And more recent papers instead either devised more complex models or shift to stochastic modelling (\citealp{DOdorico2006probabilistic}; \citealp{Beckage2011grass}). Such models did not allow much analytical exploration of their fundamental properties.    
\par
On the other hand, criticism against the ODE modelling (\citealp{Accatino2016a}; \citealp{Accatino2016trees}) pertinently argued that fires do not exert a continuous forcing on vegetation through time but rather destroy large shares of biomass through punctual events.  \citealp{Tchuinte2017} showed that the impulsive differential (IDE) framework can provide an interesting trade-off by modelling fires as punctual events albeit preserving the tractability of ODE equations that governs vegetation growth in inter-fire periods. They devised an IDE model that not only acknowledged punctual fires, but also reworked the way in which grass-fire feedback and fire impact on trees were modelled. Notably, fire intensity became a sigmoidal function of grass biomass (\citealp{Tchuinte2016}; \citealp{Yatat2016}), while a decreasing non-linear function of woody biomass was introduced to express fire impact on woody plants. The behavior of this IDE model proved immune to the increasing relationship between fire frequency and woody biomass that crippled earlier modelling efforts (\citealp{Tchuinte2017}). But the relationship between modelling results and rainfall remained indirect through models parameters that changed according to three main climatic zones (arid, mesic, humid) in the savanna biome. 
\par 
In the present paper, we will explicitly express the growth of both woody and herbaceous vegetation as functions of the mean annual rainfall, with the aim to study model predictions in direct relation to rainfall and fire frequency gradients. We will show that the two-variable ODE framework is not irrelevant per se and that an appropriate formulation of the grass-fire-tree retroaction loop makes it able to provide meaningful predictions while remaining fully analytically tractable. Through simulations, we will also show that the analogous ODE and IDE versions of our model qualitatively agree regarding stability domains, though diverging in several particular aspects. Through the present contribution we aim at proposing a framework for modelling vegetation in the savanna biome, that is both minimal (in terms of state variables and parameters) and generic in the sense that it does not pertain to particular locations (\citealp{Accatino2016a}). In so doing, we aim to show that our minimalistic tree-grass dynamical model is able to provide at  broad scales, a more consistent array of sensible predictions of possible vegetation physiognomies than previous ODE tree--grass frameworks. For instance, 
to our knowledge, existing models with only two vegetation state variables do not allow to capture the well-known fact that decreasing fire return period never leads to an increase of woody biomass whatever the place along a climatic gradient and especially in mesic and humid climatic conditions.
\par 
We aim to account for a wide range of physiognomies and dynamical outcomes of the system at both regional and continental scales by relying on a simple model that explicitly address some essential processes that are: (i) limits put by rainfall on vegetation biomass and growth (ii) asymmetric interactions between woody and herbaceous plant life forms, (iii) positive feedback between grass biomass and fire intensity and decreased fire impact with tree height.
\par 
Through integration of rainfall in plant growth and carrying capacity, the new IDE model introduced here is an extension of our previous impulsive model (\citealp{Tchuinte2017}). An idiosyncrasy of our minimalistic tree-grass model is that we considered the fire induced mortality on woody biomass by mean of two independent non-linear functions, namely $\omega$ (see (\ref{omega_fction})) and $\vartheta$ (see (\ref{theta_fction})). Considering these two functions (\citealp{Tchuinte2017}; \citealp{PhDTchuinte2017}), we showed that the previous model substantially improve previously published results on tree-grass dynamical systems. For example we showed that increasing fire return period systematically leads the system to switches from grassland or savanna to forest (forest encroachment). This result is entirely consistent with field observations (\citealp{Bond2005global}; \citealp{Bond2010beyond}; \citealp{Favier2012abrupt}; \citealp{Jeffery2014}). The novelty in the present IDE model is that we combine pulsed fire and rainfall in order to study how the  frequency of pulsed fires change the vegetation along a rainfall gradient. Assuming that fires are periodic in time and that growths and carrying capacities of grass and tree vary as nonlinear functions of rainfall, we aim to reproduce the whole set of physiognomies observable along the rainfall gradients leading in Africa from forests to deserts. 
\par

This paper is organized as follows. Section \ref{section2}  presents  modelling framework of the ODE and IDE models.  Section \ref{section3} gives the main theoretical results  of the ODE and IDE models. 
In section \ref{section4}   a 2D-bifurcation diagram in the rainfall-fire frequency
space is given and numerical simulations are also provided to discuss some important ecological scenarios. Finally, the paper ends with a conclusion and acknowledgements.

\section{Modelling frameworks: ODE and IDE}\label{section2}

The model considers the following assumptions.

\begin{itemize}
	\item [(A1)] Within the framework of tree-grass interactions, most models used vegetation state variables that corresponded to cover fractions that sum to one or less (\citealp{DeMichele2008minimal}; \citealp{Baudena2010}; \citealp{Accatino2010tree}; \citealp{Staver2011tree}; \citealp{DeMichele2011};\\ \citealp{YuDOdoricco2014ecohydrological}). It implicitly means assuming that vegetation components are mutually exclusive. Many field studies in fact showed that shrub and grass biomasses often exist under a tree crown (e.g.,  \citealp{Scholes1997}; \citealp{Abbadie2006lamto}; \citealp{Moustakas2013facilitation}). \\   In this work as well as in our previous studies (\citealp{Tchuinte2014}; \citealp{Tchuinte2016}; \citealp{Tchuinte2017}) we use biomasses as state variables since it allows a more straightforward link with field measurements (especially for grasses) and with emerging sources of remotely-sensed data as woody biomass estimates from radar backscattering (\citealp{Mermoz2015decrease}).

	\item [(A2)] The annual productions of grass and trees are assumed to be non-linear and saturating functions of the mean annual precipitation (MAP), here noted $\textbf{W}$ (in millimeter per year, mm.yr$^{-1}$). Following  \citealp{vandeKoppel1997}, \citealp{Higgins2010stability} and  \citealp{Nes2014tipping} a Monod equation is judged an adequate form to describe relationships between growth and rainfall since it is consistent with empirical data (e.g.,  \citealp{Whittaker1975}, see also fig. 4.6.3, p 191 in  \citealp{PenningDjiteye1982}).  We assume that  $\displaystyle\frac{\gamma_{G}\textbf{W}}{b_{G}+\textbf{W}}$ and $\displaystyle\frac{\gamma_{T}\textbf{W}}{b_{T}+\textbf{W}}$ are annual productions of grass and tree biomasses respectively, where $\gamma_{G}$  and $\gamma_{T}$ (in yr$^{-1}$) express   maximal growths  of grass and tree biomasses respectively,  half saturations  $b_{G}$ and $b_{T}$ (in mm.yr$^{-1}$) determine how quickly growth increase with water availability.\par 
	The estimate value   $\gamma_{G}=2$ is obtained using data from  \citealp{PenningDjiteye1982} (see figure 4.6.3 (a), p 191). The value $b_{G}=501$ mm.yr$^{-1}$ is deduced using data from  \citealp{UNESCO1981} (see fig 7, p 599).   $\gamma_{T}=0.15$ and $0.533$ are estimated using data from  \citealp{Chidumayo1990} and  \citealp{MenautCesar1979} respectively.  The value $b_{T}=1192$ mm.yr$^{-1}$ is deduced from  \citealp{Abbadie2006lamto}. Finding $b_{G}< b_{T}$  is not unrealistic but has been barely mentioned and discussed in literature. Though  \citealp{Scholes1993african} (see fig 1.1, page 3) suggested that grass and tree production may diverge along moisture gradients. \citealp{Accatino2010tree}  considered that vegetation growths are linear functions of soil moisture, however, the nonlinear relationship between soil-water and biomass production is widely observed in the field (\citealp{Mordelet1993influence}; \citealp{LeRoux1998seasonal}; \citealp{Simioni2003tree}; \citealp{House2003conundrums}).

	\item [(A3)] We assume that carrying capacities  of grass $K_{G}(\textbf{W})$ and tree $K_{T}(\textbf{W})$ are  increasing and bounded functions of water availability $\textbf{W}$. There are empirical data sets (e.g.  \citealp{UNESCO1981}; \citealp{Sankaran2005determinants};  \citealp{BuciniHanan2007}; \citealp{Staver2011tree}; \citealp{Staver2011b};\\  \citealp{Favier2012abrupt}; \citealp{Lewis2013AGBspatial}) which showed that maximum potential tree biomass increases with rainfall. For example,  \citealp{Sankaran2005determinants} analysed tree cover for $850$ field locations in African savannas along a range of increasing rainfall. They found that maximum tree cover increases from $0$ to  $80\%$  as mean annual precipitation (MAP) increases from $\sim 100$ mm to $\sim 650$ mm, value above which closed canopy  ($80\%$ cover) can be observed in the absence of disturbance. Other studies have been done to explain the changes of tree cover against MAP at a continental scale (see  \citealp{BuciniHanan2007} and  \citealp{Favier2012abrupt}).  \citealp{BuciniHanan2007}  used MODIS (Moderate Resolution Imaging Spectroradiometer) data  and they showed that  average tree cover increases asymptotically  with rainfall. 
	
	From those previous studies,  relationship between maximum  tree biomass vs. rainfall can be expressed as increasing and saturating functions (see fig 2 (a) in  \citealp{Favier2012abrupt}). To determine the form of $K_T$, we use field plot data from  \citealp{Higgins2010stability} and  \citealp{Lewis2013AGBspatial} (see also Fig. \ref{swv_fig1})  and consider that $K_T$ might follow a sigmoid  function. To fit the data we use the following function $K_{T}(\textbf{W})=\dfrac{c_T}{1+d_{T}e^{-a_{T}\textbf{W}}}$, where $c_T$ (in t.ha$^{-1}$) stands for maximum value of the tree biomass carrying capacity, $a_{T}$ (mm$^{-1}$yr) controls the steepness of the curve, and $d_{T}$ controls the location where is the inflection point. Note that the function $K_T$ is rationally analogue to the Holling type III functional response. We used the nonlinear quantile regression (\citealp{KoenkerPark1996}), as implemented in the "quantreg" library of the R program and we have $c_T=498.6$ t.ha$^{-1}$,  $d_{T}=106.7$, and $a_{T}=0.0045$ mm$^{-1}$yr.  
	Concerning the  grass biomass standing crop, $K_G$, we used empirical field data from  \citealp{Braun1972a} and \citealp{Braun1972b} (see Fig. \ref{swv_fig1} and see also  \citealp{UNESCO1981} page 599),  \citealp{MenautCesar1979}, and  \citealp{Abbadie2006lamto},  and  we consider  the following function: $K_{G}(\textbf{W})=\dfrac{c_G}{1+d_{G}e^{-a_{G}\textbf{W}}}$, where $c_G$ (in t.ha$^{-1}$) denotes the maximum value of the grass biomass carrying capacity,  $a_{G}$ (mm$^{-1}$yr) controls the steepness of the curve, and $d_{G}$ controls the location where is the inflection point. We have the following values: $c_G=17.06$ t.ha$^{-1}$, $d_{G}=14.73$, and $a_{G}=0.0029$ mm$^{-1}$yr. For the first time to our knowledge we aimed at characterizing the maximal possible woody biomass for both savanna and forest contexts.
	
	\begin{figure}[H]
		\centering
		\subfloat[][]{\includegraphics[scale=0.5]{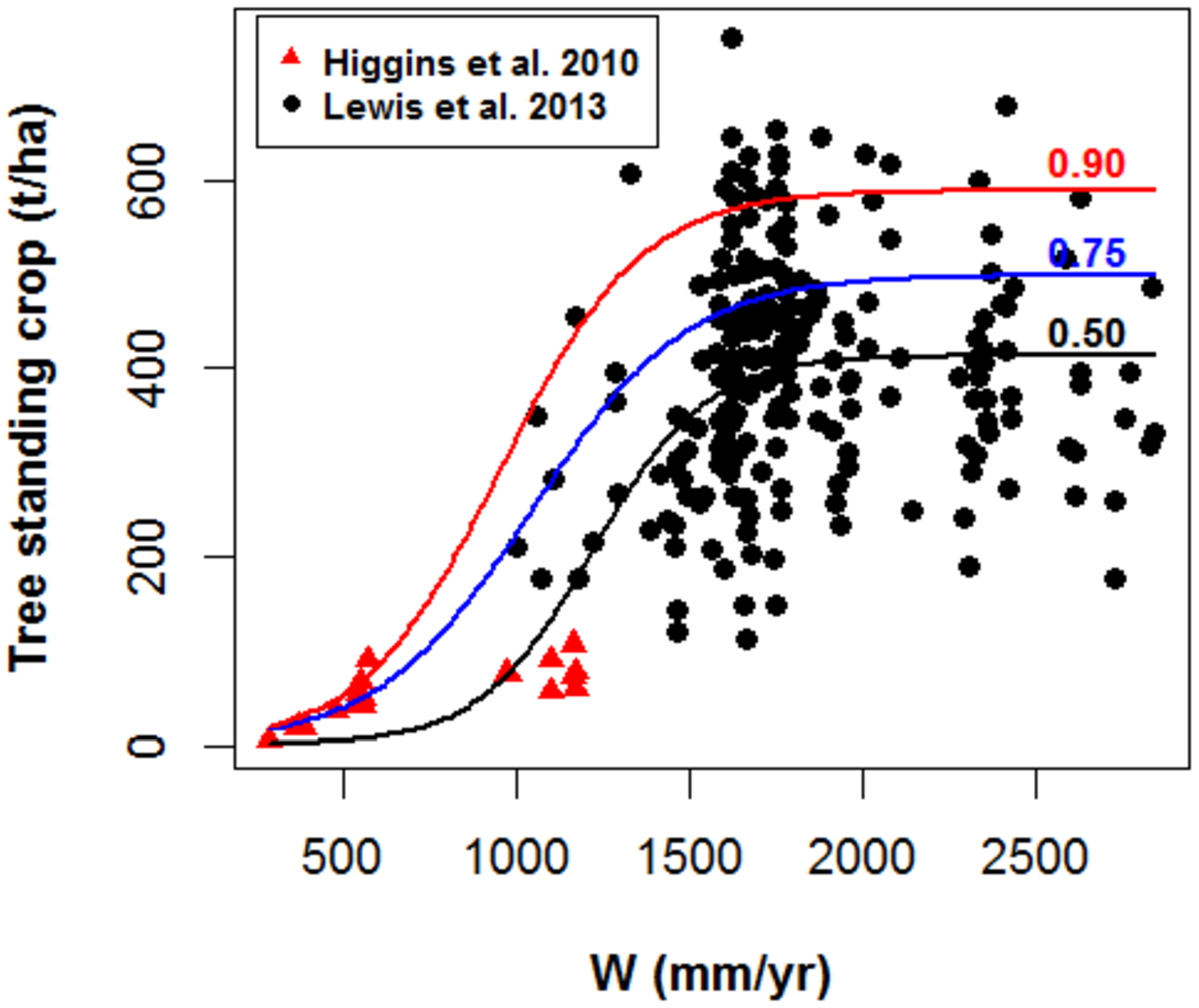}}
		\subfloat[][]{\includegraphics[scale=0.5]{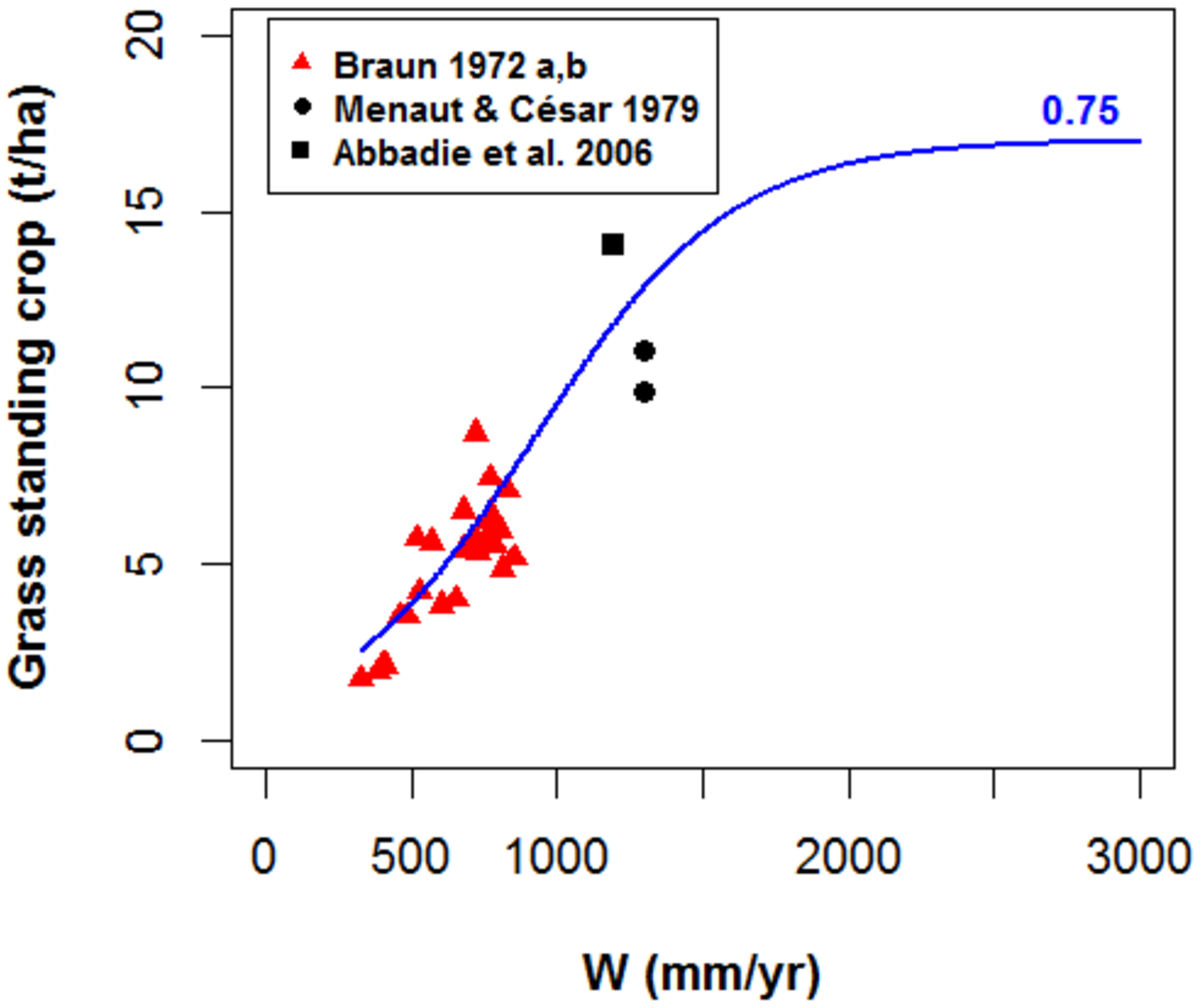}}	
		\caption{{\scriptsize (a) Maximum tree biomass production $K_{T}$  versus rainfall.  Data are from  \citealp{Higgins2010stability} and  \citealp{Lewis2013AGBspatial}. (b) Maximum grass biomass production $K_{G}$ versus rainfall. Data are  from  \citealp{MenautCesar1979}; \citealp{Braun1972a}; \citealp{Braun1972b} (see also  \citealp{UNESCO1981} page 599)  and  \citealp{Abbadie2006lamto}.}}
		\label{swv_fig1}
	\end{figure}
	
	\item [(A4)] Like in our previous works (\citealp{Tchuinte2014}; \citealp{Tchuinte2016};\\ \citealp{Tchuinte2017}), we assume that the fire intensity noted $\omega$ is an increasing and bounded function of the grass biomass given as follows:\\
	\begin{equation}
	\omega(G)=\dfrac{G^{\theta}}{G^{\theta}+\alpha^{\theta}},
	\label{omega_fction}
	\end{equation}
	where, $G$ in tons per hectare (t.ha$^{-1}$) is  grass biomass,  $\alpha$ is the value takes by $G$ when fire intensity is half its maximum, and the integer $\theta$ determines the steepness of the sigmoid.  The nonlinear form of $\omega$ is in agreement with recent models of savanna systems such as  \citealp{Staver2011tree} (for a Holling type III form) and the works of  \citealp{Tchuinte2014}; \citealp{Yatat2014}; \citealp{Tchuinte2016}; \citealp{Yatat2016}; \citealp{Tchuinte2017} (for  generic sigmoidal forms).  The parameter $\alpha$ is estimated as $\alpha=1.5$ t.ha$^{-1}$ (see fig 3 in  \citealp{Govender2006}). In the rest of this article, we set $\theta=2$.\par 
	
	\citealp{Accatino2010tree} along with other authors assumed that fire intensity increases linearly with fuel load (i.e., grass biomass). But, nonlinear forms (e.g., sigmoidal shape) of fire intensity functions agree with real ecological situations.  Although  \citealp{Staver2011tree} suggested that the nonlinear response function is a sufficient mechanism to let forest and savanna be alternative stable states, they made some peculiar assumptions (e.g. both saplings and trees die and revert to grass in proportion to their cover).  And their model is not able to predict as equilibria the full set of vegetation physiognomies that are observed along the whole climatic gradient, that features desert and wet tropical forest in addition to grassland and savanna. The central issue is that most savanna models (including ecohydrologic ones) are specific to certain climate zones. As a consequence they can account neither for all the physiognomies which are observed nor for all the multiple stable states which can be hypothesized considering the contrasted physiognomies observed under similar climatic conditions.

	\item [(A5)] Fire-induced tree/shrub mortality, noted $\vartheta$ is assumed to be a decreasing, non-linear function of tree biomass. In this paper as in   \citealp{Tchuinte2017},  we assume that fires affect differently large and small trees  since fires with high intensity (flame length $>2$m) cause greater mortality of shrubs and topkill of trees while fires of lower intensity  (flame length $<2$m)  kill only shrubs and subshrubs (\citealp{Trollope2002topkill}; \citealp{Hoffmann2003topkill}; \citealp{Higgins2007topkill}). It is evident that tree biomass and height are linked by increasing relationships. Therefore, as   in  \citealp{Tchuinte2017},  $\vartheta$ is expressed as follows: 
	
	\begin{equation}
	\vartheta(T)=\lambda_{fT}^{min} + (\lambda_{fT}^{max}-\lambda_{fT}^{min})e^{-pT},
	\label{theta_fction}
	\end{equation}	
	where, $T$ in tons per hectare (t.ha$^{-1}$) stands for  tree biomass, $\lambda_{fT}^{min}$ (in yr$^{-1}$) is minimal  loss of tree biomass due to fire in systems with	a very large tree biomass, $\lambda_{fT}^{max}$ (in yr$^{-1}$) is maximal  loss of tree/shrub biomass due to fire in open vegetation (e.g. for an isolated woody individual having its crown within the flame zone), $p$ (in t$^{-1}$.ha) is proportional to the inverse of biomass suffering an intermediate level of mortality. To parametrize $\vartheta$, we used experimental data from  \citealp{Trollope2002topkill} on the effects of height on the topkill rate of individual trees and shrubs in the Kruger National Park in South Africa, and in the central highlands of Kenya. We estimated biomass from height using   an allometric equation provided by \citealp{West2009} stating that at individual tree level, the total above-ground biomass (t/ha) is a power function of height (m) with an exponent of 4.  The estimated parameters of $\vartheta$ are: $\lambda_{fT}^{min}=0.05$ yr$^{-1}$, $\lambda_{fT}^{max}=0.9$ yr$^{-1}$ and $p=0.03$ t$^{-1}$.ha. Notice that  \citealp{Higgins2007topkill} considered that  $\lambda_{fT}^{min}<0.05$ and $\lambda_{fT}^{max}>0.9$. 
\end{itemize}

The above considerations motivate us to introduce a minimalistic eco-hydrological model under the framework of the following set of nonlinear ordinary differential equations :

\begin{equation}
\left\{
\begin{array}{l}
\displaystyle \frac{dG}{dt}=\displaystyle\frac{\gamma_{G}\textbf{W}}{b_{G}+\textbf{W}}G\left(1-\displaystyle\frac{G}{K_{G}(\textbf{W})}\right)-\delta_{G}G-\eta_{TG}TG -\lambda_{fG}fG,\\
\\
\displaystyle\frac{dT}{dt}=\displaystyle\frac{\gamma_{T}\textbf{W}}{b_{T}+\textbf{W}}T\left(1-\displaystyle\frac{T}{K_{T}(\textbf{W})}\right)-\delta_{T}T-f\vartheta(T)\omega(G)T,\\
\\
G(0)=G_{0}, T(0)=T_{0},
\end{array}
\right.
\label{swv_eq1}
\end{equation}

where,

\begin{itemize}
	\item[\textbullet]  $G$ and $T$  (in t.ha$^{-1}$) stand for grass and tree biomasses respectively;
	\item[\textbullet] \textbf{W} (in mm.yr$^{-1}$) is the mean annual precipitation MAP parameter.
	\item[\textbullet]   $G_{0}$ and $T_{0}$  (in t.ha$^{-1}$) are initial conditions of grass and tree biomasses respectively;
	\item[\textbullet]  $\delta_{G}$ and $\delta_{T}$ express respectively the rates of grass and tree biomasses loss by herbivores (grazing and/or browsing) or by human action; 
	\item[\textbullet] $\eta_{TG}$ denotes the asymmetric influence of trees on grass for light (shading) and resources (water, nutrients) which result in competitive or facilitative influences;
	\item[\textbullet] $f=\dfrac{1}{\tau}$ is the fire frequency, where $\tau$ is the fire return period;
	\item[\textbullet] $\lambda_{fG}$ is the specific loss of grass biomass due to fire. 
\end{itemize}

In  savanna ecosystems, it is widely observed that disturbances such as fires act in punctual way such that they consume tree/shrub and grass biomasses  over a short lapse of time (\citealp{House2003conundrums}; \citealp{Scheiter2009Grasstree}). In our previous works (\citealp{Tchuinte2016}; \citealp{Tchuinte2017}), we proposed  a tree--grass savanna model that considers fire occurrence and
impact as pulse periodic events through impulsive differential equations (IDE). The IDE framework  (\citealp{Lakshmikantham1989}; \citealp{Bainov1993}) is more realistic though keeping amenable to  in-deep analytical investigations of the models (\citealp{Tchuinte2016};  \citealp{Yatat2016}; \citealp{Tchuinte2017}). Moreover, we  showed that modelling fire as pulse events allow more diverse and interesting outcomes than modelling fire as a time-continuous external forcing. For instance, we showed that approximating fire regime as periodic let analytically investigating the possibility of periodic solutions  and various bistabilities. Assuming periodic pulsed fires in system (\ref{swv_eq1}), we obtain the following impulsive differential equation:

\begin{equation}
\left\{
\begin{array}{l}
\left.
\begin{array}{l}
\displaystyle \frac{dG}{dt}=\displaystyle\frac{\gamma_{G}\textbf{W}}{b_{G}+\textbf{W}}G\left(1-\displaystyle\frac{G}{K_{G}(\textbf{W})}\right)-\delta_{G}G-\eta_{TG}TG,\\
\\
\displaystyle\frac{dT}{dt}=\displaystyle\frac{\gamma_{T}\textbf{W}}{b_{T}+\textbf{W}}T\left(1-\displaystyle\frac{T}{K_{T}(\textbf{W})}\right)-\delta_{T}T,\\
\end{array}
\right\},  t\neq t_{n},\\
\\
\left.
\begin{array}{l}
G(t_{n}^{+})-G(t_{n})=-\lambda_{fG}G(t_{n}),\\
\\
T(t_{n}^{+})-T(t_{n})=-\vartheta(T(t_{n}))\omega(G(t_{n}))T(t_{n}),\\
\end{array}
\right\}, t= t_{n},  n=1,2,...,N_{\tau},\\

\end{array}
\right.
\label{pulsed_swv_eq1}
\end{equation}

with the  initial conditions 

\begin{equation}
G(t_{0})=G_{0}, T(t_{0})=T_{0},
\label{pulsed_swv_eq2}
\end{equation}

where,  $\tau=\frac{1}{f}$ is the period between two consecutive fires and $f$ is the frequency of fire, $N_{\tau}$ is a countable number of fire occurrences, $t_{n}=n\tau$, $n=1,2,...,N_{\tau}$, are called moments of impulsive effects of fire, and satisfy $0\leq t_{1}<t_{2}<...<t_{N_{\tau}}$,  $G(t_{n}^{+})$ and $T(t_{n}^{+})$ are  grass and tree biomasses instantly after an impulsive fire.\par 

\begin{rmq}
	\begin{itemize}
		\item[(i)] 
		It is important to note that the impulsive differential system (\ref{pulsed_swv_eq1}) completes our previous IDE model developed and studied in  \citealp{Tchuinte2017}. The novelty is that, here we consider that annual productions  and carrying capacities of grass and tree biomasses  vary according to the mean annual precipitation $\textbf{W}$ while in  \citealp{Tchuinte2017},  they were assumed to be constant.  Indeed, as mentioned in assumptions (A2) and (A3), there are empirical data sets which support the evidence of relationships between rainfall and annual productions  and carrying capacities of grass and tree biomasses respectively (e.g., for biomass productions see  \citealp{Whittaker1975}, see fig 7 of page 599 in  \citealp{UNESCO1981},  see also fig. 4.6.3 of page 191 in  \citealp{PenningDjiteye1982} and for maximum potential biomasses see  \citealp{UNESCO1981} page 599;  \citealp{Abbadie2006lamto}; \citealp{Higgins2010stability}; \citealp{Lewis2013AGBspatial}).
		
		\item[(ii)] Since system (\ref{pulsed_swv_eq1}) and the model developed and studied in  \citealp{Tchuinte2017} are almost similar,  the  readers are referred to appendices in  \citealp{Tchuinte2017} for the proofs of theoretical results. Accordingly, the demonstrations are deduced by  replacing the constant parameters of productions ($\gamma_{G}$ and $\gamma_{T}$) and carrying capacities  ($K_{G}$ and $K_{T}$) by non-linear, increasing and bounded functions of MAP ($\textbf{W}$) as given above in (A2) and (A3) respectively. However,  the ODE model (\ref{swv_eq1}) is proposed and studied here for the first time.  
	\end{itemize}
	\label{rmq1}
\end{rmq}

\section{Analytical results} \label{section3}

\subsection{The ODE model}\label{math_swv_model}

The right-hand side of system (\ref{swv_eq1}) is $\mathcal{C}^{1}$ i.e., continuously differentiable. Then, from the Cauchy-Lipschitz theorem,  system (\ref{swv_eq1}) has a unique maximal solution. In the ecological point of view, since the variables  of system (\ref{swv_eq1}) represent the biomasses,   each variable must stay positive and must be bounded during the time evolution (i.e., the system is said to be biologically well-behaved). Note that a solution with initial conditions in $\mathbf{R}^{2}_{+}$ stays in $\mathbf{R}^{2}_{+}$ since it can not cut the y-axis (vertical null line) and the x-axis (horizontal null line). Set 

\begin{displaymath}
\Gamma=\left\{(G,T)\in \mathbf{R}^{2}_{+} : G\leq K_{G}(\textbf{W}), T\leq K_{T}(\textbf{W})\right\},
\end{displaymath}

It is easy to verify that the sub-set $\Gamma$ is positively invariant and attracting with respect to system (\ref{swv_eq1}). It means that all trajectories of system (\ref{swv_eq1}) that start in $\Gamma$ remain positive and they do not tend to infinity with increasing time.\par


Set \begin{equation}
\left\{
\begin{array}{l}
g_{G}(\textbf{W})=\displaystyle\frac{\gamma_{G}\textbf{W}}{b_{G}+\textbf{W}},\\
g_{T}(\textbf{W})=\displaystyle\frac{\gamma_{T}\textbf{W}}{b_{T}+\textbf{W}},
\end{array}
\right.
\label{swv_growths}
\end{equation}

\begin{equation}
\left\{
\begin{array}{l}
\mathcal{R}^{1}_{\textbf{W}}=\dfrac{g_{T}(\textbf{W})}{\delta_{T}},\\
\mathcal{R}^{2}_{\textbf{W}}=\dfrac{g_{G}(\textbf{W})}{\delta_{G}+\lambda_{fG}f},
\end{array}
\right.
\label{swv_R0_desert}
\end{equation}
and

\begin{equation}
\left\{
\begin{array}{l}
T^{*}=K_{T}(\textbf{W})\left(1-\dfrac{1}{\mathcal{R}^{1}_{\textbf{W}}}\right),\\
G^{*}=K_{G}(\textbf{W})\left(1-\dfrac{1}{\mathcal{R}^{2}_{\textbf{W}}}\right).
\end{array}
\right.
\label{swv_T_G}
\end{equation}

System (\ref{swv_eq1}) has the following trivial equilibria: 

\begin{itemize}
	\item a bare soil equilibrium $\textbf{E}_{0}=(0,0)$.
	\item a forest equilibrium $\textbf{E}_{F}=(0,T^{*})$ which exists when $\mathcal{R}^{1}_{\textbf{W}}>1$. 
	\item a grassland equilibrium $\textbf{E}_{G}=(G^{*},0)$ which exists when $\mathcal{R}^{2}_{\textbf{W}}>1$.
\end{itemize}

Let us set:

$a=\dfrac{g_{T}(\textbf{W})}{K_{T}(\textbf{W})}T^{*}$, $b=\dfrac{g_{G}(\textbf{W})g_{T}(\textbf{W})}{\eta_{TG}K_{G}(\textbf{W})K_{T}(\textbf{W})}G^{*}$, $c=\dfrac{b}{G^{*}}$, $d=f\lambda_{fT}^{min}$,    $\lambda=f(\lambda_{fT}^{max}-\lambda_{fT}^{min})\times e^{-p\dfrac{g_{G}(\textbf{W})G^{*}}{\eta_{TG}K_{G}(\textbf{W})}}$ and $\alpha=p\dfrac{g_{G}(\textbf{W})}{\eta_{TG}K_{G}(\textbf{W})}$,  where $G^{*}$ is given by (\ref{swv_T_G}).\par

The existence of the positive savanna equilibrium is given in theorem \ref{al_thm1}.\par

\begin{thm} (Existence of the savanna equilibrium)\par 
	Table \ref{swv_tab_1} summarizes the conditions of existence the savanna equilibrium.
	
	\begin{table}[H]
		\begin{center}
			{\normalsize
				\centering
				\caption{Existence of the savanna equilibrium}
				\renewcommand{\arraystretch}{1}
				\begin{tabular}{|l|c|c|c|}
					\hline
					$c-\lambda\alpha$	& $a-b-d-\lambda$ & $a-b$ & Number of savanna equilibria \\
					\hline
					\multirow{2}{1cm}{$<0$}	& 
					\multirow{2}{1cm}{$>0$ or $<0$}
					& $>0$ & $0$ or $1$\\
					\cline{3-4}
					& &$<0$ & $0$ or $2$\\
					\hline
					\multirow{4}{1cm}{$<0$} &\multirow{2}{1cm}{$>0$}& $>0$ & $0$ or $1$\\
					\cline{3-4}
					& &$<0$ & $0$ or $2$\\
					\cline{2-4}
					& \multirow{2}{1cm}{$<0$}& $>0$ & $0$, $1$ or $3$\\
					\cline{3-4}
					& &$<0$ & $0$ or $2$\\
					\hline
				\end{tabular}
				\label{swv_tab_1}
			}
		\end{center}	
	\end{table}	
	\label{al_thm1}
\end{thm}
$\mbox{\bf{Proof:}}$ See appendix A, page \pageref{al_AppendixA}.\par

We now study the stability of the previous equilibria. Concerning the stability of the trivial equilibria $\textbf{E}_{0}$,  $\textbf{E}_{G}$, and $\textbf{E}_{F}$,  theorem \ref{al_thm2} holds. Set

\begin{equation}
\mathcal{R}_{F}=\dfrac{g_{G}(\textbf{W})}{\eta_{TG}T^{*}+\delta_{G}+\lambda_{fG}f}, \hspace{0.5cm}\mbox{and}\hspace{0.5cm} \mathcal{R}_{G}=\dfrac{g_{T}(\textbf{W})}{\delta_{T}+\lambda_{fT}^{max}f\omega(G^{*})}.
\label{swv_thresholds_F_G}
\end{equation}

Straightforward computations lead to the following theorem:
\begin{thm} (Stability of trivial equilibria)\par 
	\begin{itemize}
		\item[(1)] The desert equilibrium $\textbf{E}_{0}=(0, 0)$ is locally asymptotically stable (LAS) when $\mathcal{R}^{1}_{\textbf{W}}<1$ and $\mathcal{R}^{2}_{\textbf{W}}<1$.
		\item[(2)] The grassland equilibrium $\textbf{E}_{G}=(G^{*},0)$  is  LAS when $\mathcal{R}_{G}<1$.
		\item[(3)] The forest equilibrium  $\textbf{E}_{F}=(0, T^{*})$  is  LAS when $\mathcal{R}_{F}<1$.
	\end{itemize}
	
	\label{al_thm2}
\end{thm}

\begin{rmq}
	It is to be noted that the existence of $\textbf{E}_{F}$ or $\textbf{E}_{G}$ destabilizes the desert equilibrium $\textbf{E}_{0}$.
	\label{swv_rmq_1}
\end{rmq}

Let $\textbf{E}_{S}=(G_{*}, T_{*})$  a savanna equilibrium. Define

\begin{equation}
\left\{
\begin{array}{l}
\mathcal{R}^{1}_{*}=\dfrac{\dfrac{g_{G}(\textbf{W})}{K_{G}(\textbf{W})}G^{*}+\dfrac{g_{T}(\textbf{W})}{K_{T}(\textbf{W})}T^{*}-\vartheta^{'}(T_{*})f\omega(G_{*})T_{*}}{2\left(\dfrac{g_{G}(\textbf{W})}{K_{G}(\textbf{W})}G^{*}+\dfrac{g_{T}(\textbf{W})}{K_{T}(\textbf{W})}T^{*}\right)+\eta_{TG}T_{*}+f\omega(G_{*})\vartheta(T_{*})},\\
\\
\mathcal{R}^{2}_{*}=\dfrac{\dfrac{g_{G}(\textbf{W})g_{T}(\textbf{W})}{K_{G}(\textbf{W})K_{T}(\textbf{W})}G^{*}T^{*}+A_{*}B_{*}-\vartheta^{'}(T_{*})\dfrac{g_{G}(\textbf{W})}{K_{G}(\textbf{W})}G^{*}f\omega(G_{*})T_{*}}{\dfrac{g_{G}(\textbf{W})}{K_{G}(\textbf{W})}G^{*}B_{*}+\dfrac{g_{T}(\textbf{W})}{K_{T}(\textbf{W})}T^{*}A_{*}+\eta_{TG}f\omega^{'}(G_{*})\vartheta(T_{*})T^{2}_{*}-\vartheta^{'}(T_{*})A_{*}f\omega(G_{*})T_{*}},
\end{array}
\right.
\label{swv_thresholds_S}
\end{equation}

where, 

\begin{equation}
\left\{
\begin{array}{l}
A_{*}=2\dfrac{g_{G}(\textbf{W})}{K_{G}(\textbf{W})}G^{*}+\eta_{TG}T_{*},\\
\\
B_{*}=2\dfrac{g_{T}(\textbf{W})}{K_{T}(\textbf{W})}T^{*}+f\omega(G_{*})\vartheta(T_{*}).
\end{array}
\right.
\label{swv_thresholds_AB}
\end{equation}

\par 
Concerning the stability of the savanna equilibrium, the following theorem holds:

\begin{thm} (Stability of the savanna equilibrium)\par
	The savanna equilibrium $\textbf{E}_{S}=(G_{*}, T_{*})$  is locally asymptotically stable if and only if $\mathcal{R}^{1}_{*}<1$ and $\mathcal{R}^{2}_{*}>1$.	
	\label{al_thm3}
\end{thm}
$\mbox{\bf{Proof:}}$ See appendix B, page \pageref{al_AppendixB}.\par

\begin{rmq}
	Ecological meaning of some thresholds of model (\ref{swv_eq1}) are given below:
	\begin{itemize}
		\item[(i)] $\mathcal{R}^{1}_{\textbf{W}}=\dfrac{g_{T}(\textbf{W})}{\delta_{T}}$: is the primary production of tree biomass relative to tree biomass loss by herbivory (browsing) or human action.
		\item[(ii)] $\mathcal{R}^{2}_{\textbf{W}}=\dfrac{g_{G}(\textbf{W})}{\delta_{G}+\lambda_{fG}f}$: represents the primary production of grass biomass relative to fire-induced biomass loss and additional loss due to herbivory (grazing) or human action.
		\item[(iii)] $\mathcal{R}_{F}=\dfrac{g_{G}(\textbf{W})}{\eta_{TG}T^{*}+\delta_{G}+\lambda_{fG}f}$: denotes the primary production of grass biomass, relative to fire-induced grass biomass loss and additional loss due to tree/grass interactions and  to herbivory (grazing) or human action at the close forest equilibrium.
		\item[(iv)] $\mathcal{R}_{G}=\dfrac{g_{T}(\textbf{W})}{\delta_{T}+\lambda_{fT}^{max}f\omega(G^{*})}$: is the primary production of tree biomass relative   to fire-induced biomass loss at the grassland equilibrium and additional loss due to herbivory (browsing) or human action.
	\end{itemize}
	\label{thresholds_ecolo_meaning_1}
\end{rmq}

The long-term behavior of system (\ref{swv_eq1}) is summarizes in Table \ref{swv_tab_2}. The dynamic can change according to the previous thresholds.

\begin{table}[H]
	{\normalsize
		\begin{center}
			\caption{Long term dynamic of the deterministic system (\ref{swv_eq1})}
			\renewcommand{\arraystretch}{1.2}
			\begin{tabular}{lccccccc}
				\cline{1-8}
				\multicolumn{5}{c}{\bf Thresholds} &  \multirow{2}{1.3cm}{\bf Stable} & \multirow{2}{1.3cm}{\bf Unstable} & \multirow{2}{0.7cm}{\bf Case}\\
				\cline{1-5}
				$\mathcal{R}^{1}_{\textbf{W}}$ ($\mathcal{R}^{2}_{\textbf{W}}$) &  $\mathcal{R}_{G}$ &  $\mathcal{R}_{F}$ & $\mathcal{R}_{*}^{1}$& $\mathcal{R}_{*}^{2}$ & & &  \\
				\hline
				$<1(<1)$ & ND & ND & NN & NN &$ \textbf{E}_{0}$  &  & $\textbf{I}$ \\
				\hline
				\multirow{7}{2cm}{$>1(>1)$} & $>1$ & $<1$ & \multirow{3}{0.65cm}{$<1$} & \multirow{3}{0.65cm}{$<1$} & $\textbf{E}_{F}$ & $\textbf{E}_{0}$, $\textbf{E}_{G}$ & $\textbf{II}$\\
				\cline{2-3} \cline{6-8} & $<1$ & $>1$ &  &  & $\textbf{E}_{G}$ & $\textbf{E}_{0}$, $\textbf{E}_{F}$ & $\textbf{III}$\\
				\cline{2-3} \cline{6-8} & $<1$ & $<1$ &  &  & $\textbf{E}_{G}$, $\textbf{E}_{F}$  & $\textbf{E}_{0}$  & $\textbf{IV}$\\
				\cline{2-8}
				&  $>1$ &$<1$ & \multirow{4}{0.65cm}{$<1$} & \multirow{4}{0.65cm}{$>1$} & $\textbf{E}_{F}$, $\textbf{E}_{S}$ & $\textbf{E}_{0}$, $\textbf{E}_{G}$ & $\textbf{V}$\\
				\cline{2-3} \cline{6-8}
				&  $<1$ &$>1$ &  &  & $\textbf{E}_{G}$, $\textbf{E}_{S}$ & $\textbf{E}_{0}$, $\textbf{E}_{F}$ & $\textbf{VI}$\\
				\cline{2-3} \cline{6-8}
				&  $>1$ &$>1$ &  &  &  $\textbf{E}_{S}$ & $\textbf{E}_{0}$, $\textbf{E}_{G}$, $\textbf{E}_{F}$ & $\textbf{VII}$\\
				\cline{2-3} \cline{6-8}
				&  $<1$ &$<1$ &  &  &  $\textbf{E}_{F}$, $\textbf{E}_{S}$, $\textbf{E}_{G}$ & $\textbf{E}_{0}$  & $\textbf{VIII}$\\
				\hline
			\end{tabular}	
			\label{swv_tab_2}
		\end{center}
	}
\end{table}

In Tables \ref{swv_tab_2} and \ref{pulsed_Sum_table}, the notation 'ND' stands for undefined and 'NN'  denotes  not necessary. It means that the threshold may take any value.

\subsection{The IDE model}
According to the theory of impulsive differential equation (e.g. \citealp{Bainov1995}, Theorem 1.1, page 3),
system (\ref{pulsed_swv_eq1})-(\ref{pulsed_swv_eq2}) has a unique positive solution. It is straightforward to show that the compact subset $\Gamma$ is positively invariant and attracting by (\ref{pulsed_swv_eq1}). It guarantees that model (\ref{pulsed_swv_eq1}) is  well-posed mathematically and ecologically realistic.\par 
Providing $g_{G}(\textbf{W})=\gamma_{G}$, $g_{T}(\textbf{W})=\gamma_{T}$, $K_{G}(\textbf{W})=K_{G}$ and $K_{T}(\textbf{W})=K_{T}$, the system (\ref{pulsed_swv_eq1}) is similar to the model developed and studied in  \citealp{Tchuinte2017}. Thus, the demonstrations of results are the same and therefore are omitted here. 
Below   some qualitative results of system (\ref{pulsed_swv_eq1}) are given. There are two constant equilibria and two periodic solutions. As in subsection \ref{math_swv_model}, the desert equilibrium $\textbf{E}_{0}=(0,0)$ always exists and there is a forest equilibrium $\textbf{E}_{F}=(0,T^{*})$, when $\mathcal{R}^{1}_{\textbf{W}}>1$, where

\begin{equation}
T^{*}=K_{T}(\textbf{W})\left(1-\dfrac{1}{\mathcal{R}^{1}_{\textbf{W}}}\right),
\label{T_etoil}
\end{equation}

\begin{equation}
\hspace{0.2cm}\mathcal{R}^{1}_{\textbf{W}}=\dfrac{g_{T}(\textbf{W})}{\delta_{T}}, 
\label{RW1}
\end{equation}
and $g_{T}(\textbf{W})=\displaystyle\frac{\gamma_{T}\textbf{W}}{b_{T}+\textbf{W}}$.\par

Concerning the existence of the periodic grassland solution, let us set:

\begin{equation}
\mathcal{R}_{0,pulse}^{\bar{G}}=\dfrac{r_{G}(\textbf{W})}{\dfrac{1}{\tau}\ln\left(\dfrac{1}{1-\lambda_{fG}}\right)},
\label{pulsed_swv_eq4}
\end{equation} 

and

\begin{equation}
\bar{G}_{per}(t)=\dfrac{\bar{G}(\textbf{W})[(1-\lambda_{fG})e^{r_{G}(\textbf{W})\tau}-1]e^{r_{G}(\textbf{W})(t-n\tau)}}{[(1-\lambda_{fG})e^{r_{G}(\textbf{W})\tau}-1]e^{r_{G}(\textbf{W})(t-n\tau)}+\lambda_{fG}e^{r_{G}(\textbf{W})\tau}}, \hspace{0.25cm} t\in [n\tau, (n+1)\tau[, \hspace{0.25cm}n=0,1,2...,
\label{pulsed_swv_eq5}
\end{equation} 

where,   $\bar{G}(\textbf{W})=\left(1-\dfrac{1}{\bar{\mathcal{R}}^{2}_{\textbf{W}}}\right)K_{G}(\textbf{W})$,  

\begin{equation}
\bar{\mathcal{R}}^{2}_{\textbf{W}}=\dfrac{g_{G}(\textbf{W})}{\delta_{G}}, 
\label{RW2_bar}
\end{equation}

$ g_{G}(\textbf{W})=\displaystyle\frac{\gamma_{G}\textbf{W}}{b_{G}+\textbf{W}}$ and $ r_{G}(\textbf{W})=\left(1-\dfrac{1}{\bar{\mathcal{R}}^{2}_{\textbf{W}}}\right)g_{G}(\textbf{W})$.

The following theorem holds:

\begin{thm} (Existence of the periodic grassland solution)\par
	When $ \mathcal{R}_{0,pulse}^{\bar{G}}>1$,  system (\ref{pulsed_swv_eq1}) has a periodic grassland solution $\textbf{E}_{G,per}=(\bar{G}_{per}(t), 0)$, where $ \mathcal{R}_{0,pulse}^{\bar{G}}$ and $\bar{G}_{per}(t)$ are given by (\ref{pulsed_swv_eq4}) and (\ref{pulsed_swv_eq5}) respectively.
	\label{pulsed_svw_thm1}
\end{thm}
$\mbox{\bf{Proof:}}$ The proof of Theorem \ref{pulsed_svw_thm1} is similar to the proof of Theorem 3.1 given in  \citealp{Tchuinte2016} (see  Appendix B, page 18). \par

By employing the continuation theorem of the coincidence degree theory (\citealp{Gaines1977}, p 40), one can establish the existence of at least one positive periodic savanna solution of system (\ref{pulsed_swv_eq1}). Set 

\begin{equation}
\mathcal{R}_{\lambda^{max}_{fT}}^{*}=\dfrac{r_{T}(\textbf{W})}{\dfrac{1}{\tau}\ln\left(\dfrac{1}{1-\lambda^{max}_{fT}\omega(G^{*}_{s})}\right)},
\label{pulsed_swv_eq6}
\end{equation}

and 

\begin{equation}
\mathcal{R}_{\lambda^{max}_{fT}}^{**}=\dfrac{r_{T}(\textbf{W})}{\dfrac{1}{\tau}\ln\left(\dfrac{1}{1-\lambda^{max}_{fT}\vartheta(T^{*}_{s})\omega(G^{*}_{s})}\right)},
\label{pulsed_swv_eq7}
\end{equation} 

where, $r_{T}(\textbf{W})=\left(1-\dfrac{1}{\mathcal{R}^{1}_{\textbf{W}}}\right)g_{T}(\textbf{W})$, and

\begin{equation}
G^{*}_{s}=\left(1-\dfrac{\delta_{G}}{g_{G}(\textbf{W})}\right)\left(1-\dfrac{1}{ \mathcal{R}_{0,pulse}^{\bar{G}}}\right)K_{G}(\textbf{W}),
\label{pulsed_swv_eq8}
\end{equation} 

where $\mathcal{R}_{0,pulse}^{\bar{G}}$ is given by (\ref{pulsed_swv_eq4}) and $T^{*}_{s}$ is the  positive solution of (\ref{pulsed_swv_eq9})

\begin{equation}
(g_{T}(\textbf{W})-\delta_{T})-\dfrac{g_{T}(\textbf{W})}{K_{T}(\textbf{W})}T^{*}_{s} +\dfrac{1}{\tau}\ln\left(1-\vartheta(T^{*}_{s})\omega(G^{*}_{s})\right)=0.
\label{pulsed_swv_eq9}
\end{equation}

Similarly as in  \citealp{Tchuinte2017} (see Theorem 3.2, p 270), we claim the following result.

\begin{thm} (Existence of a periodic savanna solution).
	Set the following conditions:
	\begin{itemize}
		\item[$(C1)$] $\mathcal{R}_{0,pulse}^{\bar{G}}>1$ (similar as in  \citealp{Tchuinte2016}; \citealp{Tchuinte2017}),
		\item[$(C2)$] $\mathcal{R}_{\lambda^{max}_{fT}}^{*}>1$, 
		\item[$(C3)$] $\mathcal{R}_{\lambda^{max}_{fT}}^{**}>1$.
	\end{itemize}	
	If $(C1)$, $(C2)$ and $(C3)$ hold, then system (\ref{pulsed_swv_eq1}) has at least one positive periodic savanna solution $\textbf{E}_{S,per}=(G^{*}_{per}(t), T^{*}_{per}(t))$.
	\label{pulsed_svw_thm2}
\end{thm}
$\mbox{\bf{Proof:}}$ Proof is entirely similar as the proof of Theorem 3.2 in  \citealp{Tchuinte2017} (see Appendix A, p 281) and is thus omitted.\par

We will now examine the stability of the constant equilibria and the periodic solutions of (\ref{pulsed_swv_eq1}). Set

\begin{equation}
\mathcal{R}^{\bar{G}}_{0,pulse}=\dfrac{g_{G}(\textbf{W})\left(1-\dfrac{1}{\bar{\mathcal{R}}_{\textbf{W}}^{2}}\right)}{\dfrac{1}{\tau}\ln\left(\dfrac{1}{1-\lambda_{fG}}\right)},
\label{RE0_pulse}
\end{equation}

\begin{equation}
\bar{\mathcal{R}}_{F}=\dfrac{g_{G}(\textbf{W})}{\eta_{TG}T^{*}+\delta_{G}},
\label{RF_bar}
\end{equation}

and 

\begin{equation}
\mathcal{R}^{F}_{0,pulse}=\dfrac{g_{G}(\textbf{W})\left(1-\dfrac{1}{\bar{\mathcal{R}}_{F}}\right)}{\dfrac{1}{\tau}\ln\left(\dfrac{1}{1-\lambda_{fG}}\right)},
\label{RF_pulse}
\end{equation}

where $T^{*}$ and $\bar{\mathcal{R}}_{\textbf{W}}^{2}$ are given by (\ref{T_etoil}) and (\ref{RW2_bar}) respectively.
\par 
Concerning the stability of  constant equilibria, we show Theorems \ref{pulsed_svw_thm3} and \ref{pulsed_svw_thm4}.

\begin{thm} (Stability of the desert  equilibrium).
	\begin{itemize}
		\item[(i)] If $\mathcal{R}_{\textbf{W}}^{1}<1$ and $\bar{\mathcal{R}}_{\textbf{W}}^{2}<1$, then the desert equilibrium $\textbf{E}_{0}=(0, 0)$ is locally asymptotically stable (LAS).
		\item[(ii)] If $\mathcal{R}_{\textbf{W}}^{1}>1$ or $\mathcal{R}^{\bar{G}}_{0,pulse}>1$, then  the desert equilibrium  is unstable.
		\item[(iii)] If $\mathcal{R}_{\textbf{W}}^{1}<1$, $\bar{\mathcal{R}}_{\textbf{W}}^{2}>1$, and $\mathcal{R}^{\bar{G}}_{0,pulse}<1$, then the desert equilibrium is LAS.
	\end{itemize}
	\label{pulsed_svw_thm3}
\end{thm}

\begin{thm} (Stability of the forest  equilibrium).
	\begin{itemize}
		\item[(i)] If $\mathcal{R}_{\textbf{W}}^{1}<1$, then the forest equilibrium $\textbf{E}_{F}=(0, T^{*})$ is unstable.
		\item[(ii)] If $\mathcal{R}_{\textbf{W}}^{1}>1$ and $\bar{\mathcal{R}}_{F}<1$, then  $\textbf{E}_{F}$ is LAS.
		\item[(iii)] If $\mathcal{R}_{\textbf{W}}^{1}>1$, $\bar{\mathcal{R}}_{F}>1$, and $\mathcal{R}^{F}_{0,pulse}<1$, then $\textbf{E}_{F}$ is LAS.
		\item[(iv)] If $\mathcal{R}_{\textbf{W}}^{1}>1$, $\bar{\mathcal{R}}_{F}>1$, and $\mathcal{R}^{F}_{0,pulse}>1$, then $\textbf{E}_{F}$ is unstable.
	\end{itemize}
	\label{pulsed_svw_thm4}
\end{thm}

$\mbox{\bf{Proof:}}$ 
The proofs of Theorems \ref{pulsed_svw_thm3} and \ref{pulsed_svw_thm4} are similar to the proof of Theorem 3.3  given in  \citealp{Tchuinte2017} (see  Appendix B, page 25). \par

The local stability of the periodic solutions are obtained by computing the Floquet multipliers of the monodromy matrix of (\ref{pulsed_swv_eq1}). Set $(G_{per}(t), T_{per}(t))$ a periodic solution of (\ref{pulsed_swv_eq1}). Consider the behaviors of small perturbations of solutions $u(t)=G(t)-G_{per}(t)$ and $v(t)=T(t)-T_{per}(t)$. The linearized system is given by:

\begin{equation}
\begin{pmatrix}
u(t)\\
v(t)
\end{pmatrix}=\Phi(t)
\begin{pmatrix}
u(0)\\
v(0)
\end{pmatrix},
\label{monodromy_eq1}
\end{equation} 
where $\Phi(t)$	is a fundamental matrix which satisfies 

\begin{equation}
\left\{
\begin{array}{l}
\dfrac{d\Phi(t)}{dt}=\begin{pmatrix}
\mathcal{A}_{11}(t)&\mathcal{A}_{12}(t)\\
\mathcal{A}_{21}(t)&\mathcal{A}_{22}(t)\\
\end{pmatrix}\Phi(t),\\
\\
\Phi(0)=I_{2},
\end{array}
\right.
\label{monodromy_eq2}
\end{equation}
where, $I_{2}$ is the identity matrix of $\mathcal{M}_{2}(\mathbf{R})$,

\begin{equation}
\left\{
\begin{array}{lcl}
\mathcal{A}_{11}(t)=g_{G}(\textbf{W})\left(1-\dfrac{1}{\bar{\mathcal{R}}_{\textbf{W}}^{2}}\right)-\dfrac{2g_{G}(\textbf{W})}{K_{G}(\textbf{W})}G_{per}(t)-\eta_{TG}T_{per}(t),\\
\mathcal{A}_{21}(t)=0,\\
\mathcal{A}_{12}(t)=-\eta_{TG}G_{per}(t),\\
\mathcal{A}_{22}(t)=g_{T}(\textbf{W})\left(1-\dfrac{1}{\mathcal{R}_{\textbf{W}}^{1}}\right)-\dfrac{2g_{T}(\textbf{W})}{K_{T}(\textbf{W})}T_{per}(t).
\end{array}
\right.
\label{monodromy_eq3}
\end{equation}	
Concerning the  stability of the periodic grassland and the periodic savanna solutions, let us set

\begin{equation}
\mathcal{R}^{\bar{G}_{per}}_{0,pulse}=\dfrac{g_{T}(\textbf{W})\left(1-\dfrac{1}{\mathcal{R}_{\textbf{W}}^{1}}\right)}{\dfrac{1}{\tau}\ln\left(\dfrac{1}{1-\lambda^{max}_{fT}\omega(\bar{G}_{per}(\tau))}\right)},
\label{RGbar_pulse}
\end{equation}

\begin{equation}
\mathcal{R}^{*}_{S,pulse}=\dfrac{1}{\mathcal{R}^{\bar{G}}_{0,pulse}}+\dfrac{2}{\bar{G}(\textbf{W})}\left(\dfrac{1}{\tau}\int\limits_{0}^{\tau}G^{*}_{per}(s)ds\right)+\dfrac{1}{\mathcal{R}^{*}_{S}}\left(\dfrac{1}{\tau}\int\limits_{0}^{\tau}T^{*}_{per}(s)ds\right),
\label{R*S_pulse}
\end{equation}

and 

\begin{equation}
\mathcal{R}^{**}_{S,pulse}=\dfrac{1}{\mathcal{R}^{\vartheta\omega}_{0,stable}}+\dfrac{2}{T^{*}}\left(\dfrac{1}{\tau}\int\limits_{0}^{\tau}T^{*}_{per}(s)ds\right),
\label{R**S_pulse}
\end{equation}

where, $\bar{G}_{per}(\tau)$ is the grass biomass   at the periodic grassland solution when $t=\tau$ (see (\ref{pulsed_swv_eq5})), $\mathcal{R}^{\bar{G}}_{0,pulse}$ is given by (\ref{RE0_pulse}), $\bar{G}(\textbf{W})=\left(1-\dfrac{1}{\bar{\mathcal{R}}^{2}_{\textbf{W}}}\right)K_{G}(\textbf{W})$,   $T^{*}=K_{T}(\textbf{W})\left(1-\dfrac{1}{\mathcal{R}_{\textbf{W}}^{1}}\right)$, $\mathcal{R}^{*}_{S}=\dfrac{g_{G}(\textbf{W})\left(1-\dfrac{1}{\bar{\mathcal{R}}_{\textbf{W}}^{2}}\right)}{\eta_{TG}}$, and $\mathcal{R}^{\vartheta\omega}_{0,stable}=\dfrac{g_{T}(\textbf{W})\left(1-\dfrac{1}{\mathcal{R}_{\textbf{W}}^{1}}\right)}{\dfrac{1}{\tau}\ln\left(\dfrac{1}{1-\vartheta(T^{*}_{per}(\tau))\omega(G^{*}_{per}(\tau))}\right)}$, where $G^{*}_{per}(\tau)$ and $T^{*}_{per}(\tau)$ are grass biomass and tree biomass at the periodic savanna solution when $t=\tau$.\par 
We show  the following results.

\begin{thm} (Stability of the periodic grassland solution).
	\begin{itemize}
		\item[(i)] If $\mathcal{R}^{\bar{G}}_{0,pulse}>1$ and  $\mathcal{R}^{\bar{G}_{per}}_{0,pulse}<1$, then the periodic grassland solution $\textbf{E}_{G,per}=(\bar{G}_{per}(t), 0)$ is LAS.
		\item[(ii)] If $\mathcal{R}^{\bar{G}}_{0,pulse}>1$ and  $\mathcal{R}^{\bar{G}_{per}}_{0,pulse}>1$, then the periodic grassland solution is unstable.
	\end{itemize}
	\label{pulsed_svw_thm5}
\end{thm}

$\mbox{\bf{Proof:}}$
The proof of Theorem \ref{pulsed_svw_thm5} is similar to the proof of Theorem 3.4  given in  \citealp{Tchuinte2017} (see  Appendix C, page 26). \par

\begin{thm} (Stability of the periodic savanna solution).
	\begin{itemize}
		\item[(i)] If $\mathcal{R}^{*}_{S,pulse}>1$ and $\mathcal{R}^{**}_{S,pulse}>1$, then the periodic savanna solution $\textbf{E}_{S,per}=(G^{*}_{per}(t), T^{*}_{per}(t))$ is LAS.
		\item[(ii)] If $\mathcal{R}^{*}_{S,pulse}<1$ or $\mathcal{R}^{**}_{S,pulse}<1$, then $\textbf{E}_{S,per}$ is unstable.
	\end{itemize}	
	\label{pulsed_svw_thm6}
\end{thm}

$\mbox{\bf{Proof:}}$ The proof of Theorem \ref{pulsed_svw_thm6} is similar to the proof of Theorem 3.5  given in  \citealp{Tchuinte2017} (see  Appendix D, page 27). \par

Based on the previous results, we summarize the long-term behaviors of system (\ref{pulsed_swv_eq1}) in   Table \ref{pulsed_Sum_table}.

\begin{table}[H]
	{\footnotesize
		\begin{center}
			\caption{Long term dynamic of the semi-discrete system (\ref{pulsed_swv_eq1})}
			\renewcommand{\arraystretch}{1.2}
			\begin{tabular}{lccccccccc}
				\cline{1-10}
				\multicolumn{7}{c}{\bf Thresholds} &  \multirow{2}{1.3cm}{\bf Stable} & \multirow{2}{1.5cm}{\bf Unstable} & \multirow{2}{0.7cm}{\bf Case}\\
				\cline{1-7}
				$\mathcal{R}^{1}_{\textbf{W}}$ $(\bar{\mathcal{R}}^{2}_{\textbf{W}})$ &  $\mathcal{R}^{\bar{G}}_{0,pulse}$ &  $\bar{\mathcal{R}}_{F}$ & $\mathcal{R}^{F}_{0,pulse}$  & $\mathcal{R}^{\bar{G}_{per}}_{0,pulse}$  & $\mathcal{R}^{*}_{S,pulse}$& $\mathcal{R}^{**}_{S,pulse}$ & & &  \\
				\hline
				\multirow{2}{1.5cm}{$<1$ (NN)} &  \multirow{2}{1cm}{$<1$} &  \multirow{2}{1cm}{ND} &  \multirow{2}{1cm}{ND} & \multirow{2}{1cm}{NN} & \multirow{2}{1cm}{NN}  & \multirow{2}{1cm}{NN}  & \multirow{2}{0.5cm}{$\textbf{E}_{0}$}  & \multirow{2}{0.5cm}{}  & \multirow{2}{0.5cm}{$\textbf{I}$} \\
				& & & & & & & &  & \\ 
				\hline
				\multirow{10}{1.5cm}{$>1$$(>1)$} & 	\multirow{10}{1cm}{$>1$} & \multirow{2}{1cm}{$<1$} & \multirow{2}{1cm}{NN} & \multirow{2}{1cm}{$>1$} & \multirow{5}{1cm}{$<1$} & \multirow{5}{1cm}{$<1$} & \multirow{2}{0.5cm}{$\textbf{E}_{F}$} & $\textbf{E}_{0}$, $\textbf{E}_{G,per}$ & \multirow{2}{0.5cm}{$\textbf{II}$} \\ 
				& & & & & & & & $\textbf{E}_{S,per}$& \\
				\cline{3-5}\cline{8-10}	& & \multirow{4}{1cm}{$>1$} & \multirow{2}{1cm}{$>1$} & \multirow{3}{1cm}{$<1$} & & & \multirow{2}{0.5cm}{$\textbf{E}_{G,per}$} & $\textbf{E}_{0}$, $\textbf{E}_{F}$ & \multirow{2}{0.5cm}{$\textbf{III}$}\\
				& & & & & & & & $\textbf{E}_{S,per}$& \\  
				\cline{4-4}\cline{8-10} & & & \multirow{3}{1cm}{$<1$} & & & & $\textbf{E}_{F}$, $\textbf{E}_{G,per}$ & $\textbf{E}_{0}$, $\textbf{E}_{S,per}$ & $\textbf{IV}$ \\
				\cline{5-10}	& & & & $>1$  & \multirow{5}{1cm}{$>1$} & \multirow{5}{1cm}{$>1$} &$\textbf{E}_{F}$, $\textbf{E}_{S,per}$  &$\textbf{E}_{0}$, $\textbf{E}_{G,per}$ & $\textbf{V}$ \\
				\cline{5-5}  \cline{8-10}	& & & & \multirow{2}{1cm}{$<1$}  &  &  &$\textbf{E}_{F}$, $\textbf{E}_{S,per}$, $\textbf{E}_{G,per}$  &$\textbf{E}_{0}$  & $\textbf{VI}$ \\
				\cline{3-4}\cline{8-10}& & \multirow{3}{1cm}{$<1$} & \multirow{3}{1cm}{$>1$} &  & & &$\textbf{E}_{G,per}$,   $\textbf{E}_{S,per}$& $\textbf{E}_{0}$, $\textbf{E}_{F}$ & $\textbf{VII}$ \\
				\cline{5-5}\cline{8-10}	& & & & \multirow{2}{1cm}{$>1$} & & &  \multirow{2}{0.5cm}{$\textbf{E}_{S,per}$}& $\textbf{E}_{0}$, $\textbf{E}_{F}$  & \multirow{2}{0.5cm}{$\textbf{VIII}$} \\  
				& & & & & & & & $\textbf{E}_{G,per}$ & \\ 			
				\hline
			\end{tabular}	
			\label{pulsed_Sum_table}
		\end{center}
	}
\end{table}

\section{Numerical simulations and discussion} \label{section4}

Like in our  previous works (\citealp{Tchuinte2014}; \citealp{Tchuinte2017}) we  use the nonstandard finite difference (NSFD) scheme    to solve systems (\ref{swv_eq1}) and (\ref{pulsed_swv_eq1}) numerically and  provide simulations that strictly comply with the properties of the systems. The nonstandard approach has shown to be very effective to solve dynamical systems in biosciences (see for instance  \citealp{Anguelov2012} and references therein). Parameter values used for numerical simulations are given in Table \ref{params_figR2_fig1}.

\begin{table}[H]
	{\footnotesize 
		\begin{center}
			\caption{ {\small Parameter values related to Fig. \ref{swv_fig2}-(b) and Figs. \ref{bistability_fig}$-$\ref{fig2_bif_W}. 
					.}}
			\renewcommand{\arraystretch}{1.2}
			\begin{tabular}{cccccc}
				\hline
				$c_{G}$,  t.ha$^{-1}$   & $c_{T}$,  t.ha$^{-1}$  & $b_{G}$, mm.yr$^{-1}$ & $b_{T}$, mm.yr$^{-1}$  &  $a_{G}$, yr$^{-1}$ & $a_{T}$, yr$^{-1}$ \\
				\hline
				$20$ & $450$  & $501$ & $1192$ & $0.0029$ & $0.0045$ \\
				\hline
				$d_{G}$, $-$	& $d_{T}$, $-$ & $\gamma_{G}$, yr$^{-1}$ & $\gamma_{T}$, yr$^{-1}$ & $\delta_{G}$, yr$^{-1}$ & $\delta_{T}$, yr$^{-1}$\\
				\hline
				$14.73$		& $106.7$ & $2.5$ & $1$ & $0.01$ & $0.1$\\
				\hline
				$\lambda_{fG}$, $-$ & $\lambda_{fT}^{min}$, $-$ & $\lambda_{fT}^{max}$, $-$ & $p$, t$^{-1}$ha & $\alpha$, t.ha$^{-1}$ &$\eta_{TG}$, ha.t$^{-1}$yr$^{-1}$\\
				\hline
				$0.3$ & $0.05$ & $0.7$ & $0.01$ & $1$ & $0.01$  \\
				\hline
			\end{tabular}
			\label{params_figR2_fig1}
		\end{center}}
	\end{table}

	\subsection{Bifurcation map}
	
\begin{figure}[H]
		\centering
		\subfloat[][]{\includegraphics[scale=0.45]{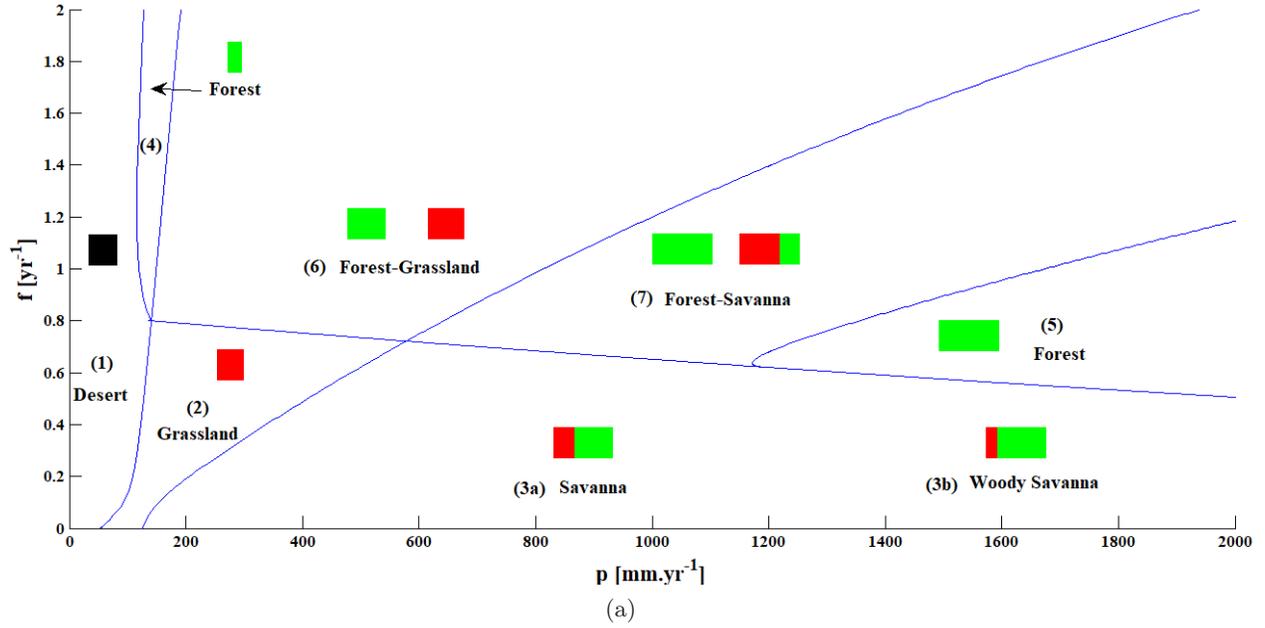}}
		\vspace{0.5cm}
		\subfloat[][]{\includegraphics[scale=0.45]{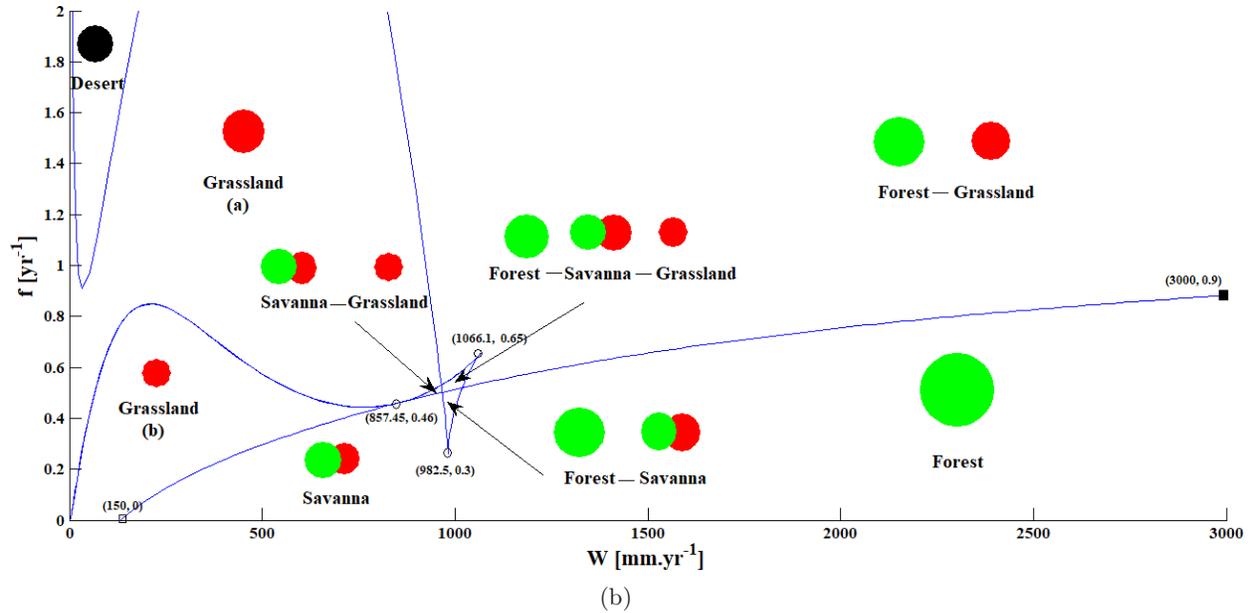}}	
		\caption{{\scriptsize Bifurcation diagrams using Matcont.   (a) reproduces the diagram of the   \citealp{Accatino2010tree} model (see figure 2 in  \citealp{Accatino2010tree} for parameter values used).  (b) is the diagram of system (\ref{swv_eq1}).
				Single red, green and black rectangles in panel (a) (dots in panel (b))  stand for  grassland, forest and desert  respectively. Twinned red and green symbols stand for savanna (coexistence state). Size of the symbols qualitatively denote grass and tree covers in panel (a) and biomass levels in panel (b).  The  parameter values used for Fig. \ref{swv_fig2}-(b) are given in Table \ref{swv_tab_2}.
				(Color in the online version).}}
		\label{swv_fig2}
	\end{figure}

	A bifurcation is a  qualitative change of the behaviors of the system when some parameters cross particular values. 
	Following  \citealp{Accatino2010tree}, we use a graphical Matlab   toolbox (\citealp{Matlab}) called Matcont to build the bifurcation diagrams of our model. Matcont is a useful software which allows computing numerical continuation of equilibria, limit points, Hopf points, branch points
	of equilibria, limit cycles, fold, flip, torus and branch point bifurcation
	points of limit cycles, and homoclinic orbits of dynamical systems. It is freely available online according to
	http://www.matcont.ugent.be/. One can use Matcont to compute a codimension 1 (i.e., one control parameter) or codimension 2 bifurcations.
	Here, we perform a bifurcation diagram of codimension 2 in the environmental space  defined by the mean annual precipitation (MAP) $\textbf{W}$ and  fire frequency $f$ (see Fig. \ref{swv_fig2}). 
	The readers can refer to  \citealp{Dhooge2003Matcont}; \citealp{Dhooge2006Matcont};  \citealp{Govaerts2007Matcont} and references therein for more details and examples on numerical methods  for two-parameter bifurcation analysis in Matcont. See Appendix C, page \pageref{al_AppendixC} for a condensed overview of explanations of how  model (\ref{swv_eq1}) is implemented into MatCont.
	\par

	The bifurcation diagram of model (\ref{swv_eq1}) is provided in Fig. \ref{swv_fig2}-(b) in the two dimensional parameter space ($f$ vs. $\textbf{W}$)  using Matcont. Depending on those parameters, the system involves both monostable and multi-stable situations. Once a bifurcation point in the parameter space (see e.g., cusp points represented by circles) is found, Matcont produces the entire bifurcation curve passing through that point.  Blue curves which partition the parameter space into subregions are bifurcation curves. Trajectories starting in the same subregion have qualitatively the same dynamics. It is also important to note that when the mean annual precipitation (MAP) increases, we logically observe a gradual increase of tree biomass for any given level of $f$. For instance in Fig. \ref{swv_fig2}-(b), considering low values of $f$ (say $< 0.3$) the savanna cede pace to forest as W increases. On the other hand,  several scenarios are possible when the fire frequency increases depending on the range of \textbf{W} values. Fig. \ref{swv_fig2}-(a) shows the complete dynamics of the   \citealp{Accatino2010tree} model and Fig. \ref{swv_fig2}-(b) shows the complete long term dynamics of system (\ref{swv_eq1}) in specified (rainfall, fire frequency) parameter ranges. There are two main observations: (i) Fig. \ref{swv_fig2}-(b) shows a bistability between grassland  and savanna (see also Fig. \ref{bistability_fig}-(b) for an illustration) and a tristability between forest, savanna and grassland (see also Fig. \ref{al_bifurc_S_G_F_ode}-(a)) while these subregions do not exist in Fig. \ref{swv_fig2}-(a); (ii)  in Fig. \ref{swv_fig2}-(a), when fire frequency decreases the system does not favor tree  expansion whatever the   context (semi-arid, mesic and humid), while  an opposite trend is observed in Fig. \ref{swv_fig2}-(b). For instance in Fig. \ref{swv_fig2}-(a) between ca. 600--1200 mm.yr$^{-1}$  the  \citealp{Accatino2010tree} system can shift from a bistability between forest and savanna (see the subregion (7)) to a monostability of savanna (see the subregion (3a)). Moreover above ca. 1200 mm.yr$^{-1}$ their system can shifts from forest (see the subregion (5)) to  savanna woodland (see the subregion (3b)). On the contrary, in Fig. \ref{swv_fig2}-(b), decreasing the fire frequency favors the tree expansion whatever the climatic context. This is in agreement with what is observed in the field as suggested by several empirical studies (\citealp{Mitchard2009measuring}; \citealp{Bond2010beyond}; \citealp{Favier2012abrupt}; \citealp{Mitchard2013woody} for a review). For instance in  Fig. \ref{swv_fig2}-(b), between ca. $600--1200$ mm.yr$^{-1}$ model (\ref{swv_eq1}) predicts three scenarios: In scenario 1 the system can shift from grassland to savanna, in scenario 2 the system can shift from  grassland to monostability of a savanna woodland passing through a bistability between grassland and savanna (see also Fig. \ref{bistability_fig} for trajectories illustrating this transition). In scenario 3, the system can shift from a bistability between forest and grassland to forest passing through a tristability between forest, savanna and grassland and a bistability between forest and savanna (see also Fig. \ref{al_bifurc_S_G_F_ode} for an illustration of this transition). Above ca. 1200 mm.yr$^{-1}$   the system can shift from a bistability between forest and grassland to a monostability of forest. 
	
	\subsection{Illustrations and discussion}
	Throughout this subsection, the simulations provided  with our minimalistic ODE model (\ref{swv_eq1}) and IDE model (\ref{pulsed_swv_eq1}) show the well-known fact that increasing fire return period systematically implies an increase in woody biomass. This is pivotal when one describes the tree-grass dynamics in fire-prone savanna ecosystems (\citealp{Bond2005global}; \citealp{Bond2010beyond}; \cite{Mitchard2013woody}).	 As far as we know,
	it is a completely new feature for   'minimalistic'  tree-grass ODE  models, that  has not been reported in previous models with only two state variables.\par 
	Though the model aims to be qualitatively relevant for a large swath of African situation, we ground our simulations in a selected north-south gradient  located at the $16$ \textdegree E of longitude, and between ca. $6$ and $10$ \textdegree N of latitude (i.e., between ca. $900$ to $1500$ mm.yr$^{-1}$ of MAP). The area goes from the Adamawa region to the Centre region in Cameroon and it spans the main vegetation physiognomies of  Central Africa.  Using longitude and latitude data, the MAP data were extracted from BIO12 (http://www.worldclim.org/bioclim, see also \citealp{Hijmans2005}) using the "raster" package of RStudio, version $1.1.383$.

	\begin{figure}[H]
		\centering
		\subfloat[][]{	\includegraphics[scale=0.38]{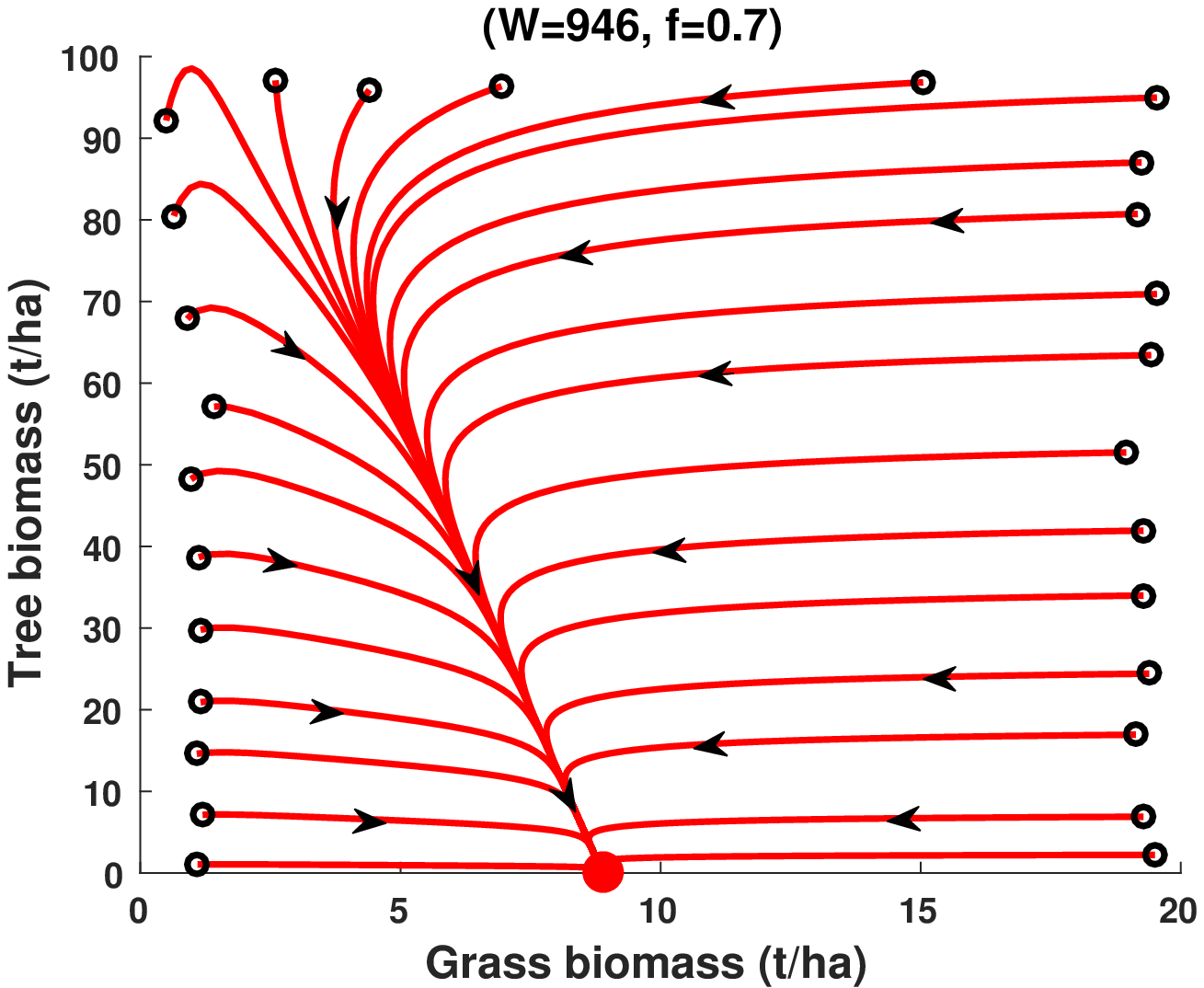}}
		\subfloat[][]{	\includegraphics[scale=0.38]{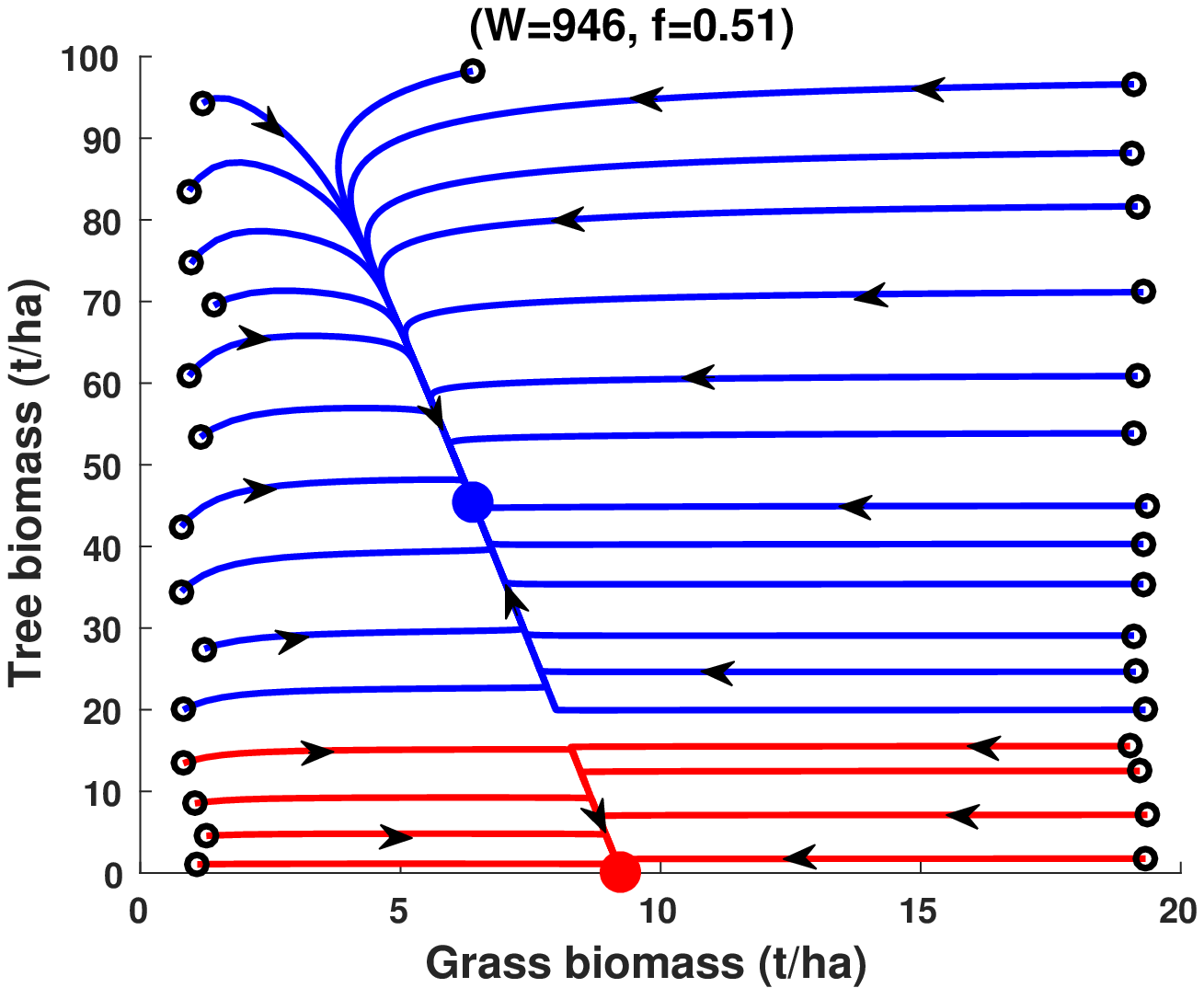}}
		\subfloat[][]{	\includegraphics[scale=0.38]{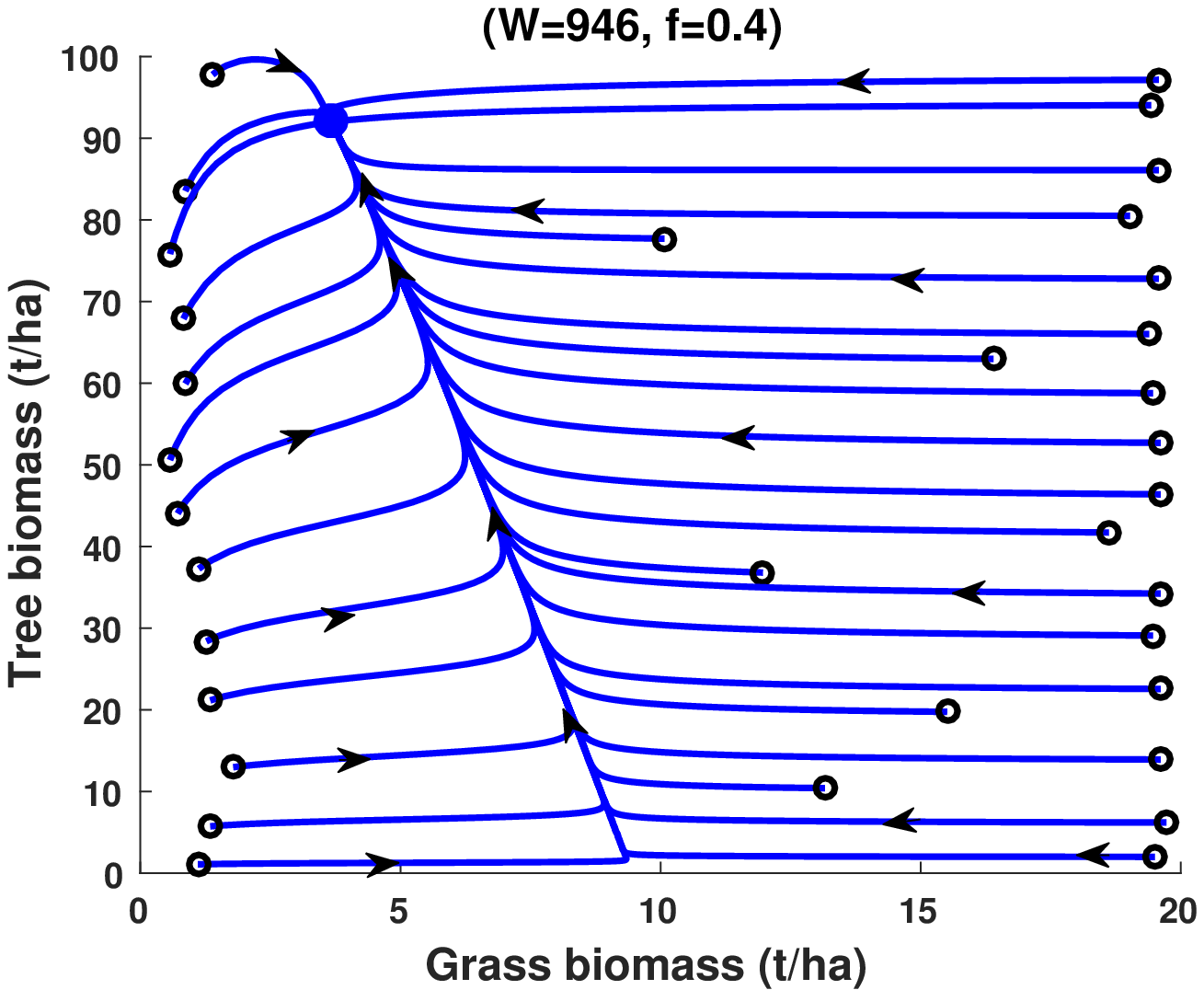}}	
		\caption{{\scriptsize Illustration of a bifurcation due to the fire frequency $f$ with the continuous model (\ref{swv_eq1}). Here,  $\textbf{W}=946$ mm.yr$^{-1}$. Open black circles are initial conditions. Blue and red dots are savanna and grassland equilibria respectively. (Color in the online version). }}
		\label{bistability_fig}
	\end{figure}

	In Fig. \ref{bistability_fig}, a decrease in the fire frequency $f$ from $0.7$ to $0.51$, then to $0.4$, leads  the ODE system  to shift from grassland state to a savanna-grassland bistable state, and then to a savanna woodland state under a constant $\textbf{W}$ (MAP) of ca. 950 mm.yr$^{-1}$.  
	\par

	\begin{figure}[H]
		\centering
		\subfloat[][]{	\includegraphics[scale=0.38]{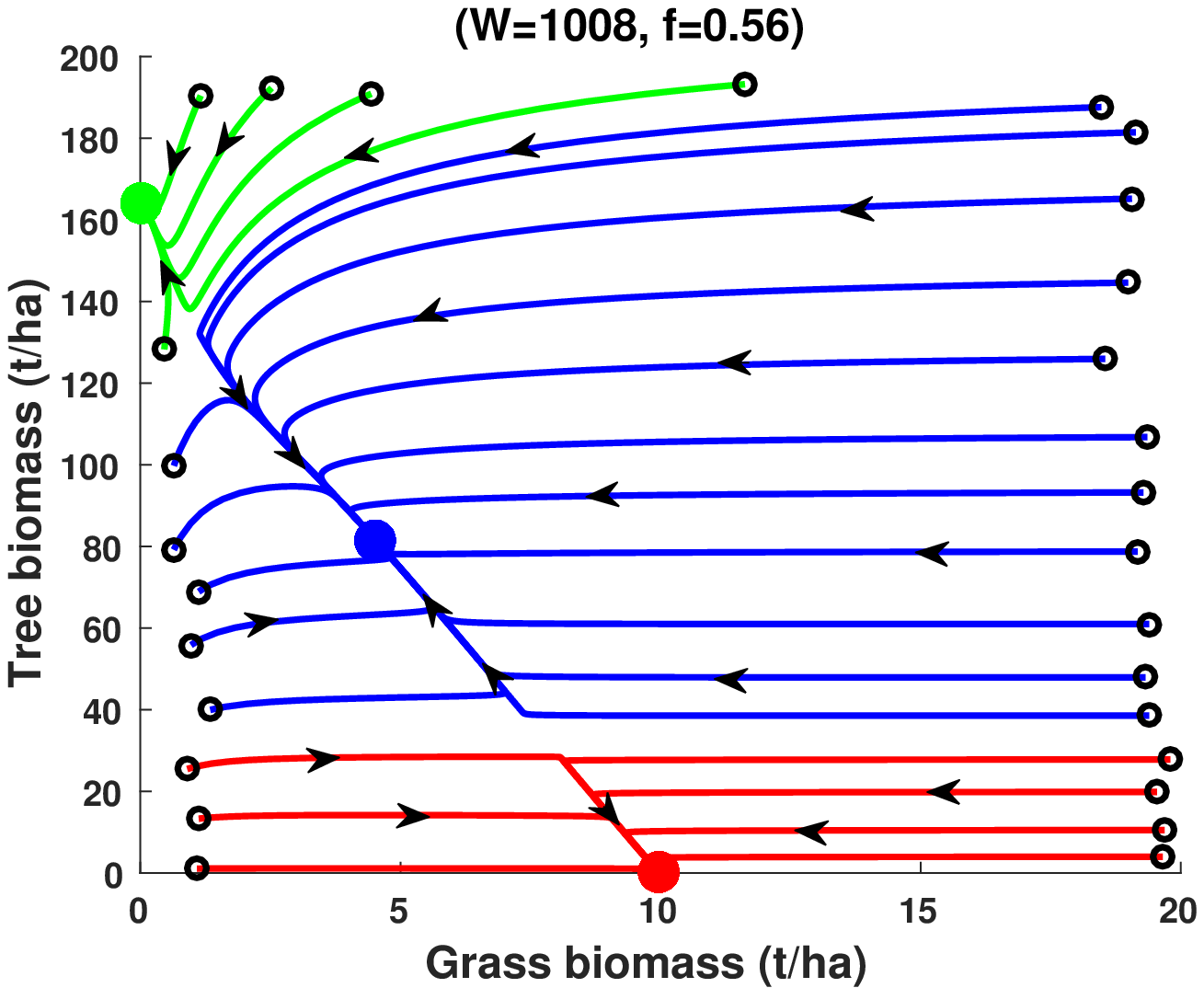}}
		\subfloat[][]{	\includegraphics[scale=0.38]{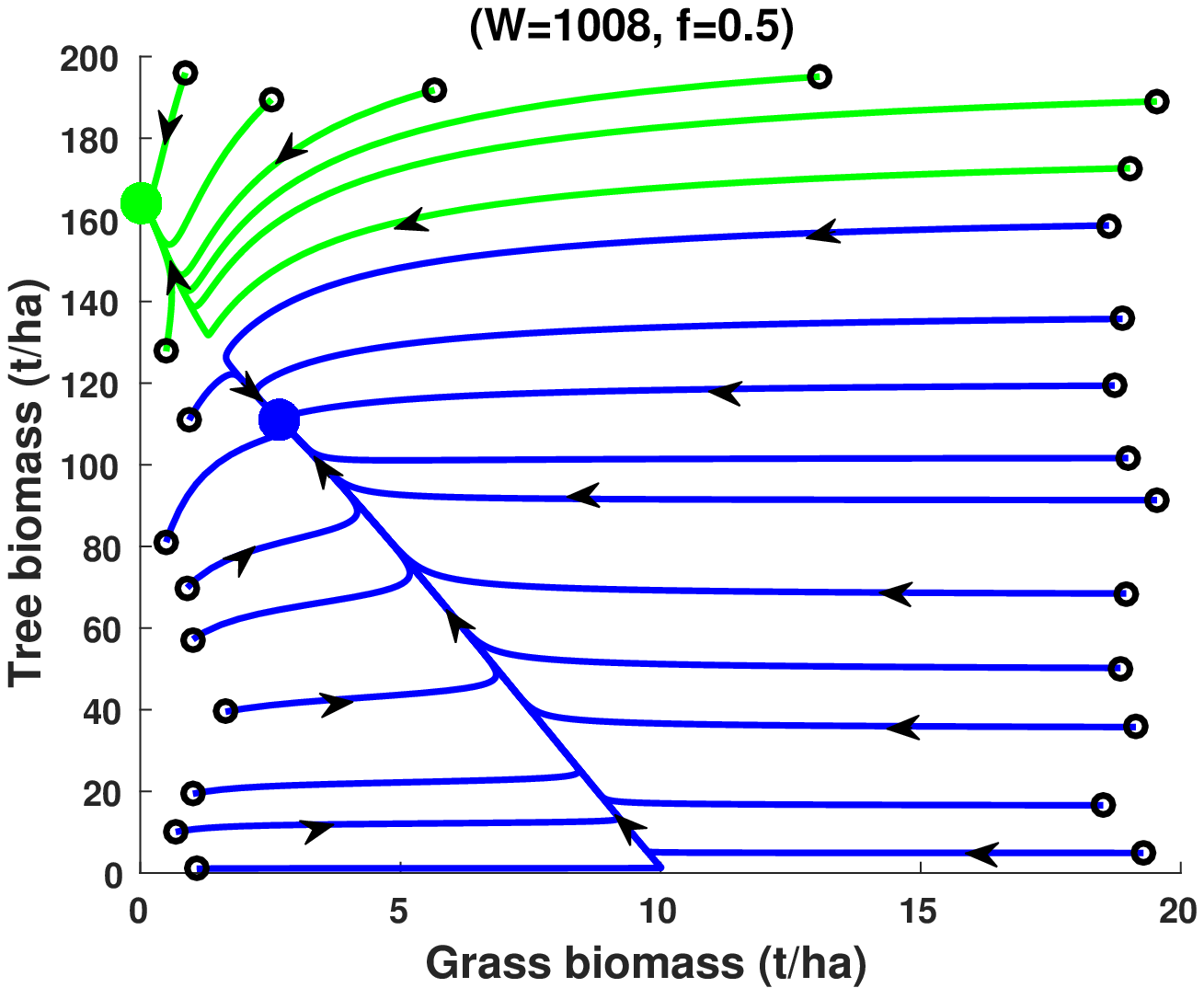}}
		\subfloat[][]{	\includegraphics[scale=0.38]{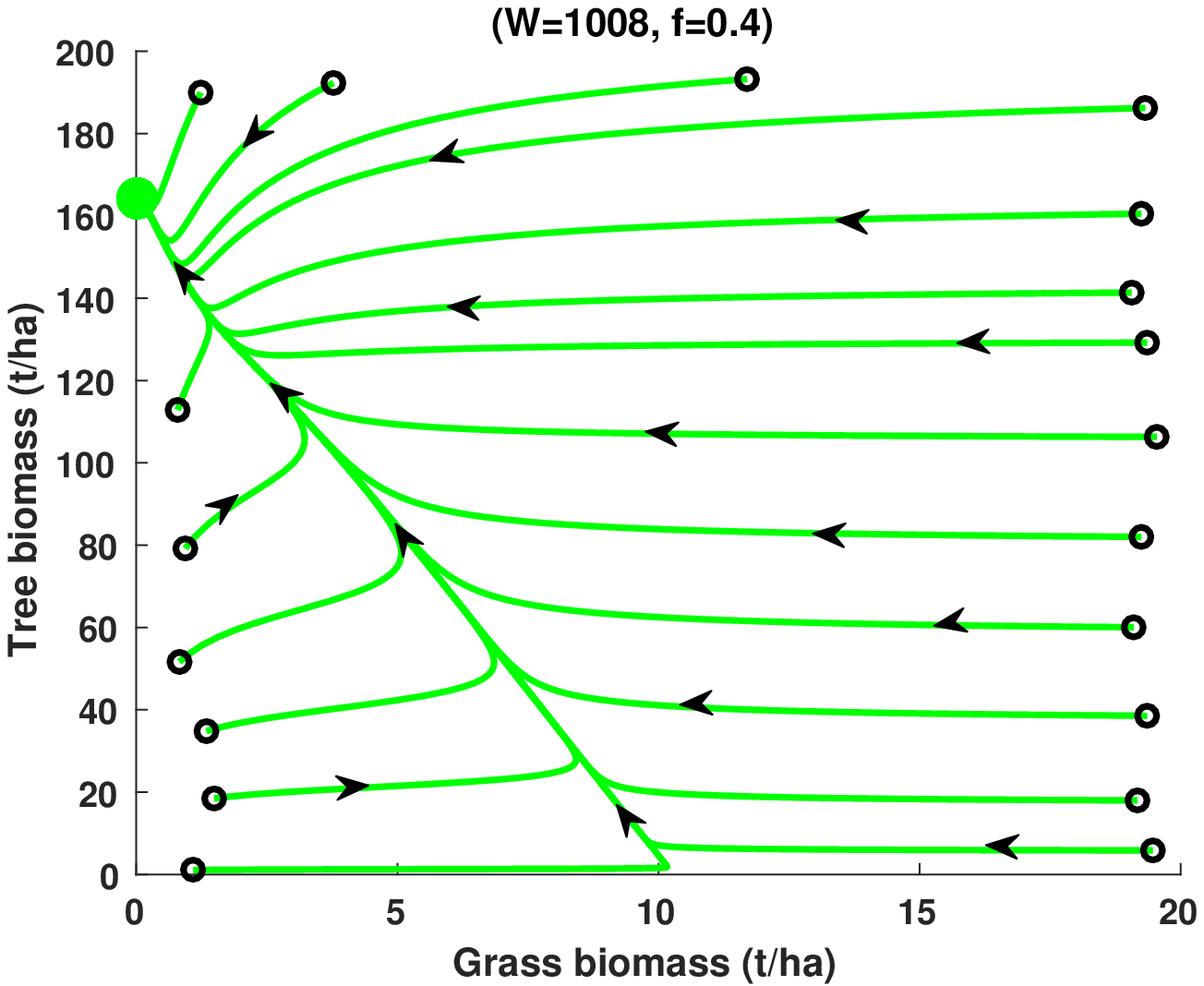}}
		\caption{{\scriptsize Bifurcation due to the fire frequency $f$ with the continuous model (\ref{swv_eq1}) for rainfall slightly above $1008$ mm.yr$^{-1}$.}}
		\label{al_bifurc_S_G_F_ode}
	\end{figure}

	Fig. \ref{al_bifurc_S_G_F_ode} illustrates a  bifurcation from a forest-savanna-grassland tristable state to a forest state passing through a forest-savanna bistable state,  for a constant MAP $\textbf{W}$ of ca. 1000 mm.yr$^{-1}$ under a decrease of fire frequency $f$. Domains of stability for grassland, savanna and forest are in red, blue and green, respectively.  Forest expansion has been regularly  observed at local and regional scales when the fire frequency decreases (see  \citealp{Mitchard2013woody} for a review).\par 
	
\begin{figure}[H]
		\centering
		\subfloat[][]{\includegraphics[scale=0.5]{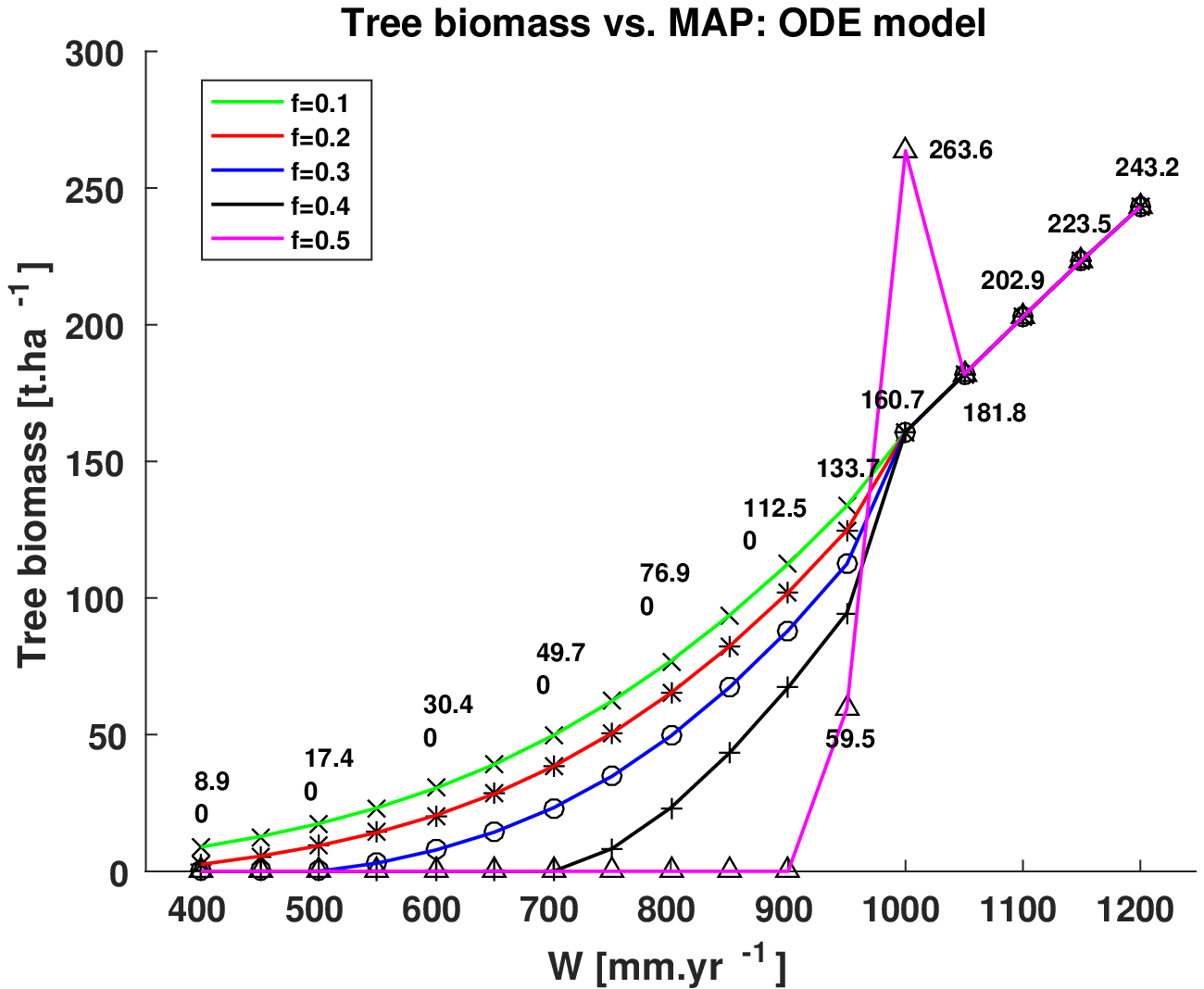}}
		\subfloat[][]{\includegraphics[scale=0.5]{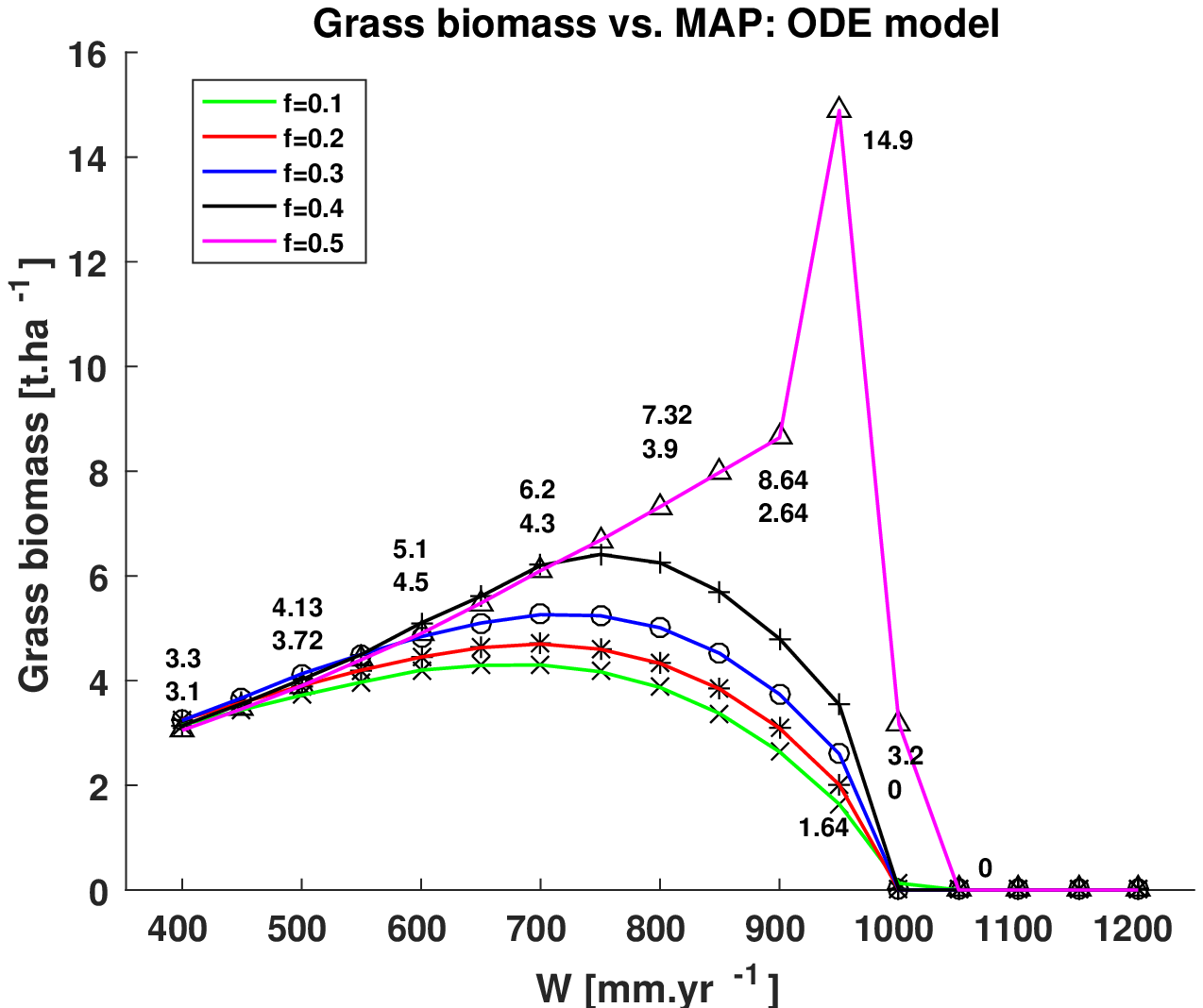}}
		
		\caption{{\scriptsize Expected tree (a) and grass (b) biomasses  as functions of MAP (\textbf{W}) and fire frequencies ($f$, yr$^{-1}$) with the ODE model.  Expected maximum and minimum biomass values are mentioned for a set of \textbf{W} values used in simulations. The peak values of the magenta curves in panels (a) and (b) are the expected  biomasses at the forest-savanna bistable state and  savanna-grassland bistable state respectively. (Color in the online version). 
			}}
			\label{biomass_fig}
		\end{figure}

	With the MAP values chosen for Fig. \ref{bistability_fig} and Fig. \ref{al_bifurc_S_G_F_ode}, we aimed to document the critical parameter space area where the multi-stability between the main vegetation physiognomies occur. It is indeed a stretch of the rainfall gradient where notable differences in biomasses are observed in relation to variations in fire frequencies (see Fig. \ref{biomass_fig}).
	
\begin{figure}[H]
			\centering
			\subfloat[][]{\includegraphics[scale=0.5]{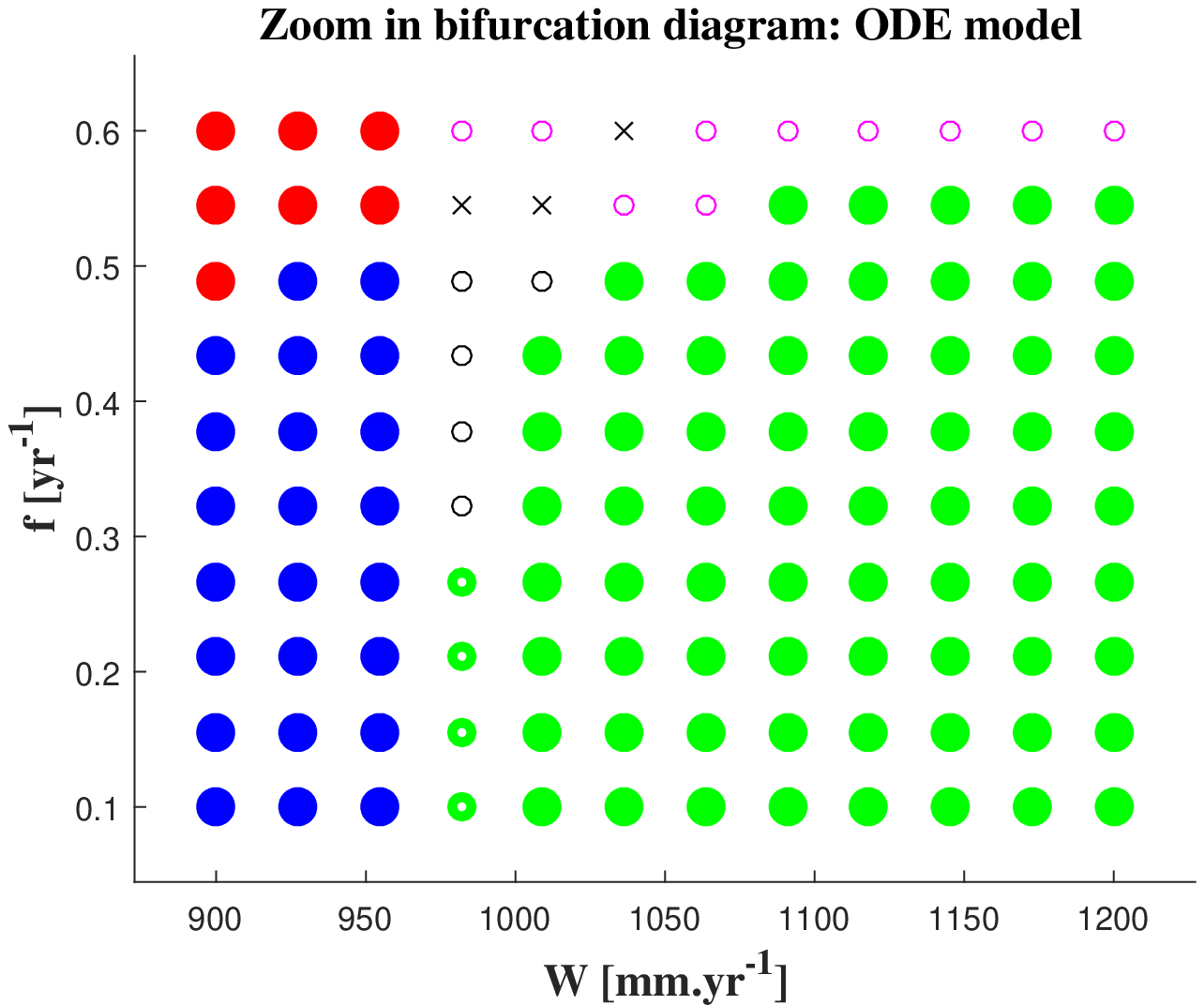}}
			\subfloat[][]{\includegraphics[scale=0.5]{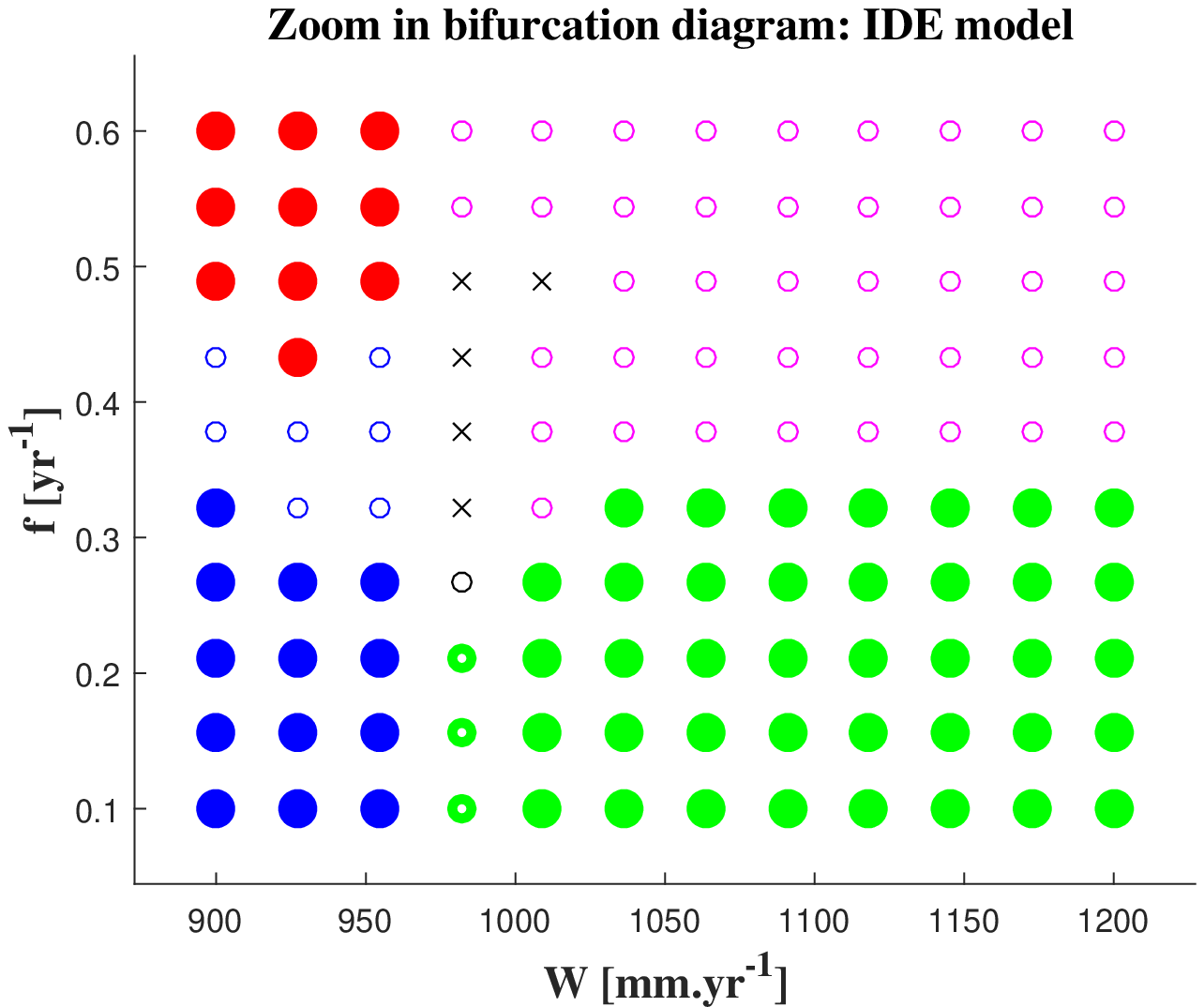}}	
			\caption{{\scriptsize Zoom in the outcomes of models (\ref{swv_eq1}) and (\ref{pulsed_swv_eq1}) evaluated along gradients of MAP ($\textbf{W}$) and fire frequency $f$. The red, blue and green dots stand for grassland, savanna and forest respectively. The magenta and blue circles stand for bistability between forest and grassland, and bistability between savanna and grassland respectively.  The black and green circles stand for bistability between forest and savanna; and  monostability of the woody savanna respectively. The symbol of the cross denotes the tristability situation. (Color in the online version). 
				}}
				\label{zoom_bif}
			\end{figure}

		Fig. \ref{biomass_fig} illustrates  the expected tree and grass biomasses (at the  equilibrium) versus rainfall  for $f=0.1$, $0.2$, $0.3$, $0.4$ and $0.5$ fires per year. It is related to Fig. \ref{zoom_bif}-(a) and it shows a progressive increase of the tree biomass and a decrease  of the grass biomass when the mean annual precipitation (MAP) $\textbf{W}$ increases.



\begin{figure}[H]
				\centering
				\subfloat[][]{	\includegraphics[scale=0.38]{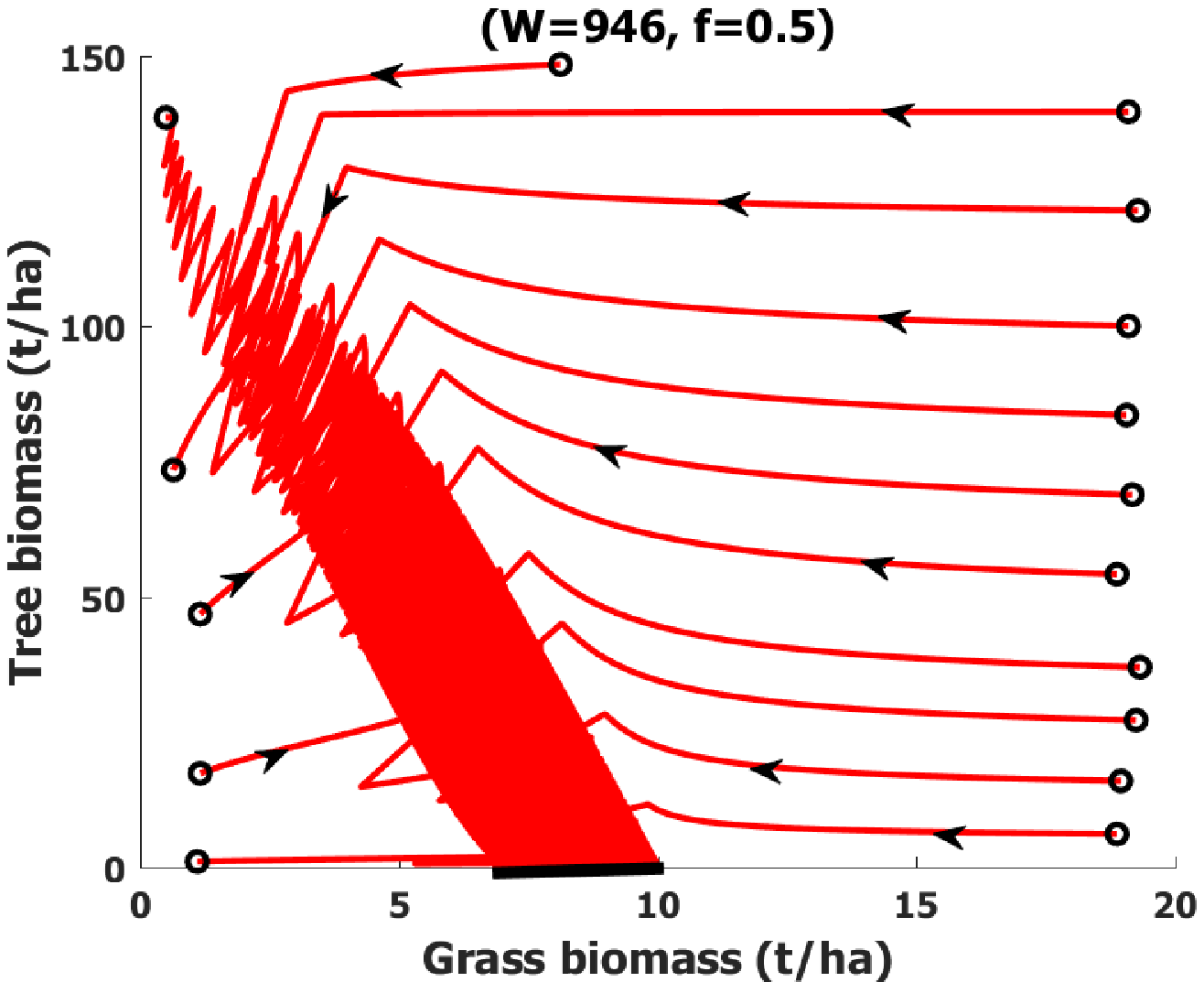}}
				\subfloat[][]{	\includegraphics[scale=0.38]{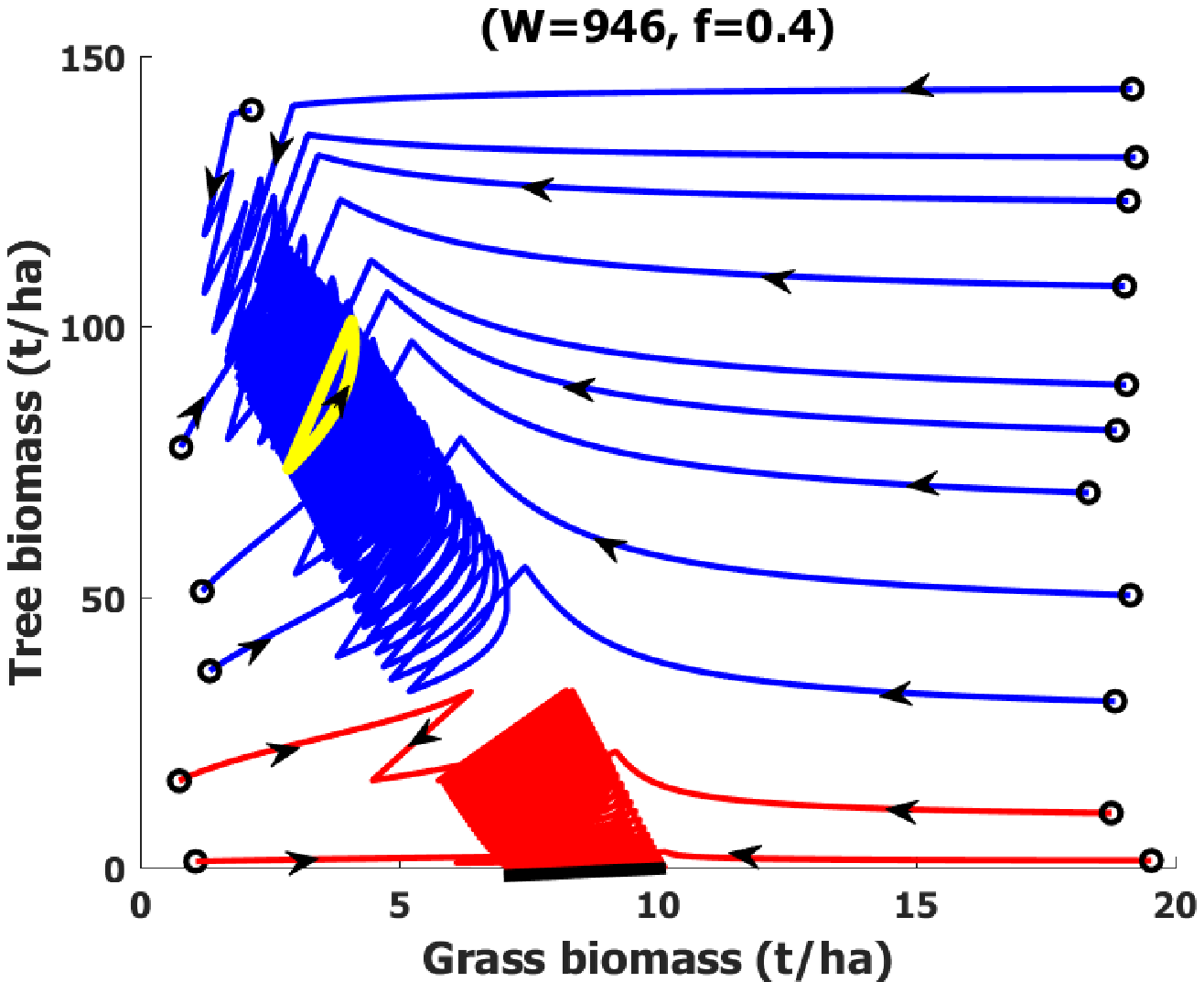}}
				\subfloat[][]{	\includegraphics[scale=0.38]{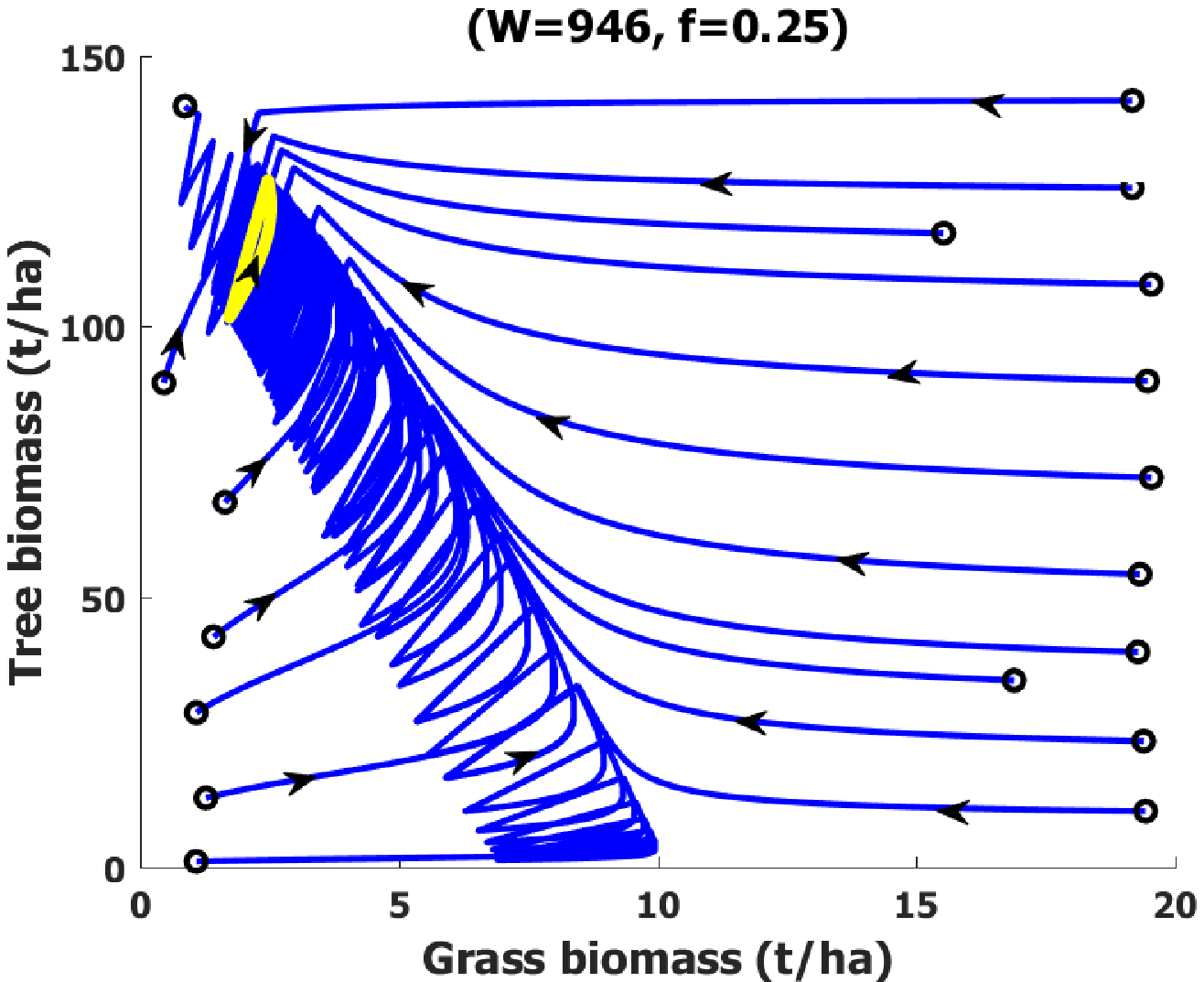}}
				\caption{{\scriptsize  Illustration of  a bifurcation due  to  fire frequency $f$ with the  impulsive model  (\ref{pulsed_swv_eq1}). Fig. \ref{bistability_fig} and this figure have same parameters with varying fire frequencies. The yellow curve and black segment denote the periodic savanna (in fact woodland) and grassland solutions respectively.}}
				\label{al_bifurc_S_G_ide}
			\end{figure}

			Fig. \ref{zoom_bif} shows the model outputs for system (\ref{pulsed_swv_eq1}) depending on fire frequency and mean annual precipitation. It is an analogue pulsed version of the bifurcation diagram given in Fig. \ref{swv_fig2}. Fig. \ref{zoom_bif}-(a) is a zoom of the bifurcation diagram given in Fig. \ref{swv_fig2}-(b) for $(f,\textbf{W})\in [0, 0.6]\times[900, 1200]$ and Fig. \ref{zoom_bif}-(b) is the analogous IDE version of Fig. \ref{zoom_bif}-(a). Indeed Figs. \ref{zoom_bif}-(a) and (b) qualitatively show the same  model outcomes, but it seems that the transversal bifurcation curve of Fig. \ref{swv_fig2}-(b) (see the blue line linking the empty and bold squares) shifts down with the IDE analogue. It means that  at the same level of precipitation using the IDE model (\ref{pulsed_swv_eq1}), bifurcations are predicted at lower fire frequencies than with the ODE model (\ref{swv_eq1})   (see for example Figs. \ref{bistability_fig} and \ref{al_bifurc_S_G_ide}). It implies that for a given level of precipitations preserving the forest-grassland bistability would necessitate a smaller fire frequency according to the IDE model than with the ODE model.

			Both Figs. \ref{al_bifurc_S_G_ide} and  \ref{bistability_fig} plotted respectively for the IDE  model (\ref{pulsed_swv_eq1}) and  ODE model (\ref{swv_eq1})   show qualitatively similar behaviors corresponding to a decrease in fire frequency for constant rainfall. In Fig. \ref{al_bifurc_S_G_ide}, when  $f$ decreases, the system shifts from a   monostable periodic grassland (see panel (a)) to a   monostable periodic savanna (see panel (c)) passing through a bistability between  periodic savanna and  periodic grassland  solutions (see panel (b)).

			\begin{figure}[H]
				\centering
				\subfloat[][]{	\includegraphics[scale=0.38]{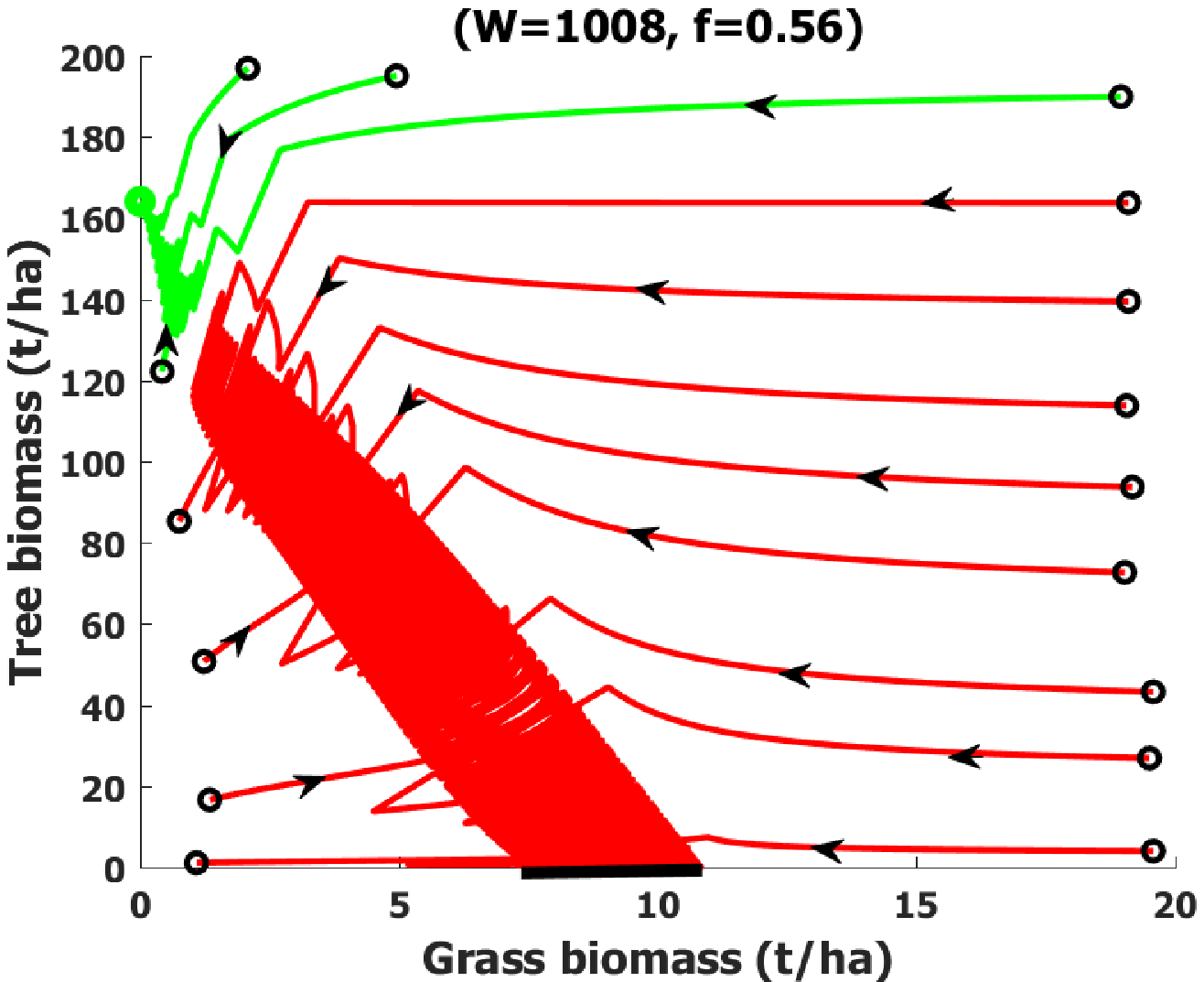}}
				\subfloat[][]{	\includegraphics[scale=0.38]{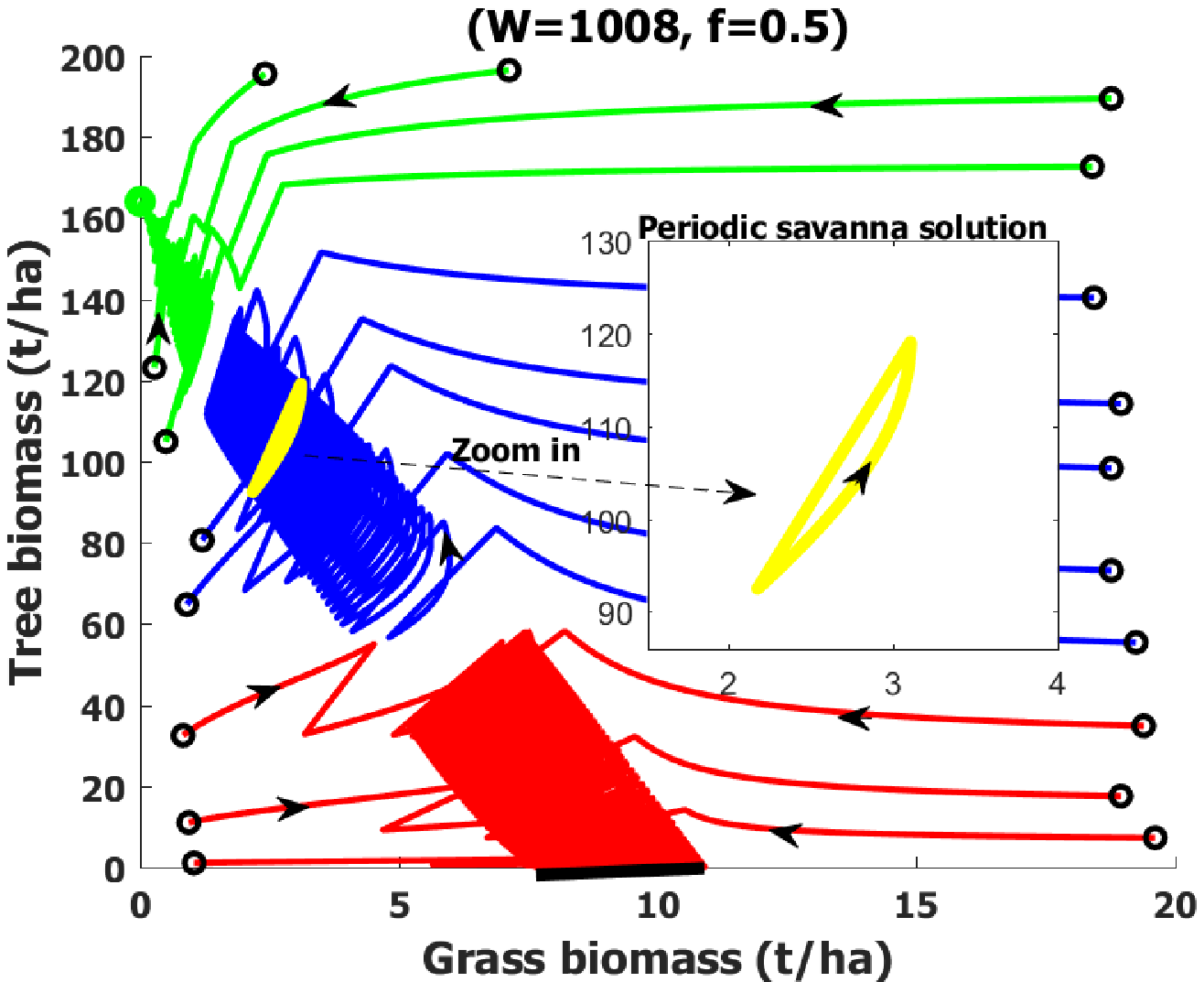}}
				\subfloat[][]{	\includegraphics[scale=0.38]{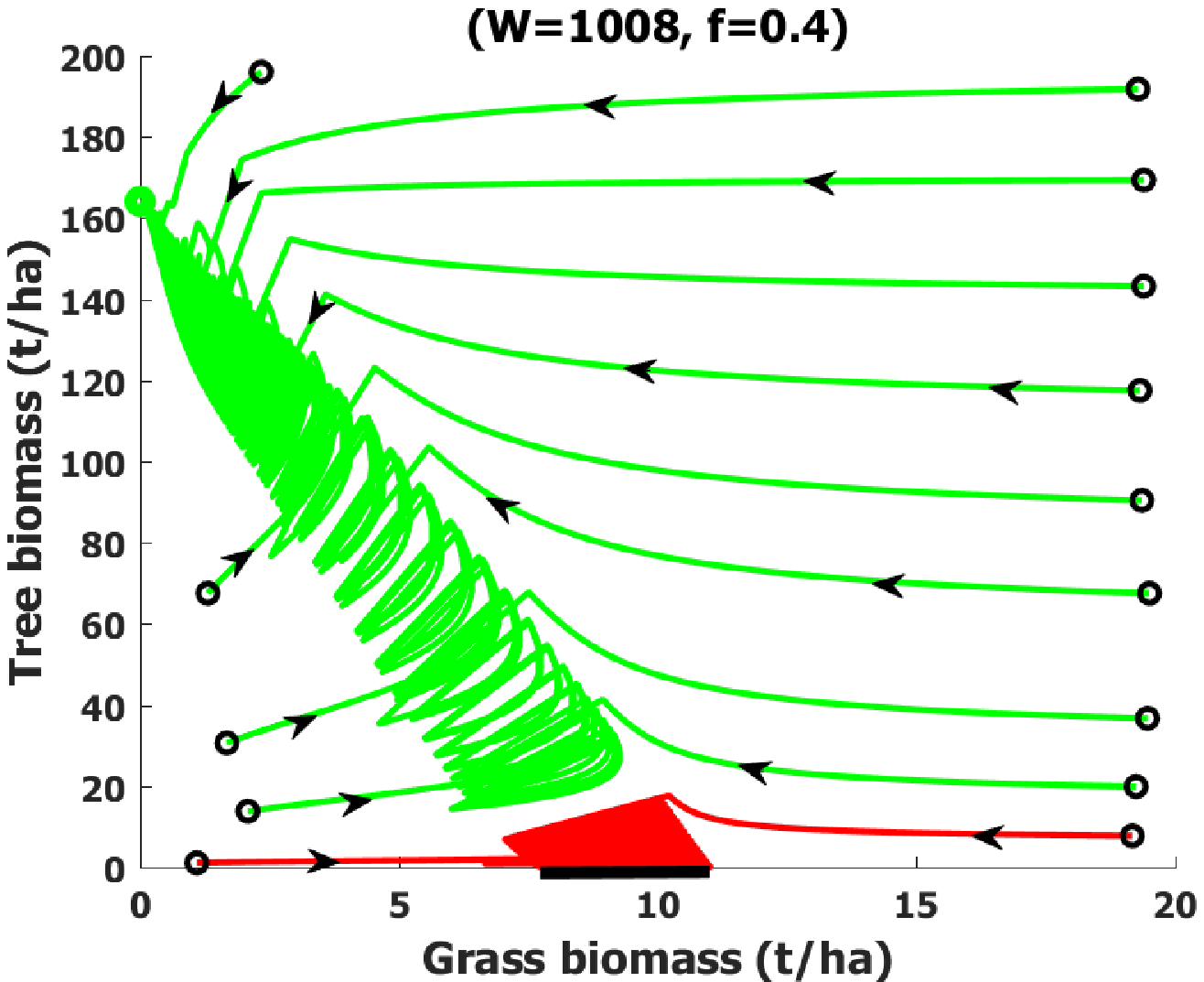}}
				\caption{{\scriptsize  Illustration of  a bifurcation due  to a decrease in fire frequency $f$ with the  IDE model  (\ref{pulsed_swv_eq1}). Both  Fig. \ref{al_bifurc_S_G_F_ode} (ODE) and this figure have same MAP $\textbf{W}=1008$ mm.yr$^{-1}$ and same sequence of $f$ values. The yellow curve and black segment denote the periodic savanna and grassland solutions respectively.}}
				\label{al_bifurc_S_F_G_ide_1}
			\end{figure}
			
			Fig. \ref{al_bifurc_S_F_G_ide_1} shows a bifurcation due to $f$ using the IDE model (\ref{pulsed_swv_eq1}). It illustrates the expansion of forest into savanna and grassland when $f$ decreases. For $f=0.56$, panel (a) in Fig. \ref{al_bifurc_S_F_G_ide_1} shows  a bistability associating forest  and a periodic grassland (having a big basin of attraction). When $f$ decreases from $0.56$ to $0.5$ the system bifurcates to a tristability between forest and two  periodic solutions: savanna and   grassland (see panel (b) in Fig. \ref{al_bifurc_S_F_G_ide_1}). Further decrease of $f$ from $0.5$ to $0.4$ leads the system to bifurcate to a bistability associating forest and a periodic grassland solution with a small basin of attraction (see panel (c) in Fig. \ref{al_bifurc_S_F_G_ide_1}).

			\begin{figure}[H]
				\centering
				\subfloat[][]{	\includegraphics[scale=0.38]{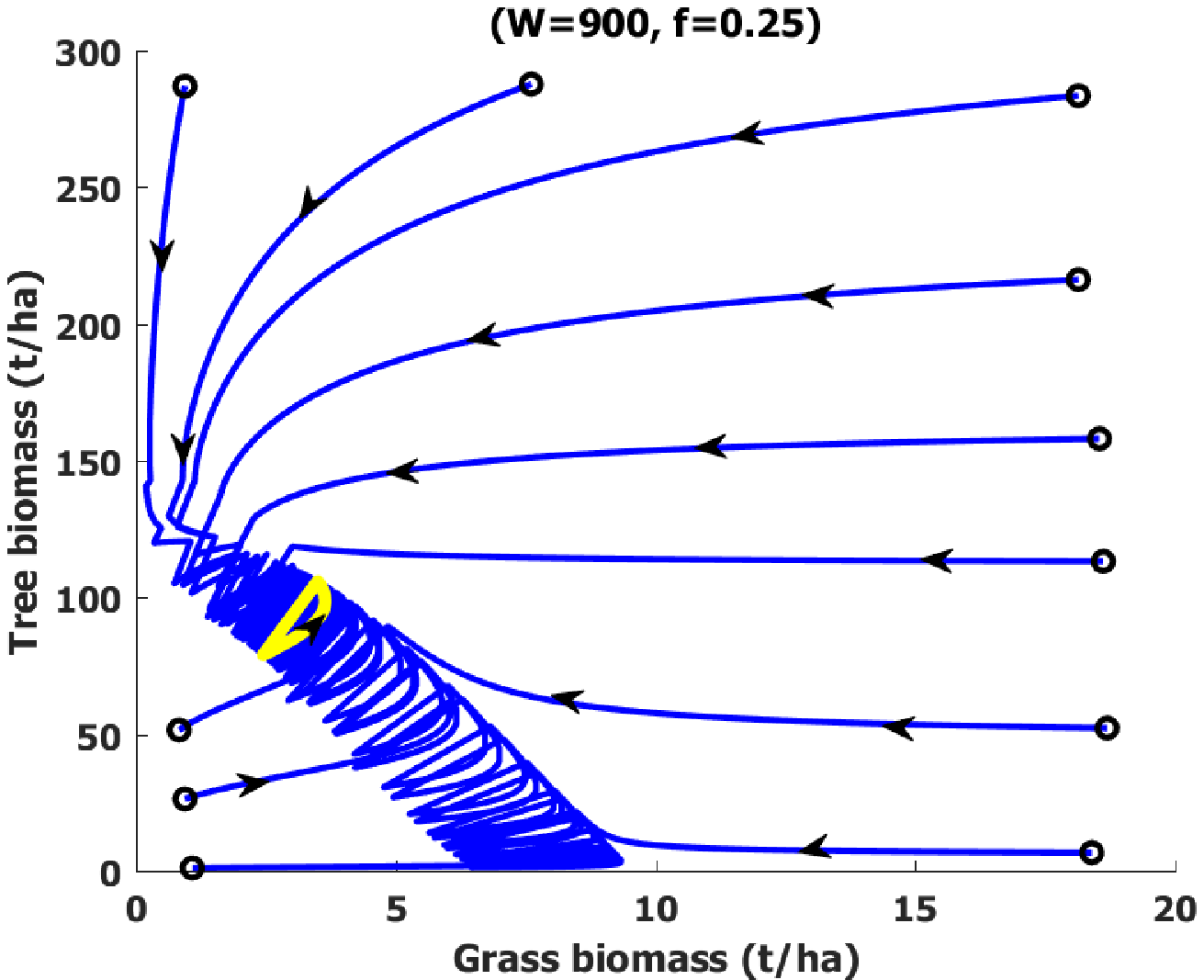}}
				\subfloat[][]{	\includegraphics[scale=0.38]{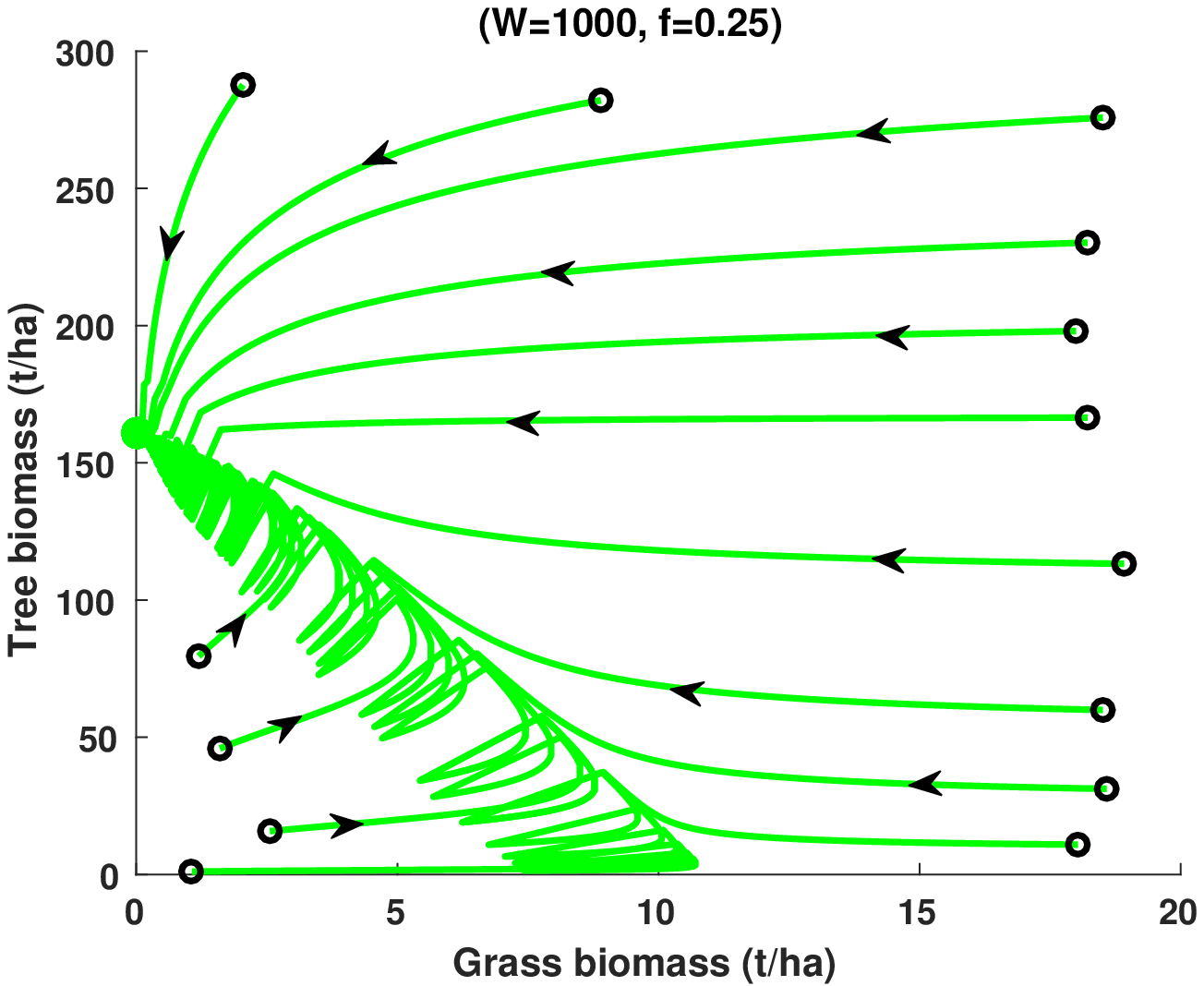}}
				\subfloat[][]{	\includegraphics[scale=0.38]{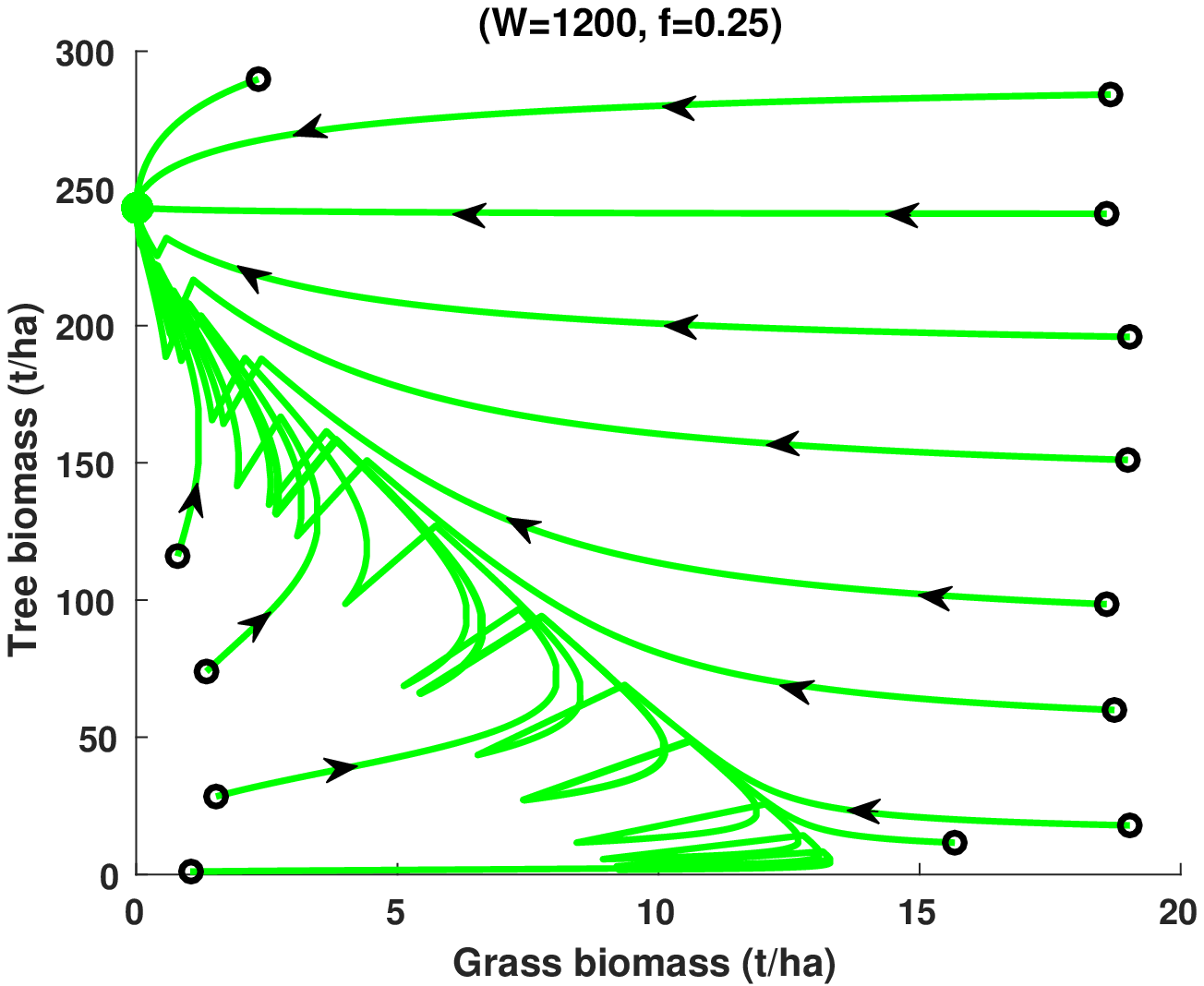}}
				\caption{{\scriptsize  This figure shows  a bifurcation from periodic savanna to forest  with increasing MAP $\textbf{W}$ for a rather low annual fire frequency value i.e.,  $f=0.25$ with the IDE model (\ref{pulsed_swv_eq1}).}}
				\label{fig1_bif_W}
			\end{figure}

			Fig. \ref{fig1_bif_W} illustrates a bifurcation due to $\textbf{W}$ using the IDE model (\ref{pulsed_swv_eq1}). An increase of the MAP $\textbf{W}$ leads the system to shift from a periodic savanna solution to a forest state under a  constant fire frequency value $f=0.25$.			
			
			\begin{figure}[H]
				\centering
				\subfloat[][]{	\includegraphics[scale=0.38]{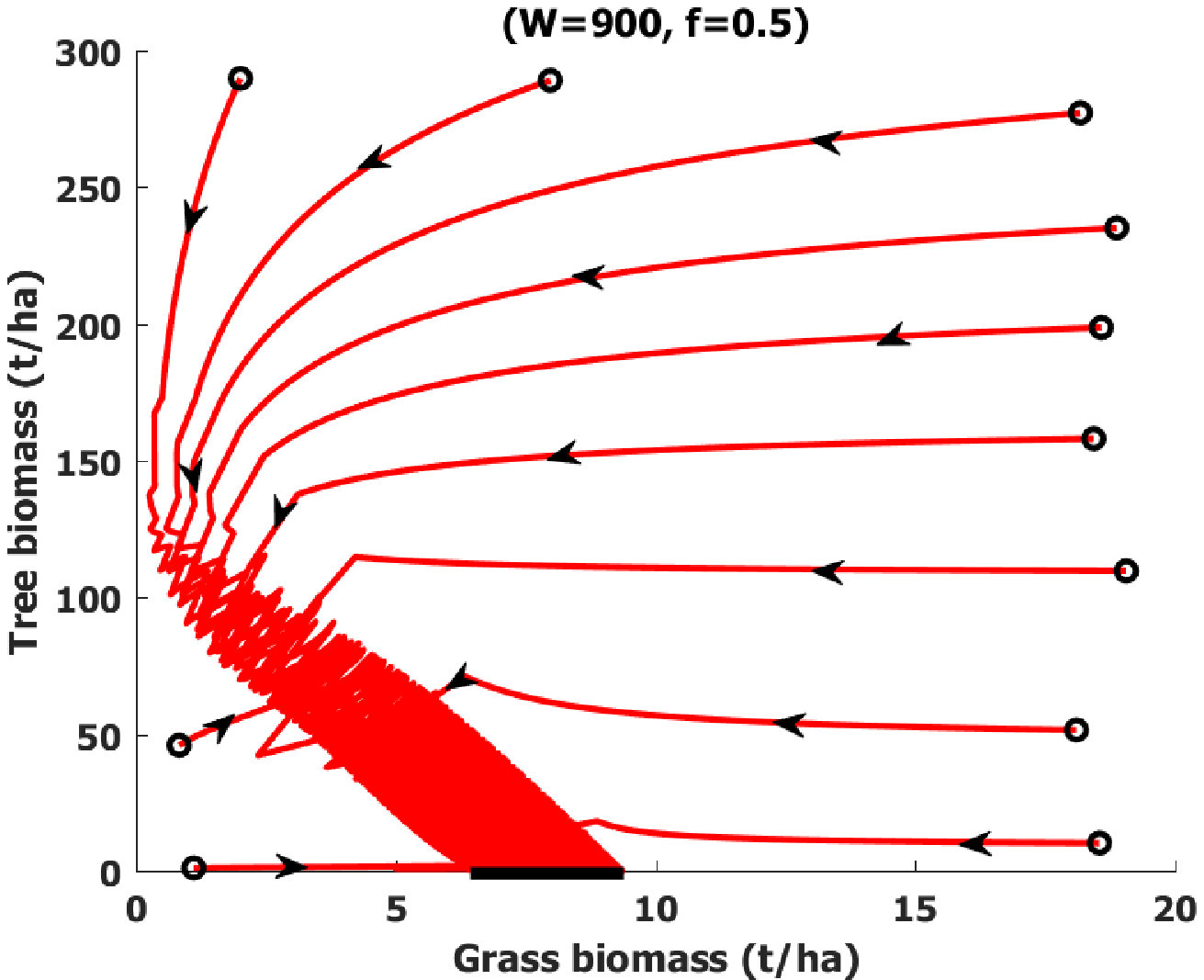}}
				\subfloat[][]{	\includegraphics[scale=0.38]{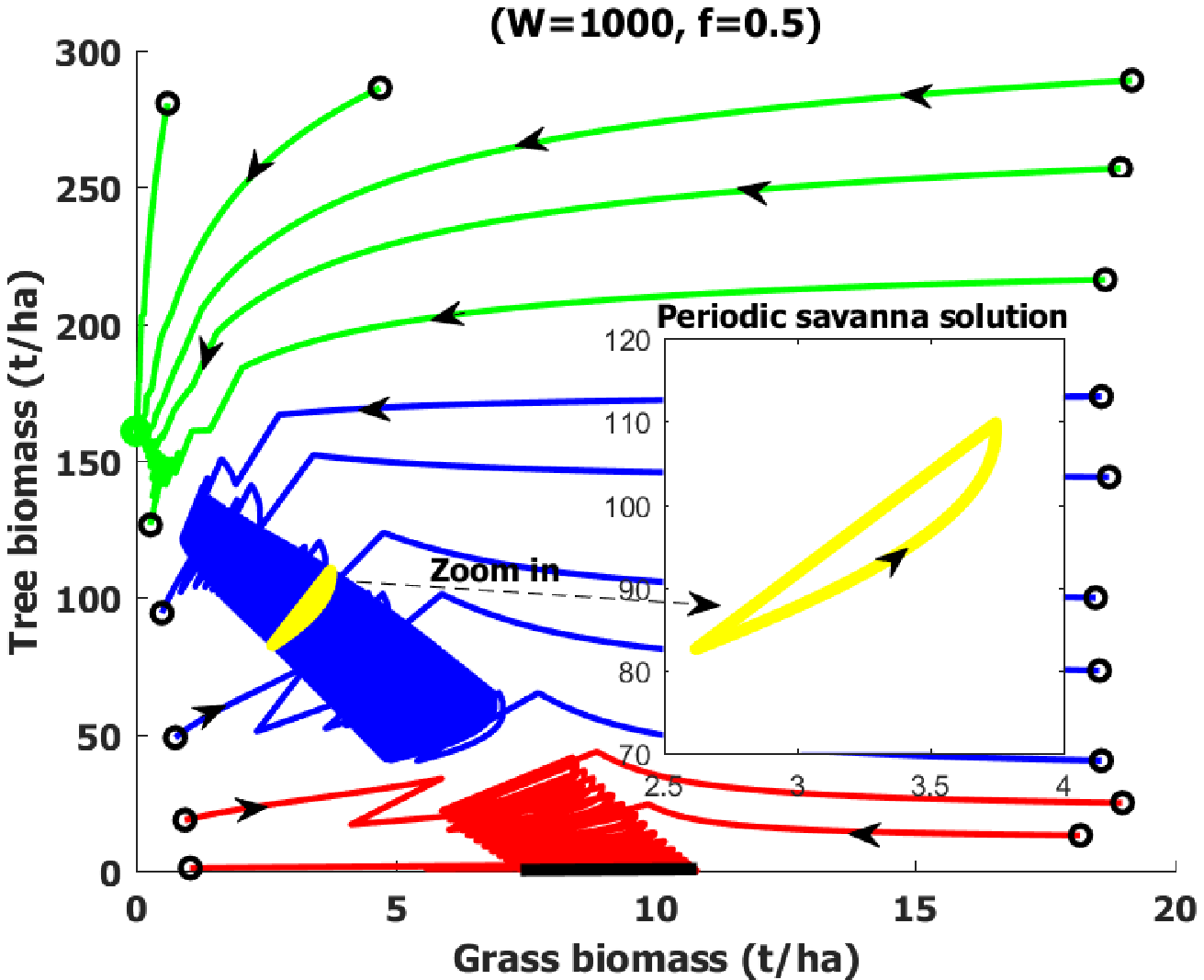}}
				\subfloat[][]{	\includegraphics[scale=0.38]{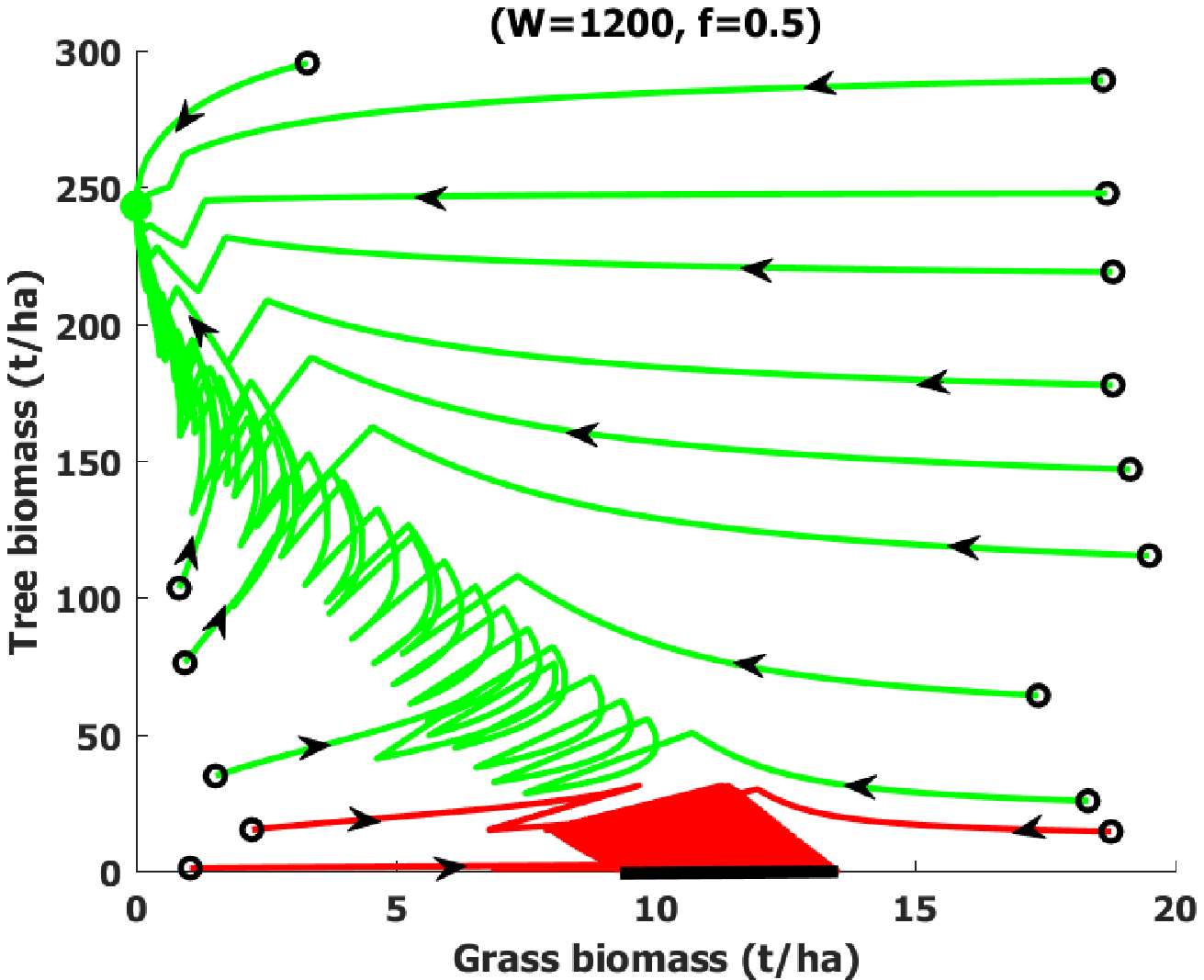}}
				\caption{{\scriptsize  Illustration of  a bifurcation due  to an increase in the MAP $\textbf{W}$ with the IDE model (\ref{pulsed_swv_eq1}). Here $f=0.5$.}}
				\label{fig2_bif_W}
			\end{figure}

			In Fig. \ref{fig2_bif_W}, when the MAP $\textbf{W}$ increases in presence of a higher fire frequency (compared to Fig. \ref{fig1_bif_W}), the system bifurcates from a periodic grassland solution (see panel (a)) to a bistability between forest and a periodic grassland solution (see  panel (c)) passing through a tristability associating forest,  periodic savanna and grassland solutions (see panel (b)).	\par

			In the literature most  tree-grass dynamical systems based on differential equations were  analysed focusing on the stable vegetation states, but never on the trajectories. This  has been criticized recently by \citealp{Accatino2016a} who compared two modelling approaches of tree-grass savanna dynamics they used and that differ in several aspects, notably with respect to the way  in which fire is modelled.   The first modelling approach which is referred to equilibrium model (EM) considers  fire occurrence as a constant parameter, whereas the second modelling approach which is referred as non-equilibrium model (NEM) considers  fire occurrence as a stochastic event whose probability  depends on the amount of dry grass biomass which is built-up in the preceding rainy seasons. According to \citealp{Accatino2016a}, one fundamental difference between EM and NEM resides in how their two models  can be analysed. While  EM analyses are focused on steady states,   NEM analyses also extent  to trajectories. Indeed, the \citealp{Accatino2016a} study shows that their EM modelling has some limitations  concerning predictions of woody cover variation along a rainfall gradient, while several aspects simultaneously contribute to  more satisfactory results obtained with  NEM. Here, we have  illustrated the fact that predictions that are qualitatively satisfactory can be obtained by directly improving the ODE based 'EM' framework, notably regarding the  fire induced  mortality on woody biomass expressed through two independent non-linear  functions $\omega(G)$ (see  (\ref{omega_fction})) and $\vartheta(T)$ (see (\ref{theta_fction})). Nonlinear shape of $\omega(G)$ was also retained by some other authors (\citealp{Scheiter2007};  \citealp{Higgins2010stability}; \citealp{Staver2011tree}; \citealp{Nes2014tipping}; \citealp{YuDOdoricco2014ecohydrological}) to model the fire mediated feedback of
			grass onto tree dynamics. The function $\vartheta(T)$ was introduced in our previous tree-grass model (\cite{Tchuinte2017}) to take into account the response of trees to fires of a given intensity. With our ODE based 'EM' framework, we illustrated how stable vegetation equilibria varied in function of MAP and fire frequency (see Fig. \ref{zoom_bif}-(a)), and for five values of $f$: 0.1, 0.2, 0.3, 0.4 and 0.5, we also illustrated the long-term trajectories expected for tree and grass biomasses at the equilibrium for different levels  along a rainfall gradient (see Fig. \ref{biomass_fig}). The analysis of these trajectories would be meaningful per se.
			\par 
			
			Most savanna fires burn due to human ignition (\citealp{Favier2004modelling}; \citealp{Govender2006}; \citealp{Archibald2009}), but it is believed that these systems are seldom ignition limited, and more often limited by available fuel (\citealp{Archibald2010}; \citealp{vanLeeuwen2014}).
			\citealp{Accatino2016a} underlined the questionable assumption according to which the parameter $f$ of fire frequency should be independent of vegetation characteristics. According to \cite{Accatino2016a}, fire frequency should be considered an emergent property of the dynamical system. This is in contrast with many EM which consider fire as a priori determined and constant parameter. Here,  in our ODE based EM framework, fire frequency $f$ is kept as constant multiplier of $\omega(G)$ (non-linear and bounded function of grass biomass, see (\ref{omega_fction})), but we interpret it as a man-induced "targeted" fire frequency (as for instance in a fire management plan), which will not translate into actual frequency of fires of notable impact (because of $\omega(G)$ being in its low branch) as long as grass biomass if not of sufficient quantity.   We thereby split fire frequency from fire impact. This modelling choice is based on the known fact that
			grass biomass controls fire spread (\citealp{Govender2006}; \citealp{vanLeeuwen2014}). The $\omega(G)$ function is also expected to take into account the
			difficult spreading of fire due to fuel of average
			low quantity keeping in mind natural spatial variability
			of grass biomass. Low grassy fuel  results in lower-fire intensities that  barely propagate leaving a large share of the area unburnt, as frequently observed in the field (see \citealp{Diouf2012}).  This makes the difference between: (i) fire frequency $f$ to be seen as an external forcing upon the tree-grass system (think about a targeted fire regime in a managed area such as a ranch or a national park); and (ii) actual yearly fire probability (or frequency) of occurrence in any arbitrary piece of land when (i) has been set. Moreover,  relationship between the  proportion of the area burnt and grass abundance is likely to be sharply nonlinear, as suggested by the impressive results reported by \citealp{McNaughton1992} at the scale of the entire Serengeti National Park  (Tanzania). According to \citealp{McNaughton1992}, herbivory can produce extensive firebreaks creating a  landscape mosaic which drastically limits fire spreading.  Local fire frequency dwindled over a decade following grass biomass suppression by soaring herbivore populations, while the ignition regime by communities dwelling around the park likely remained more or less the same (\citealp{McNaughton1992}).\par 
			
			Using a very simple ODE model, \citealp{Staver2011tree} concluded that fire cannot influence vegetation but for MAP above $1000$ mm.yr$^{-1}$. 
			Referring to a more complicated simulation "gap" model, \citealp{Lehsten2016} reached results that suggest significant fire influence from $400$ mm MAP upwards. Experimental data from the Kruger National Park (\citealp{Govender2006}; \citealp{Lehsten2016}) suggest notable fire influence below $700$ mm.yr$^{-1}$. In Fig. \ref{biomass_fig} we reach the conclusion that fire frequency considerably change the expected biomasses for MAP values under $1000$ mm.yr$^{-1}$, while our Fig. \ref{swv_fig2} suggests that fire may be able to contribute to the savanna vs. grassland bifurcation for MAP values in the range $500-1000$ mm.yr$^{-1}$

\section{Conclusion}
Here, we have developed and presented a 'minimalistic' tree-grass model that considers interactions of fire and water availability in tree-grass ecosystems. Our model integrates the mean annual precipitation (MAP) as a parameter which shapes growth rates and carrying capacities of tree and grass biomasses. It is to our best knowledge the first time that consistent responses curves (Fig. \ref{swv_fig1}) are assessed from existing information all along the rainfall gradient. The ODE version of the model is fully mathematically tractable as are most of the results relating to the IDE analogue. Compared to  \citealp{Tchuinte2017}, the novelty in the present contribution is that we combine pulsed fire and precipitation-explicit vegetation growth in order to study how the frequency of pulsed fires shapes the vegetation along the rainfall gradient. Both ODE and IDE versions differ fundamentally from existing tree-grass models in the literature in the way that water availability is considered. Some  existing models considered    additional  soil moisture variables  (e.g.,  \citealp{Accatino2010tree}; \citealp{YuDOdoricco2014ecohydrological}) leading to more complex set of parameters and more complex mathematical systems. But, there is no need of an additional  equation about soil moisture since its dynamic is very rapid compared to change in vegetation (fast and slow variables) (\citealp{Barbier2008}; \citealp{Martinez2013spatial}).  
According to our bifurcation diagram (see Fig. \ref{swv_fig2}-(b)), we were able to account for a wide range of physiognomies and dynamical outcomes of the tree-grass system at regional-continental scales by relying on a simple model that explicitly address some essential processes that are: (i) limits put by rainfall on vegetation growth and standing biomass (ii) asymmetric interactions between woody and herbaceous plant life forms, (iii) positive feedback between grass biomass and fire intensity and decreased fire impact with tree height (in fact cumulated woody biomass in the model). For tree-grass  ODE frameworks,  two very general bifurcation diagrams ('big pictures')  were produced  to predict  the vegetation dynamics in the parameters space defined by rainfall and fire frequency: Fig. \ref{swv_fig2}-(a) from    \citealp{Accatino2010tree}  and   Fig. \ref{swv_fig2}-(b) from our model (\ref{swv_eq1}).  Fig. \ref{swv_fig2}-(b) improves the results found in Fig. \ref{swv_fig2}-(a) since it  predicts more
diverse possible situations. For instance it shows a domain where grassland and savanna are bistable and
a domain where  forest, savanna and grassland are tristable. These vegetation outcomes  agree with results found by  remotely-sensed studies (\citealp{Hirota2011}; \citealp{Favier2012abrupt}). Along a general transect over Central Africa,  according to \citealp{Favier2012abrupt} (Figure 2-(e)),  there is a bistability associating sparsely wooded savanna and grassland at ca. $800$ mm.yr$^{-1}$  and tristability is observed very locally around latitude of $7-10$ \textdegree N  (north) (i.e., between ca. $1031$ and $1475$ mm.yr$^{-1}$ of MAP).\par
Under its impulsive form,  model (\ref{pulsed_swv_eq1}) presented here is an extension of our previous IDE model where we already considered the fire induced mortality on woody biomass by mean of two independent non-linear functions, namely $\omega$ (see \ref{omega_fction}) and $\vartheta$ (see \ref{theta_fction}; \citealp{Tchuinte2017}; \citealp{PhDTchuinte2017}). In the present paper we showed that the introduction of these two functions was in fact decisive since even the ODE version of the model proved able to provide sensible results. Notably, we showed that increasing fire return period systematically leads the system to switch from grassland or savanna to forest (forest encroachment). This result is entirely consistent with field observations (\citealp{Bond2005global}; \citealp{Bond2010beyond}; \citealp{Favier2012abrupt}; \citealp{Jeffery2014}). But to our knowledge, it is not established that any other model with only two-state variables is able to retrieve this fundamental behavior. Some more complex models may also fail to display the right behavior in some particular ecological contexts. This should be considered as benchmarking test for existing models.

\begin{acknowledgements}
{\scriptsize A. Tchuint\'e Tamen gratefully acknowledges   financial supports  of the French Embassy in Yaound\'e (Cameroon) through the SCAC fund 2015, and  the French National Institute for Research for Sustainable Development (IRD) in Yaound\'e (Cameroon)  during the preparation of this manuscript. He thanks Prof. S. Bowong and Prof. J.J. Tewa for  their helps and fruitful discussions throughout this work. He also thanks researchers of the UMR-AMAP Laboratory in Montpellier (France) for useful interactions during his internship at AMAP Lab. }
\end{acknowledgements}

\section*{Appendix A: Proof of Theorem \ref{al_thm1} (Existence of a savanna equilibrium)}
\label{al_AppendixA}

From system (\ref{swv_eq1}),  a savanna equilibrium $\textbf{E}_{S}=(G_{*},T_{*})$ satisfies

\begin{equation}
\left\{
\begin{array}{lcl}
g_{G}(\textbf{W})\left(1-\displaystyle\frac{G_{*}}{K_{G}(\textbf{W})}\right)-(\delta_{G}+\lambda_{fG}f)-\eta_{TG}T_{*}=0,\\
\\
g_{T}(\textbf{W})\left(1-\displaystyle\frac{T_{*}}{K_{T}(\textbf{W})}\right)-\delta_{T}-f\vartheta(T_{*})\omega(G_{*})=0.\\
\end{array}
\right.
\label{app_eq1}
\end{equation}

Using the first equation of (\ref{app_eq1}), we have

\begin{equation}
T_{*}=\dfrac{1}{\eta_{TG}}\left(g_{G}(\textbf{W})-(\delta_{G}+\lambda_{fG}f)-\dfrac{g_{G}(\textbf{W})}{K_{G}(\textbf{W})}G_{*}\right)=\dfrac{g_{G}(\textbf{W})}{\eta_{TG}K_{G}(\textbf{W})}(G^{*}-G_{*}).
\label{app_eq2}
\end{equation}

From (\ref{app_eq2}) note that 
one of the conditions to have a plausible savanna equilibrium is: 
\begin{equation}
G^{*}>G_{*}.
\label{app_eq2*}
\end{equation}

Substituting (\ref{app_eq2}) in the second equation of (\ref{app_eq1}) gives

\begin{equation}
\dfrac{(g_{T}(\textbf{W})-\delta_{T})-\dfrac{g_{G}(\textbf{W})g_{T}(\textbf{W})}{\eta_{TG}K_{G}(\textbf{W})K_{T}(\textbf{W})}(G^{*}-G_{*})}{\omega(G_{*})}=f\vartheta(T_{*}).
\label{app_eq3}
\end{equation}

From (\ref{app_eq3}), introducing the expression of $\omega(G)$, we have 

\begin{equation}
\dfrac{\dfrac{g_{T}(\textbf{W})}{K_{T}(\textbf{W})}T^{*}-\dfrac{g_{G}(\textbf{W})g_{T}(\textbf{W})}{\eta_{TG}K_{G}(\textbf{W})K_{T}(\textbf{W})}G^{*}+\dfrac{g_{G}(\textbf{W})g_{T}(\textbf{W})}{\eta_{TG}K_{G}(\textbf{W})K_{T}(\textbf{W})}G_{*}}{\dfrac{G_{*}^{2}}{G_{*}^{2}+\alpha^{2}}}=f\vartheta(T_{*}),
\label{app_eq4}
\end{equation}
where

\begin{equation}
f\vartheta(T_{*})=f\lambda_{fT}^{min}+f(\lambda_{fT}^{max}-\lambda_{fT}^{min})\times e^{-p\dfrac{g_{G}(\textbf{W})G^{*}}{\eta_{TG}K_{G}(\textbf{W})}}\times e^{p\dfrac{g_{G}(\textbf{W})G_{*}}{\eta_{TG}K_{G}(\textbf{W})}}.
\label{app_eq5}
\end{equation}

From (\ref{app_eq4}) and (\ref{app_eq5}) we have:

\begin{equation}
(a-b+cG_{*})\left(1+\dfrac{\alpha^{2}}{G_{*}^{2}}\right)=d+\lambda e^{\alpha G_{*}},
\label{app_eq6}
\end{equation}
where,\\

$a=\dfrac{g_{T}(\textbf{W})}{K_{T}(\textbf{W})}T^{*}$, $b=\dfrac{g_{G}(\textbf{W})g_{T}(\textbf{W})}{\eta_{TG}K_{G}(\textbf{W})K_{T}(\textbf{W})}G^{*}$, $c=\dfrac{b}{G^{*}}$, $d=f\lambda_{fT}^{min}$,    $\lambda=f(\lambda_{fT}^{max}-\lambda_{fT}^{min})\times e^{-p\dfrac{g_{G}(\textbf{W})G^{*}}{\eta_{TG}K_{G}(\textbf{W})}}$ and $\alpha=p\dfrac{g_{G}(\textbf{W})}{\eta_{TG}K_{G}(\textbf{W})}$.\par 

From equation (\ref{app_eq6}) we have

\begin{equation}
cG_{*}^{3}-\lambda G_{*}^{2}e^{\alpha G_{*}} +(a-b-d)G_{*}^{2}+c\alpha^{2}G_{*}+(a-b)\alpha^{2}=0.
\label{app_eq7}
\end{equation}

Set    $H(G_{*})=cG_{*}^{3}-\lambda G_{*}^{2}e^{\alpha G_{*}} +(a-b-d)G_{*}^{2}+c\alpha^{2}G_{*}+(a-b)\alpha^{2}$.  $H$ is a function of one variable $G_{*}\in ]0,+\infty[$.  To find the number of real positive roots of $H(G_{*})$, we will use the intermediate value theorem  which is generally good for investigating real roots of  differentiable and monotonous functions.\par 
We have

\begin{equation}
\left\{
\begin{array}{lcl}
\lim\limits_{G_{*}\longrightarrow 0}H(G_{*})=(a-b)\alpha^{2},\\
\lim\limits_{G_{*}\longrightarrow +\infty}H(G_{*})=-\infty.
\end{array}
\right.
\label{app_eq8}
\end{equation}
The derivative of $H$ is $H^{'}(G_{*})=3cG_{*}^{2}-\lambda(\alpha G_{*}^{2}+2G_{*})e^{\alpha G_{*}}+2(a-b-d)G_{*}+c\alpha^{2}$. We have

\begin{equation}
\left\{
\begin{array}{lcl}
\lim\limits_{G_{*}\longrightarrow 0}H^{'}(G_{*})=c\alpha^{2}>0,\\
\lim\limits_{G_{*}\longrightarrow +\infty}H^{'}(G_{*})=-\infty.
\end{array}
\right.
\label{app_eq9}
\end{equation}

Denote by $H_{1}$ the derivative of $H^{'}$. We have $H_{1}(G_{*})=6cG_{*}-\lambda(\alpha^{2} G_{*}^{2}+4\alpha G_{*}+2)e^{\alpha G_{*}}+2(a-b-d)$.
The limits of $H_{1}(G_{*})$ at $0$ and $+\infty$ are:

\begin{equation}
\left\{
\begin{array}{lcl}
\lim\limits_{G_{*}\longrightarrow 0}H_{1}(G_{*})=2(a-b-d-\lambda),\\
\lim\limits_{G_{*}\longrightarrow +\infty}H_{1}(G_{*})=-\infty.
\end{array}
\right.
\label{app_eq10}
\end{equation}

Denote by $H_{2}$ the derivative of $H_{1}$. We have $H_{2}(G_{*})=6c-\lambda(\alpha^{3} G_{*}^{2}+6\alpha^{2} G_{*}+6\alpha)e^{\alpha G_{*}}$ and 

\begin{equation}
\left\{
\begin{array}{lcl}
\lim\limits_{G_{*}\longrightarrow 0}H_{2}(G_{*})=6(c-\lambda\alpha),\\
\lim\limits_{G_{*}\longrightarrow +\infty}H_{2}(G_{*})=-\infty.
\end{array}
\right.
\label{app_eq11}
\end{equation}

We have $H_{2}^{'}(G_{*})=-\lambda(\alpha^{4} G_{*}^{2}+8\alpha^{3} G_{*}+12\alpha)e^{\alpha G_{*}}<0$. It implies that $H_{2}$ decreases.

\begin{enumerate}
	\item[(I)] If $c-\lambda\alpha<0$, then $H_{2}<0$. It means that $H_{1}$ decreases.
	\begin{itemize}
		\item[1)] If $a-b-d-\lambda<0$, then $H_{1}<0$. It implies that $H^{'}$ decreases. Using (\ref{app_eq9}) and the intermediate value theorem, there exists a unique  $G_{*1}\in ]0,+\infty[$ such that $H^{'}(G_{*1})=0$.
		\begin{itemize}
			\item[a)] If $H(G_{*1})<0$, then there is no plausible savanna equilibrium.
			\item[b)] If $H(G_{*1})>0$ and $a>b$, then there exists a unique savanna equilibrium $\textbf{E}_{*}=(G_{*},T_{*})$ such that $G_{*}\in]G_{*1}, G^{*}[$.
			\item[c)] If $H(G_{*1})>0$ and $a<b$, then there are two savanna equilibria: $\textbf{E}^{1}_{*}=(G^{1}_{*},T^{1}_{*})$ and $\textbf{E}^{2}_{*}=(G^{2}_{*},T^{2}_{*})$ such that $G^{1}_{*}\in]0, G_{*1}[$ and $G^{2}_{*}\in]G_{*1}, G^{*}[$. 
		\end{itemize}
		\item[2)] If $a-b-d-\lambda>0$, then using (\ref{app_eq10}) and the intermediate value theorem, there exists a unique  $G_{*2}\in ]0,+\infty[$ such that $H_{1}(G_{*2})=0$. From (\ref{app_eq9}) we have $H^{'}(G_{*2})>0$. Then using (\ref{app_eq9}) and the intermediate value theorem, there exists a unique  $G_{*3}\in ]G_{*2},+\infty[$ such that  $H^{'}(G_{*3})=0$. Similarly as in $1)$ we have $0$, $1$, or $2$ savanna equilibria.
		\begin{itemize}
			\item[a)] If $H(G_{*3})<0$, then there is no plausible savanna equilibrium.
			\item[b)] If $H(G_{*3})>0$ and $a>b$, then there exists a unique savanna equilibrium $\textbf{E}_{**}=(G_{**},T_{**})$ such that $G_{**}\in]G_{*3}, G^{*}[$.
			\item[c)] If $H(G_{*3})>0$ and $a<b$, then there are two savanna equilibria: $\textbf{E}^{1}_{**}=(G^{1}_{**},T^{1}_{**})$ and $\textbf{E}^{2}_{**}=(G^{2}_{**},T^{2}_{**})$ such that $G^{1}_{**}\in]0, G_{*3}[$ and $G^{2}_{**}\in]G_{*3}, G^{*}[$. 
		\end{itemize}
	\end{itemize}
	\item[(II)] If $c-\lambda\alpha>0$,	 then using (\ref{app_eq11}) and the intermediate value theorem, there exists a unique  $\bar{G}_{*1}\in ]0,+\infty[$ such that $H_{2}(\bar{G}_{*1})=0$.
	
	\begin{itemize}
		\item[1)] If $H_{1}(\bar{G}_{*1})<0$, then $H_{1}(G_{*})<0$. It implies that $H^{'}$ decreases. Using (\ref{app_eq9}) and the intermediate value theorem, there exists a unique  $\bar{G}_{*2}\in ]0,+\infty[$ such that $H^{'}(\bar{G}_{*2})=0$.
		\begin{itemize}
			\item[a)] If $H(\bar{G}_{*2})<0$, then  there is no plausible savanna equilibrium.
			\item[b)] If $H(\bar{G}_{*2})>0$ and $a>b$, then there exists a unique savanna equilibrium $\bar{\textbf{E}}_{*}=(\bar{G}_{*},\bar{T}_{*})$ such that $\bar{G}_{*}\in]\bar{G}_{*2}, G^{*}[$.
			\item[c)] If $H(\bar{G}_{*2})>0$ and $a<b$, then there are two savanna equilibria: $\bar{\textbf{E}}^{1}_{*}=(\bar{G}^{1}_{*},\bar{T}^{1}_{*})$ and $\bar{\textbf{E}}^{2}_{*}=(\bar{G}^{2}_{*},\bar{T}^{2}_{*})$ such that $\bar{G}^{1}_{*}\in]0, \bar{G}_{*2}[$ and $\bar{G}^{2}_{*}\in]\bar{G}_{*2}, G^{*}[$. 
		\end{itemize}
		\item[2)] If $H_{1}(\bar{G}_{*1})>0$ and $a-b-d-\lambda>0$, then using (\ref{app_eq10}) and the intermediate value theorem, there is $\bar{G}_{*3}\in ]\bar{G}_{*2},+\infty[$ such that $H_{1}(\bar{G}_{*3})=0$.  Using (\ref{app_eq9}) there exists $\bar{G}_{*4}\in ]\bar{G}_{*3},+\infty[$ such that $H^{'}(\bar{G}_{*4})=0$. 
		\begin{itemize}
			\item[a)] If $H(\bar{G}_{*4})<0$, then  there is no plausible savanna equilibrium.
			\item[b)] If $H(\bar{G}_{*4})>0$ and $a>b$, then there exists a unique savanna equilibrium $\bar{\textbf{E}}_{**}=(\bar{G}_{**},\bar{T}_{**})$ such that $\bar{G}_{**}\in]\bar{G}_{*4}, G^{*}[$.
			\item[c)] If $H(\bar{G}_{*4})>0$ and $a<b$, then there are two savanna equilibria: $\bar{\textbf{E}}^{1}_{**}=(\bar{G}^{1}_{**},\bar{T}^{1}_{**})$ and $\bar{\textbf{E}}^{2}_{**}=(\bar{G}^{2}_{**},\bar{T}^{2}_{**})$ such that $\bar{G}^{1}_{**}\in]0, \bar{G}_{*4}[$ and $\bar{G}^{2}_{**}\in]\bar{G}_{*4}, G^{*}[$. 
		\end{itemize}
		\item[3)] If $H_{1}(\bar{G}_{*1})>0$ and $a-b-d-\lambda<0$, then using (\ref{app_eq10}) and the intermediate value theorem there are $\bar{G}_{*5}\in]0, \bar{G}_{*1}[$ and $\bar{G}_{*6}\in]\bar{G}_{*1}, +\infty[$ such that $H_{1}(\bar{G}_{*5})=0=H_{1}(\bar{G}_{*6})$.
		\begin{itemize}
			\item[a)] If $H^{'}(\bar{G}_{*5})>0$ and $H^{'}(\bar{G}_{*6})>0$, then using (\ref{app_eq9}) and the intermediate value theorem there exists $\bar{G}_{*7}\in]\bar{G}_{*6}, +\infty[$ such that $H^{'}(\bar{G}_{*7})=0$.
			\begin{itemize}
				\item[1.] If  $H(\bar{G}_{*7})<0$, then there is no plausible savanna equilibrium.
				\item[2.] If $H(\bar{G}_{*7})>0$ and $a>b$, then there exists a unique savanna equilibrium $\bar{\textbf{E}}_{***}=(\bar{G}_{***},\bar{T}_{***})$ such that $\bar{G}_{***}\in]\bar{G}_{*7}, G^{*}[$.
				\item[3.] If $H(\bar{G}_{*7})>0$ and $a<b$, then there are two savanna equilibria: $\bar{\textbf{E}}^{1}_{***}=(\bar{G}^{1}_{***},\bar{T}^{1}_{***})$ and $\bar{\textbf{E}}^{2}_{***}=(\bar{G}^{2}_{***},\bar{T}^{2}_{***})$ such that $\bar{G}^{1}_{***}\in]0, \bar{G}_{*7}[$ and $\bar{G}^{2}_{***}\in]\bar{G}_{*7}, G^{*}[$.
			\end{itemize}
			
			\item[b)] If $H^{'}(\bar{G}_{*5})<0$ and $H^{'}(\bar{G}_{*6})>0$, then using (\ref{app_eq9}) and the intermediate value theorem there are $\bar{G}_{*8}\in]0, \bar{G}_{*5}[$ and $\bar{G}_{*9}\in]\bar{G}_{*6}, +\infty[$ such that $H^{'}(\bar{G}_{*8})=0=H^{'}(\bar{G}_{*9})$.
			Therefore,  using (\ref{app_eq8}) and the intermediate value theorem we have:		
			\begin{itemize}
				\item[1.] If $H(\bar{G}_{*8})>0$ and $a<b$, then  there exist two savanna equilibria: $\bar{\textbf{E}}^{1}_{****}=(\bar{G}^{1}_{****},\bar{T}^{1}_{****})$ and $\bar{\textbf{E}}^{2}_{****}=(\bar{G}^{2}_{****},\bar{T}^{2}_{****})$ such that $\bar{G}^{1}_{****}\in]0, \bar{G}_{*8}[$ and $\bar{G}^{2}_{****}\in]\bar{G}_{*9}, G^{*}[$.
				\item[2.] If ($H(\bar{G}_{*8})>0$ and $a>b$) or ($H(\bar{G}_{*8})<0$, $a<b$, and $H(\bar{G}_{*9})>0$), then  there is a unique savanna equilibrium  $\bar{\textbf{E}}_{****}=(\bar{G}_{****},\bar{T}_{****})$  such that $\bar{G}_{****}\in]\bar{G}_{*9}, G^{*}[$.
				\item[3.] If $H(\bar{G}_{*8})<0$, $a<b$, and $H(\bar{G}_{*9})<0$, then there is no plausible savanna equilibrium.
				\item[4.] If $H(\bar{G}_{*8})<0$, $a>b$, and $H(\bar{G}_{*9})<0$, then there is a unique savanna equilibrium $\bar{\textbf{E}}_{*****}=(\bar{G}_{*****},\bar{T}_{*****})$  such that $\bar{G}_{*****}\in]0, \bar{G}_{*8}[$.
				\item[5.] If $H(\bar{G}_{*8})<0$, $a>b$, and $H(\bar{G}_{*9})>0$, there are three savanna equilibria: $\bar{\textbf{E}}^{1}_{*****}=(\bar{G}^{1}_{*****},\bar{T}^{1}_{*****})$, $\bar{\textbf{E}}^{2}_{*****}=(\bar{G}^{2}_{****},\bar{T}^{2}_{*****})$, and $\bar{\textbf{E}}^{3}_{*****}=(\bar{G}^{3}_{****},\bar{T}^{3}_{*****})$  such that $\bar{G}^{1}_{*****}\in]0, \bar{G}_{*8}[$, $\bar{G}^{2}_{****}\in]\bar{G}_{*8}, \bar{G}_{*9}[$, and $\bar{G}^{3}_{****}\in]\bar{G}_{*9}, G^{*}[$.
			\end{itemize}
			\item[c)] If $H^{'}(\bar{G}_{*5})<0$ and $H^{'}(\bar{G}_{*6})<0$, then using (\ref{app_eq9}) and the intermediate value theorem there exists $\bar{G}_{*10}\in]0, G_{*5}[$ such that $H^{'}(\bar{G}_{*10})=0.$ Using (\ref{app_eq8})	 and the intermediate value theorem we have:
			\begin{itemize}
				\item[1.] If $H(\bar{G}_{*10})<0$, then  there is no plausible savanna equilibrium.
				\item[2.] If $H(\bar{G}_{*10})>0$ and $a>b$, then there exists a unique savanna equilibrium $\bar{\textbf{E}}=(\bar{G},\bar{T})$ such that $\bar{G}\in]\bar{G}_{*10}, G^{*}[$.
				\item[3.] If $H(\bar{G}_{*10})>0$ and $a<b$, then there are two savanna equilibria: $\bar{\textbf{E}}^{1}=(\bar{G}^{1},\bar{T}^{1})$ and $\bar{\textbf{E}}^{2}=(\bar{G}^{2},\bar{T}^{2})$ such that $\bar{G}^{1}\in]0, \bar{G}_{*10}[$ and $\bar{G}^{2}\in]\bar{G}_{*10}, G^{*}[$. 
			\end{itemize}
		\end{itemize}	
	\end{itemize}	
\end{enumerate}
Hence the theorem.

\section*{Appendix B: Proof of Theorem \ref{al_thm3} (Stability of the savanna equilibrium)}
\label{al_AppendixB}

From (\ref{app_eq12}), the Jacobian matrix at the savanna equilibrium $\textbf{E}_{*}=(G_{*}, T_{*})$ is given by

\begin{displaymath}
J_{*}=J(G_{*}, T_{*})=\left(
\begin{array}{ccc}
J_{*}^{11} & J_{*}^{12}\\
J_{*}^{21} & J_{*}^{22}\\
\end{array}
\right),
\end{displaymath}

where,

\begin{equation}
\left\{
\begin{array}{lcl}
J_{*}^{11}=g_{G}(\textbf{W})-(\delta_{G}+\lambda_{fG}f)-2\dfrac{g_{G}(\textbf{W})}{K_{G}(\textbf{W})}G_{*}-\eta_{TG}T_{*},\\
J_{*}^{21}=-f\vartheta(T_{*})\omega^{'}(G_{*})T_{*},\\
J_{*}^{12}=-\eta_{TG}G_{*},\\
J_{*}^{22}=g_{T}(\textbf{W})-\delta_{T}-2\dfrac{g_{T}(\textbf{W})}{K_{G}(\textbf{W})}T_{*}-f\omega(G_{*})[\vartheta(T_{*})+T_{*}\vartheta^{'}(T_{*})].
\end{array}
\right.
\label{app_eq13}
\end{equation}

The characteristic equation of $J_{*}$ is

\begin{equation}
\mu^{2}-tr(J_{*})\mu+det(J_{*})=0,
\label{app_eq14}
\end{equation}
where, $tr(J_{*})=J_{*}^{11}+J_{*}^{22}$ and $det(J_{*})=J_{*}^{11}J_{*}^{22}-J_{*}^{21}J_{*}^{12}$. It follows that all eigenvalues of characteristic equation have negative real part
if and only if $tr(J_{*})<0$ and $det(J_{*})>0$.\par

We have
\begin{align*}
tr(J_{*})&=J_{*}^{11}+J_{*}^{22}\\
&=\dfrac{g_{G}(\textbf{W})}{K_{G}(\textbf{W})}(G^{*}-2G_{*})-\eta_{TG}T_{*}+\dfrac{g_{T}(\textbf{W})}{K_{T}(\textbf{W})}(T^{*}-2T_{*})-f\omega(G_{*})[\vartheta(T_{*})+T_{*}\vartheta^{'}(T_{*})]\\
&=\left\{\dfrac{g_{G}(\textbf{W})}{K_{G}(\textbf{W})}G^{*}+\dfrac{g_{T}(\textbf{W})}{K_{T}(\textbf{W})}T^{*}-f\omega(G_{*})\vartheta^{'}(T_{*})T_{*}\right\}\\
&-\left\{2\left(\dfrac{g_{G}(\textbf{W})}{K_{G}(\textbf{W})}G^{*}+\dfrac{g_{T}(\textbf{W})}{K_{T}(\textbf{W})}T^{*}\right)+\eta_{TG}T_{*}+f\omega(G_{*})\vartheta(T_{*})\right\}\\
&=\left\{2\left(\dfrac{g_{G}(\textbf{W})}{K_{G}(\textbf{W})}G^{*}+\dfrac{g_{T}(\textbf{W})}{K_{T}(\textbf{W})}T^{*}\right)+\eta_{TG}T_{*}+f\omega(G_{*})\vartheta(T_{*})\right\}(\mathcal{R}^{1}_{*}-1), 
\end{align*}

where,\par  

$\mathcal{R}^{1}_{*}=\dfrac{\dfrac{g_{G}(\textbf{W})}{K_{G}(\textbf{W})}G^{*}+\dfrac{g_{T}(\textbf{W})}{K_{T}(\textbf{W})}T^{*}-f\omega(G_{*})\vartheta^{'}(T_{*})T_{*}}{2\left(\dfrac{g_{G}(\textbf{W})}{K_{G}(\textbf{W})}G^{*}+\dfrac{g_{T}(\textbf{W})}{K_{T}(\textbf{W})}T^{*}\right)+\eta_{TG}T_{*}+f\omega(G_{*})\vartheta(T_{*})}$.\par 

We have

\begin{align*}
det(J_{*})&=J_{*}^{11}J_{*}^{22}-J_{*}^{21}J_{*}^{12}\\
&=\left[\dfrac{g_{G}(\textbf{W})}{K_{G}(\textbf{W})}(G^{*}-2G_{*})-\eta_{TG}T_{*}\right]\left[\dfrac{g_{T}(\textbf{W})}{K_{T}(\textbf{W})}(T^{*}-2T_{*})-f\omega(G_{*})(\vartheta(T_{*})+T_{*}\vartheta^{'}(T_{*}))\right]\\
&-\eta_{TG}f\omega^{'}(G_{*})\vartheta(T_{*})T^{2}_{*}\\
&=\dfrac{g_{G}(\textbf{W})g_{T}(\textbf{W})}{K_{G}(\textbf{W})K_{T}(\textbf{W})}G^{*}T^{*}+\left(2\dfrac{g_{G}(\textbf{W})}{K_{G}(\textbf{W})}G^{*}+\eta_{TG}T_{*}\right)\left(2\dfrac{g_{T}(\textbf{W})}{K_{T}(\textbf{W})}T^{*}+f\omega(G_{*})[\vartheta(T_{*})+\vartheta^{'}(T_{*})T_{*}]\right)\\
&-\dfrac{g_{G}(\textbf{W})}{K_{G}(\textbf{W})}G^{*}\left(2\dfrac{g_{T}(\textbf{W})}{K_{T}(\textbf{W})}T^{*}+f\omega(G_{*})[\vartheta(T_{*})+\vartheta^{'}(T_{*})T_{*}]\right)-\dfrac{g_{T}(\textbf{W})}{K_{T}(\textbf{W})}T^{*}\left(2\dfrac{g_{G}(\textbf{W})}{K_{G}(\textbf{W})}G^{*}+\eta_{TG}T_{*}\right)\\
&-\eta_{TG}f\omega^{'}(G_{*})\vartheta(T_{*})T^{2}_{*}\\
&=\left(\dfrac{g_{G}(\textbf{W})}{K_{G}(\textbf{W})}G^{*}B_{*}+\dfrac{g_{T}(\textbf{W})}{K_{T}(\textbf{W})}T^{*}A_{*}+\eta_{TG}f\omega^{'}(G_{*})\vartheta(T_{*})T^{2}_{*}\right)(\mathcal{R}^{2}_{*}-1),
\end{align*}

where,\par 

$A_{*}=2\dfrac{g_{G}(\textbf{W})}{K_{G}(\textbf{W})}G^{*}+\eta_{TG}T_{*}$,  
$B_{*}=2\dfrac{g_{T}(\textbf{W})}{K_{T}(\textbf{W})}T^{*}+f\omega(G_{*})[\vartheta(T_{*})+\vartheta^{'}(T_{*})T_{*}]$, and\\ 
$$\mathcal{R}^{2}_{*}=\dfrac{\dfrac{g_{G}(\textbf{W})g_{T}(\textbf{W})}{K_{G}(\textbf{W})K_{T}(\textbf{W})}G^{*}T^{*}+A_{*}B_{*}}{\dfrac{g_{G}(\textbf{W})}{K_{G}(\textbf{W})}G^{*}B_{*}+\dfrac{g_{T}(\textbf{W})}{K_{T}(\textbf{W})}T^{*}A_{*}+\eta_{TG}f\omega^{'}(G_{*})\vartheta(T_{*})T^{2}_{*}}.$$\par 

Thus, the savanna equilibrium $\textbf{E}=(G_{*}, T_{*})$ is locally asymptotically stable if and only if $\mathcal{R}^{1}_{*}<1$ and $\mathcal{R}^{2}_{*}>1$. This ends the proof of theorem \ref{al_thm3}.

\section*{Appendix C: Brief overview of implementation of model (\ref{swv_eq1}) with MatCont}
\label{al_AppendixC}

Upon the procedure of implementation of the bifurcation in Matcont (\citealp{Dhooge2003Matcont}; \citealp{Dhooge2006Matcont}), many steps are required.  Below is the condensed overview of how one can use the software to implement the model. \par 
\begin{itemize}
	\item First, start to give numerical values to the parameters and the initial conditions in the Starter window of Matcont.  \par 
	\item  The second step is to compute the orbits to explore the dynamics and find equilibria. Matcont window indicates an orbit point noted $P\_O (1)$, meaning orbit number one which starts from the initial point. Choose the 2Dplot window in the Matcont window and put time on the abscissa and the variable T on the ordinate. 
	In the Matcont window click on Forward to start the forward computation.  One can explore the model, if necessary, by running orbits for different parameters and initial conditions.\par
	\item In step three, choose the  Initial point item via Select in the Matcont window to see the initial points window. This lists the two orbits which have been run: $P\_O (1)$ and $P\_O (2)$. Then select the last point of the last orbit that corresponds to the equilibrium of the current parameter settings.
	Now one can use the steady states found to draw the curve of the equilibrium as a function of one parameter.\par 
	\item In step four, choose Equilibrium in the Initial point item of Select Type. Note that the Starter window changes and now contains the values of the last point of the orbit.
	One can now select the parameter over which to continue the equilibrium. Here, choose the fire frequency $f$.\par 
	
	\item  In step five,  open a new 2Dplot window and draw the equilibrium curve over f. Compute both forward and backward to have the full equilibrium curve. Matcont indicates "LP (1)", "BP (1)" for forward and "LP (2)" and "BP (2)" for backward. Note that BP means "Branch point", which is the same thing as a transcritical bifurcation and LP means "Limit point", which refers to a saddle-node bifurcation. \par 
	\item In step six,  one can now continue the LP and BP bifurcations through the \textbf{W}$-$f parameter space. That is draw the location of bifurcation as a function of two parameters. Then select for example the first LP point via Select and via Initial point. The new Stater window appears, in which one should keep parameters \textbf{W} and f active. Computing both forward and backward, MatCont detects  three cusp points "CP" labelled by the small empty circles in Fig. \ref{swv_fig2}-(b).  Each CP point means that there is an equilibrium point with a zero eigenvalue, which corresponds to the centre type point in the theory of bifurcations.  Now select the first BP point. The forward computing draws the curve from the bottom CP point to the full square and the backward computing draws the curve from the same CP point to the empty square (see Fig. \ref{swv_fig2}-(b)). Notice that the second BP point can not be continued since there is an equilibrium which has positive eigenvalues, which means that it can not converge. 
	
\end{itemize}

\bibliographystyle{spmpsci}      
\bibliography{Biblio_new_model}

\begin{thebibliography}{90}
\providecommand{\natexlab}[1]{#1}
\providecommand{\url}[1]{\texttt{#1}}
\expandafter\ifx\csname urlstyle\endcsname\relax
  \providecommand{\doi}[1]{doi: #1}\else
  \providecommand{\doi}{doi: \begingroup \urlstyle{rm}\Url}\fi

\bibitem[Abbadie et~al.(2006)Abbadie, Gignoux, Roux, and
  Lepage]{Abbadie2006lamto}
L.~Abbadie, J.~Gignoux, X.~Roux, and M.~Lepage.
\newblock \emph{Lamto: structure, functioning, and dynamics of a savanna
  ecosystem}, volume 179.
\newblock Springer, 2006.

\bibitem[Accatino and De~Michele(2016)]{Accatino2016a}
F.~Accatino and C.~De~Michele.
\newblock Interpreting woody cover data in tropical and subtropical areas:
  Comparison between the equilibrium and the non-equilibrium assumption.
\newblock \emph{Eco. Comp.}, 25:\penalty0 60--67, 2016.

\bibitem[Accatino et~al.(2010)Accatino, De~Michele, Vezzoli, Donzelli, and
  Scholes]{Accatino2010tree}
F.~Accatino, C.~De~Michele, R.~Vezzoli, D.~Donzelli, and R.~J. Scholes.
\newblock Tree--grass co-existence in savanna: interactions of rain and fire.
\newblock \emph{J. Theor. Biol.}, 267\penalty0 (2):\penalty0 235--242, 2010.
\newblock \doi{10.1016/j.jtbi.2010.08.012}.

\bibitem[Accatino et~al.(2016)Accatino, Wiegand, Ward, and
  De~Michele]{Accatino2016trees}
F.~Accatino, K.~Wiegand, D.~Ward, and C.~De~Michele.
\newblock Trees, grass, and fire in humid savannas-{T}he importance of life
  history traits and spatial processes.
\newblock \emph{Ecol. Modell.}, 320:\penalty0 135--144, 2016.

\bibitem[Anguelov et~al.(2012)Anguelov, Dumont, and Lubuma]{Anguelov2012}
R.~Anguelov, Y.~Dumont, and J.~M.-S. Lubuma.
\newblock On nonstandard finite difference schemes in biosciences.
\newblock \emph{AIP Conf. Proc.}, 1487\penalty0 (1):\penalty0 212--223, 2012.
\newblock \doi{10.1063/1.4758961}.
\newblock URL
  \url{http://scitation.aip.org/content/aip/proceeding/aipcp/10.1063/1.4758961}.

\bibitem[Archibald et~al.(2009)Archibald, Roy, Brian, Wilgen, and
  Scholes]{Archibald2009}
S.~Archibald, David~P. Roy, W~Brian, Van Wilgen, and R.~J. Scholes.
\newblock What limits fire? an examination of drivers of burnt area in southern
  africa.
\newblock \emph{Global Change Biol.}, 15\penalty0 (3):\penalty0 613--630, 2009.

\bibitem[Archibald et~al.(2010)Archibald, Scholes, Roy, Roberts, and
  Boschetti]{Archibald2010}
S.~Archibald, R.~J. Scholes, D.~P. Roy, G.~Roberts, and L.~Boschetti.
\newblock Southern african fire regimes as revealed by remote sensing.
\newblock \emph{Int. J. Wildland Fire}, 19\penalty0 (7):\penalty0 861--878,
  2010.

\bibitem[Bainov and Simeonov(1993)]{Bainov1993}
D.~D. Bainov and P.~S. Simeonov.
\newblock \emph{Impulsive differential equations: periodic solutions and
  applications}, volume~66.
\newblock CRC Press, 1993.

\bibitem[Bainov and Simeonov(1995)]{Bainov1995}
D.~D. Bainov and P.~S. Simeonov.
\newblock \emph{Impulsive differential equations: asymptotic properties of the
  solutions}, volume~28.
\newblock World Scientific, 1995.

\bibitem[Barbier et~al.(2008)Barbier, Couteron, Lefever, Deblauwe, and
  Lejeune]{Barbier2008}
N.~Barbier, P.~Couteron, R.~Lefever, V.~Deblauwe, and O.~Lejeune.
\newblock Spatial decoupling of facilitation and competition at the origin of
  gapped vegetation patterns.
\newblock \emph{Ecology}, 89\penalty0 (6):\penalty0 1521--1531, 2008.

\bibitem[Baudena and Rietkerk(2013)]{Baudena2013}
M.~Baudena and M.~Rietkerk.
\newblock Complexity and coexistence in a simple spatial model for arid savanna
  ecosystems.
\newblock \emph{Theor. Ecol.}, 6\penalty0 (2):\penalty0 131--141, 2013.

\bibitem[Baudena et~al.(2010)Baudena, D'Andrea, and Provenzale]{Baudena2010}
M.~Baudena, F.~D'Andrea, and A.~Provenzale.
\newblock An idealized model for tree-grass coexistence in savannas: the role
  of life stage structure and fire disturbances.
\newblock \emph{J. Ecol.}, 98:\penalty0 74--80, 2010.

\bibitem[Baudena et~al.(2014)Baudena, Dekker, van Bodegom, Cuesta, Higgins,
  Lehsten, Reick, Rietkerk, Scheiter, Yin, Zavala, and
  Brovkin]{Baudena2014forests}
M.~Baudena, S.~C. Dekker, P.~M. van Bodegom, B.~Cuesta, S.~I. Higgins,
  V.~Lehsten, C.~H. Reick, M.~Rietkerk, S.~Scheiter, Z.~Yin, M.~A. Zavala, and
  V.~Brovkin.
\newblock Forests, savannas and grasslands: bridging the knowledge gap between
  ecology and dynamic global vegetation models.
\newblock \emph{Biogeosci. Discuss.}, 11\penalty0 (6):\penalty0 9471--9510,
  2014.
\newblock \doi{10.5194/bgd-11-9471-2014}.

\bibitem[Beckage et~al.(2009)Beckage, Platt, and Gross]{Beckage2009}
B.~Beckage, W.~Platt, and L.~Gross.
\newblock Vegetation, fire and feedbacks: a disturbance-mediated model of
  savannas.
\newblock \emph{Am. Nat.}, 174\penalty0 (6):\penalty0 805--818, 2009.

\bibitem[Beckage et~al.(2011)Beckage, Gross, and Platt]{Beckage2011grass}
B.~Beckage, L.J. Gross, and W.~J. Platt.
\newblock Grass feedbacks on fire stabilize savannas.
\newblock \emph{Ecol. Modell.}, 222\penalty0 (14):\penalty0 2227--2233, 2011.
\newblock \doi{10.1016/j.ecolmodel.2011.01.015}.

\bibitem[Bond(2008)]{Bond2008}
W.~J. Bond.
\newblock What limits trees in c4 grasslands and savannas?
\newblock \emph{Annu. Rev. Ecol. Evol. Syst.}, 39:\penalty0 641--659, 2008.

\bibitem[Bond and Parr(2010)]{Bond2010beyond}
W.~J. Bond and C.~L. Parr.
\newblock Beyond the forest edge: ecology, diversity and conservation of the
  grassy biomes.
\newblock \emph{Biol. Conserv.}, 143\penalty0 (10):\penalty0 2395--2404, 2010.

\bibitem[Bond et~al.(2005)Bond, Woodward, and Midgley]{Bond2005global}
W.~J. Bond, F.~I. Woodward, and G.~F. Midgley.
\newblock The global distribution of ecosystems in a world without fire.
\newblock \emph{New Phytol.}, 165\penalty0 (2):\penalty0 525--538, 2005.

\bibitem[Braun(1972a)]{Braun1972a}
H.~M.~H Braun.
\newblock Primary production in the serengeti: purpose, methods and some
  results of research.
\newblock In \emph{IBP regional meeting on Grasslands Research Projects (Lamto,
  Ivory Coast, 30.12.1971--3.1.1972)}, 1972a.

\bibitem[Braun(1972b)]{Braun1972b}
H.~M.~H Braun.
\newblock \emph{Botanische samenstelling van de vegetaties in de Serengeti
  Plains. Typescript report to Wageningen University and the Serengeti Research
  Institute (in Dutch)}.
\newblock 1972b.

\bibitem[Bucini and Hanan(2007)]{BuciniHanan2007}
G.~Bucini and N.~P. Hanan.
\newblock A continental-scale analysis of tree cover in african savannas.
\newblock \emph{Global Ecol. Biogeogr.}, 16:\penalty0 593--605, 2007.

\bibitem[Chidumayo(1990)]{Chidumayo1990}
E.~N Chidumayo.
\newblock Above-ground woody biomass structure and productivity in a zambezian
  woodland.
\newblock \emph{For. Ecol. Manage.}, 36\penalty0 (1):\penalty0 33--46, 1990.

\bibitem[Cuni-Sanchez et~al.(2016)Cuni-Sanchez, White, Calders, Jeffery,
  Abernethy, Burt, Disney, Gilpin, Gomez-Dans, and Lewis]{Sanchez2016african}
A.~Cuni-Sanchez, Lee J.~T. White, Kim Calders, K.~J. Jeffery, K.~Abernethy,
  Andrew Burt, Mathias Disney, Martin Gilpin, Jose~L Gomez-Dans, and Simon~L
  Lewis.
\newblock African savanna-forest boundary dynamics: A 20-year study.
\newblock \emph{PLoS One}, 11\penalty0 (6):\penalty0 e0156934, 2016.

\bibitem[De~Michele et~al.(2008)De~Michele, Vezzoli, Pavlopoulos, and
  Scholes]{DeMichele2008minimal}
C.~De~Michele, R.~Vezzoli, H.~Pavlopoulos, and R.~J. Scholes.
\newblock A minimal model of soil water--vegetation interactions forced by
  stochastic rainfall in water-limited ecosystems.
\newblock \emph{Ecol. Modell.}, 212\penalty0 (3):\penalty0 397--407, 2008.

\bibitem[De~Michele et~al.(2011)De~Michele, Accatino, Vezzoli, and
  Scholes]{DeMichele2011}
C.~De~Michele, F.~Accatino, R.~Vezzoli, and R.J. Scholes.
\newblock Savanna domain in the herbivores-fire parameter space exploiting a
  tree--grass--soil water dynamic model.
\newblock \emph{J. Theor. Biol.}, 289:\penalty0 74--82, 2011.
\newblock \doi{10.1016/j.jtbi.2011.08.014}.

\bibitem[Dhooge et~al.(2003)Dhooge, Govaerts, and Kuznetsov]{Dhooge2003Matcont}
A.~Dhooge, W.~Govaerts, and Y.~A. Kuznetsov.
\newblock {MATCONT: A MATLAB} package for numerical bifurcation analysis of
  {ODEs}.
\newblock \emph{ACM TOMS}, 29\penalty0 (2):\penalty0 141--164, 2003.

\bibitem[Dhooge et~al.(2006)Dhooge, Govaerts, Kuznetsov, Mestrom, Riet, and
  Sautois]{Dhooge2006Matcont}
A.~Dhooge, W.~Govaerts, Y.~A. Kuznetsov, W.~Mestrom, A.~M. Riet, and
  B.~Sautois.
\newblock \emph{MATCONT and CL MATCONT: Continuation toolboxes in matlab}.
\newblock Universiteit Gent, Belgium and Utrecht University, The Netherlands,
  2006.

\bibitem[Diouf et~al.(2012)Diouf, Barbier, Lykke, Couteron, Deblauwe, Mahamane,
  Saadou, and Bogaert]{Diouf2012}
A.~Diouf, N.~Barbier, A.M. Lykke, P.~Couteron, V.~Deblauwe, A.~Mahamane,
  M.~Saadou, and J.~Bogaert.
\newblock Relationships between fire history, edaphic factor and woody
  vegetation structure and composition in a semi-arid savanna landscape (niger,
  west africa).
\newblock \emph{Appl. Veg. Sci.}, 15:\penalty0 488--500, 2012.

\bibitem[D'Odorico et~al.(2006)D'Odorico, Laio, and
  Ridolfi]{DOdorico2006probabilistic}
P.~D'Odorico, F.~Laio, and L.~Ridolfi.
\newblock A probabilistic analysis of fire-induced tree-grass coexistence in
  savannas.
\newblock \emph{Am. Nat.}, 167\penalty0 (3):\penalty0 E79--E87, 2006.
\newblock \doi{10.1086/500617}.

\bibitem[Favier et~al.(2004)Favier, Chave, Fabing, Schwartz, and
  Dubois]{Favier2004modelling}
C.~Favier, J.~Chave, A.~Fabing, D.~Schwartz, and M.A. Dubois.
\newblock Modelling forest--savanna mosaic dynamics in man-influenced
  environments: effects of fire, climate and soil heterogeneity.
\newblock \emph{Ecol. Modell.}, 171\penalty0 (1):\penalty0 85--102, 2004.
\newblock \doi{10.1016/j.ecolmodel.2003.07.003}.

\bibitem[Favier et~al.(2012)Favier, Aleman, Bremond, Dubois, Freycon, and
  Yangakola]{Favier2012abrupt}
C.~Favier, J.~Aleman, L.~Bremond, M.A. Dubois, V.~Freycon, and J.-M. Yangakola.
\newblock Abrupt shifts in african savanna tree cover along a climatic
  gradient.
\newblock \emph{Global Ecol. Biogeogr.}, 21\penalty0 (8):\penalty0 787--797,
  2012.
\newblock \doi{10.1111/j.1466-8238.2011.00725.x}.

\bibitem[Frost et~al.(1986)Frost, Medina, Menaut, Solbrig, Swift, and
  Walker]{Frost1986}
P.~Frost, E.~Medina, J.~C. Menaut, O.~Solbrig, M~Swift, and B~Walker.
\newblock \emph{Responses of savannas to stress and disturbance}.
\newblock Biology International, 1986.

\bibitem[Gaines and Mawhin(1977)]{Gaines1977}
R.E. Gaines and J.L. Mawhin.
\newblock \emph{Coincidence degree, and nonlinear differential equations}.
\newblock Springer, 1977.

\bibitem[Govaerts et~al.(2007)Govaerts, Ghaziani, Kuznetsov, and
  Meijer]{Govaerts2007Matcont}
W.~Govaerts, R.~K. Ghaziani, Yu.~A. Kuznetsov, and H.~G.~E. Meijer.
\newblock Numerical methods for two-parameter local bifurcation analysis of
  maps.
\newblock \emph{SIAM J. Sci. Comput.}, 29\penalty0 (6):\penalty0 2644--2667,
  2007.

\bibitem[Govender et~al.(2006)Govender, Trollope, and Van~Wilgen]{Govender2006}
N.~Govender, W.~S.~W. Trollope, and B.~W. Van~Wilgen.
\newblock The effect of fire season, fire frequency, rainfall and management on
  fire intensity in savanna vegetation in south africa.
\newblock \emph{J. Appl. Ecol.}, 43\penalty0 (4):\penalty0 748--758, 2006.

\bibitem[Higgins et~al.(2007)Higgins, Bond, February, Bronn, Euston-Brown,
  Enslin, Govender, Rademan, O'Regan, Potgieter, Scheiter, Sowry, Trollope, and
  Trollope]{Higgins2007topkill}
S.~I. Higgins, W.~J. Bond, E.~C. February, A.~Bronn, D.~I.~W. Euston-Brown,
  B.~Enslin, N.~Govender, L.~Rademan, S.~O'Regan, A.~L.~F. Potgieter,
  S.~Scheiter, R.~Sowry, L.~Trollope, and W.~S.~W. Trollope.
\newblock Effects of four decades of fire manipulation on woody vegetation
  structure in savanna.
\newblock \emph{Ecology}, 88\penalty0 (5):\penalty0 1119--1125, 2007.

\bibitem[Higgins et~al.(2010)Higgins, Scheiter, and
  Sankaran]{Higgins2010stability}
S.~I. Higgins, S.~Scheiter, and M.~Sankaran.
\newblock The stability of african savannas: insights from the indirect
  estimation of the parameters of a dynamic model.
\newblock \emph{Ecology}, 91\penalty0 (6):\penalty0 1682--1692, 2010.

\bibitem[Higgins et~al.(2000)Higgins, Bond, and Trollope]{Higgins2000fire}
S.I. Higgins, W.J. Bond, and W.S.W. Trollope.
\newblock Fire, resprouting and variability: a recipe for grass--tree
  coexistence in savanna.
\newblock \emph{J. Ecol.}, 88\penalty0 (2):\penalty0 213--229, 2000.
\newblock \doi{10.1046/j.1365-2745.2000.00435.x}.

\bibitem[Hijmans et~al.(2005)Hijmans, Cameron, Parra, Jones, and
  Jarvis]{Hijmans2005}
R.~J. Hijmans, S.~E. Cameron, J.~L. Parra, P.~G. Jones, and A.~Jarvis.
\newblock Very high resolution interpolated climate surfaces for global land
  areas.
\newblock \emph{Int. J. Climatol.}, 25\penalty0 (15):\penalty0 1965--1978,
  2005.

\bibitem[Hirota et~al.(2011)Hirota, Holmgren, Van~Nes, and
  Scheffer]{Hirota2011}
M.~Hirota, M.~Holmgren, E.~H. Van~Nes, and M.~Scheffer.
\newblock Global resilience of tropical forest and savanna to critical
  transitions.
\newblock \emph{Science}, 334\penalty0 (6053):\penalty0 232--235, 2011.

\bibitem[Hoffmann and Solbrig(2003)]{Hoffmann2003topkill}
W.~A. Hoffmann and O.~T. Solbrig.
\newblock The role of topkill in the differential response of savanna woody
  species to fire.
\newblock \emph{For. Ecol. Manage.}, 180\penalty0 (1):\penalty0 273--286, 2003.

\bibitem[House et~al.(2003)House, Archer, Breshears, and
  Scholes]{House2003conundrums}
J.~I. House, S.~Archer, D.~D. Breshears, and R.~J. Scholes.
\newblock Conundrums in mixed woody--herbaceous plant systems.
\newblock \emph{J. Biogeogr.}, 30\penalty0 (11):\penalty0 1763--1777, 2003.

\bibitem[Jeffery et~al.(2014)Jeffery, Korte, Palla, Walters, White, and
  Abernethy]{Jeffery2014}
K.~J. Jeffery, L.~Korte, F.~Palla, G.~M. Walters, L.~White, and K.~Abernethy.
\newblock \emph{Fire management in a changing landscape: a case study from
  {L}op{\'e} {N}ational {P}ark{,} {G}abon}, volume 20.1.
\newblock International Union for Conservation of Nature and Natural Resources,
  2014.

\bibitem[Koenker and Park(1996)]{KoenkerPark1996}
R.~Koenker and B.~J. Park.
\newblock An interior point algorithm for nonlinear quantile regression.
\newblock \emph{Journal of Econometrics}, 71\penalty0 (1):\penalty0 265--283,
  1996.

\bibitem[Lakshmikantham et~al.(1989)Lakshmikantham, Bainov, and
  Simeonov]{Lakshmikantham1989}
V.~Lakshmikantham, D.~Bainov, and P.S. Simeonov.
\newblock \emph{Theory of impulsive differential equations}, volume~6.
\newblock World Scientific, 1989.

\bibitem[Le~Roux and Bariac(1998)]{LeRoux1998seasonal}
X.~Le~Roux and T.~Bariac.
\newblock Seasonal variations in soil, grass and shrub water status in a west
  african humid savanna.
\newblock \emph{Oecologia}, 113\penalty0 (4):\penalty0 456--466, 1998.

\bibitem[Lehmann et~al.(2011)Lehmann, Archibald, Hoffmann, and
  Bond]{Lehmann2011}
C.~E.~R. Lehmann, S.~A. Archibald, W.~A. Hoffmann, and W.~J. Bond.
\newblock Deciphering the distribution of the savanna biome.
\newblock \emph{New Phytol.}, 191\penalty0 (1):\penalty0 197--209, 2011.

\bibitem[Lehsten et~al.(2016)Lehsten, Arneth, Spessa, Thonicke, and
  Moustakas]{Lehsten2016}
V.~Lehsten, A.~Arneth, A.~Spessa, K.~Thonicke, and A.~Moustakas.
\newblock The effect of fire on tree--grass coexistence in savannas: a
  simulation study.
\newblock \emph{Int. J. wildland fire}, 25\penalty0 (2):\penalty0 137--146,
  2016.

\bibitem[Lewis et~al.(2013)Lewis, Sonk\'{e}, Sunderland, Begne, Lopez-Gonzalez,
  van~der Heijden, Phillips, Affum-Baffoe, Baker, Banin, Bastin, Beeckman,
  Boeckx, Bogaert, De~Canni{\`e}re, Chezeaux, Clark, Collins, Djagbletey,
  Djuikouo, Droissart, Doucet, Ewango, Fauset, Feldpausch, Foli, Gillet,
  Hamilton, Harris, Hart, de~Haulleville, Hladik, Hufkens, Huygens, Jeanmart,
  Jeffery, Kearsley, Leal, LIoyd, Lovett, Makana, Malhi, Marshall, Ojo, Peh,
  Pickavance, Poulsen, Reitsma, Sheil, Simo, Taedoumg, Talbot, Taplin, Taylor,
  Thomas, Toirambe, Verbeeck, Vleminckx, White, Willcock, Woell, and
  Zemagho]{Lewis2013AGBspatial}
S.~L. Lewis, B.~Sonk\'{e}, T.~Sunderland, S.~K. Begne, G.~Lopez-Gonzalez,
  G.~M.~F. van~der Heijden, O.~L. Phillips, K.~Affum-Baffoe, T.~R. Baker,
  L.~Banin, J.~F. Bastin, H.~Beeckman, P.~Boeckx, J.~Bogaert,
  C.~De~Canni{\`e}re, E.~Chezeaux, C.~J. Clark, M.~Collins, G.~Djagbletey,
  M.~N.~K. Djuikouo, V.~Droissart, J.~L. Doucet, C.~E.~N. Ewango, S.~Fauset,
  T.~R. Feldpausch, E.~G. Foli, J.~F. Gillet, A.~C. Hamilton, D.~J. Harris,
  T.~B. Hart, T.~de~Haulleville, A.~Hladik, K.~Hufkens, D.~Huygens,
  P.~Jeanmart, K.~J. Jeffery, E.~Kearsley, M.~E. Leal, J.~LIoyd, J.~C. Lovett,
  J.~R. Makana, Y.~Malhi, A.~R. Marshall, L.~Ojo, K.~S.-H. Peh, G.~Pickavance,
  J.~R. Poulsen, J.~M. Reitsma, D.~Sheil, M.~Simo, H.~E. Taedoumg, J.~Talbot,
  J.~R.~D. Taplin, D.~Taylor, S.~C. Thomas, B.~Toirambe, H.~Verbeeck,
  J.~Vleminckx, L.~J.~T. White, S.~Willcock, H.~Woell, and L.~Zemagho.
\newblock Aboveground biomass and structure of 260 african tropical forests.
\newblock \emph{Phil. Trans. R. Soc. B.}, 368:\penalty0 20120295, 2013.

\bibitem[Mart{\'\i}nez-Garc{\'\i}a et~al.(2013)Mart{\'\i}nez-Garc{\'\i}a,
  Calabrese, and L{\'o}pez]{Martinez2013spatial}
R.~Mart{\'\i}nez-Garc{\'\i}a, J.~M. Calabrese, and C.~L{\'o}pez.
\newblock Spatial patterns in mesic savannas: the local facilitation limit and
  the role of demographic stochasticity.
\newblock \emph{J. Theor. Biol.}, 333:\penalty0 156--165, 2013.

\bibitem[MATLAB()]{Matlab}
MATLAB.
\newblock \emph{The Mathworks Inc.}
\newblock http://www.mathworks.com.

\bibitem[McNaughton(1992)]{McNaughton1992}
S.~J. McNaughton.
\newblock The propagation of disturbance in savannas through food webs.
\newblock \emph{J. Veg. Sci.}, 3\penalty0 (3):\penalty0 301--314, 1992.

\bibitem[Menaut and Cesar(1979)]{MenautCesar1979}
J.~C. Menaut and J~Cesar.
\newblock Structure and primary productivty of {L}amto savannas, {I}vory
  {C}oast.
\newblock \emph{Ecology}, pages 1197--1210, 1979.

\bibitem[Mermoz et~al.(2015)Mermoz, R{\'e}jou-M{\'e}chain, Villard, Le~Toan,
  Rossi, and Gourlet-Fleury]{Mermoz2015decrease}
S.~Mermoz, M.~R{\'e}jou-M{\'e}chain, L.~Villard, T.~Le~Toan, V.~Rossi, and
  S.~Gourlet-Fleury.
\newblock Decrease of l-band sar backscatter with biomass of dense forests.
\newblock \emph{Remote Sens. Environ.}, 159:\penalty0 307--317, 2015.

\bibitem[Mitchard and Flintrop(2013)]{Mitchard2013woody}
E.~T.~A. Mitchard and C.~M. Flintrop.
\newblock Woody encroachment and forest degradation in sub-saharan africa's
  woodlands and savannas 1982--2006.
\newblock \emph{Phil. Trans. R. Soc. B}, 368\penalty0 (1625):\penalty0
  20120406, 2013.

\bibitem[Mitchard et~al.(2009)Mitchard, Saatchi, Gerard, Lewis, and
  Meir]{Mitchard2009measuring}
E.~T.~A. Mitchard, S.~S. Saatchi, F.~F. Gerard, S.~L. Lewis, and P.~Meir.
\newblock Measuring woody encroachment along a forest--savanna boundary in
  central africa.
\newblock \emph{Earth Interact.}, 13\penalty0 (8):\penalty0 1--29, 2009.

\bibitem[Mordelet(1993)]{Mordelet1993influence}
P.~Mordelet.
\newblock \emph{Influence des arbres sur la strate herbac{\'e}e d'une savane
  humide(Lamto, C{\^o}te d'Ivoire)}.
\newblock PhD thesis, 1993.

\bibitem[Moustakas et~al.(2013)Moustakas, Kunin, Cameron, and
  Sankaran]{Moustakas2013facilitation}
A.~Moustakas, William.~E. Kunin, Tom.~C. Cameron, and Mahesh Sankaran.
\newblock Facilitation or competition? tree effects on grass biomass across a
  precipitation gradient.
\newblock \emph{PloS One}, 8\penalty0 (2):\penalty0 e57025, 2013.

\bibitem[Penning~de Vries and Djit{\`e}ye(1982)]{PenningDjiteye1982}
F.~W.~T. Penning~de Vries and M.~A. Djit{\`e}ye.
\newblock \emph{La productivit{\'e} des p{\^a}turages sah{\'e}liens: une
  {\'e}tude des sols, des v{\'e}g{\'e}tations et de l'exploitation de cette
  ressource naturelle. (The productivity of Sahelian rangelands, a study of
  soils, vegetations, and exploitation of this natural resource.)}, volume 918.
\newblock Agric. Res. Rep. (Versl. Landbouwk. Onderz.), 1982.

\bibitem[Rietkerk et~al.(1997)Rietkerk, van~den Bosch, and van~de
  Koppel]{Rietkerk1997site}
M.~Rietkerk, F.~van~den Bosch, and J.~van~de Koppel.
\newblock Site-specific properties and irreversible vegetation changes in
  semi-arid grazing systems.
\newblock \emph{Oikos}, pages 241--252, 1997.

\bibitem[Rietkerk et~al.(2002)Rietkerk, Boerlijst, van Langevelde,
  HilleRisLambers, van~de Koppel, Kumar, Prins, and de~Roos]{Rietkerk2002self}
M.~Rietkerk, M.~C. Boerlijst, F.~van Langevelde, R.~HilleRisLambers, J.~van~de
  Koppel, L.~Kumar, H.~H.~T. Prins, and A.~M. de~Roos.
\newblock Self-organization of vegetation in arid ecosystems.
\newblock \emph{Am. Nat.}, 160\penalty0 (4):\penalty0 524--530, 2002.

\bibitem[Sankaran et~al.(2005)Sankaran, Hanan, Scholes, Ratnam, Augustine,
  Cade, Gignoux, Higgins, Le~Roux, Ludwig, Ardo, Banyikwa, Bronn, Bucini,
  Caylor, Coughenour, Diouf, Ekaya, Feral, February, Frost, Hiernaux, Hrabar,
  Metzger, Prins, Ringrose, Sea, Tews, J, and
  Zambatis]{Sankaran2005determinants}
M.~Sankaran, N.~P. Hanan, R.~J. Scholes, J.~Ratnam, D.~J. Augustine, B.~S.
  Cade, J.~Gignoux, S.~I. Higgins, X.~Le~Roux, F.~Ludwig, J.~Ardo, F.~Banyikwa,
  A.~Bronn, G.~Bucini, K.~K. Caylor, M.~B. Coughenour, A.~Diouf, W.~Ekaya,
  C.~J. Feral, E.~C. February, P.~G.~H. Frost, P.~Hiernaux, H.~Hrabar, K.~L.
  Metzger, H.~H.~T. Prins, S.~Ringrose, W.~Sea, J.~Tews, Worden. J, and
  N.~Zambatis.
\newblock Determinants of woody cover in african savannas.
\newblock \emph{Nature}, 438\penalty0 (7069):\penalty0 846--849, 2005.
\newblock \doi{10.1038/nature04070}.

\bibitem[Scheiter(2009)]{Scheiter2009Grasstree}
S.~Scheiter.
\newblock \emph{Grass--tree interactions and the ecology of African savannas
  under current and future climates}.
\newblock PhD thesis, 2009.

\bibitem[Scheiter and Higgins(2007)]{Scheiter2007}
S.~Scheiter and S.~I. Higgins.
\newblock Partitioning of root and shoot competition and the stability of
  savannas.
\newblock \emph{Am. Nat.}, 179:\penalty0 587--601, 2007.

\bibitem[Scholes(2003)]{Scholes2003convex}
R.~J. Scholes.
\newblock Convex relationships in ecosystems containing mixtures of trees and
  grass.
\newblock \emph{Environ. Resour. Econ.}, 26\penalty0 (4):\penalty0 559--574,
  2003.

\bibitem[Scholes and Archer(1997)]{Scholes1997}
R.J. Scholes and S.~R. Archer.
\newblock Tree-grass interactions in savannas.
\newblock \emph{Annu. Rev. Ecol. Evol. Syst.}, pages 517--544., 1997.
\newblock \doi{10.1146/annurev.ecolsys.28.1.517}.

\bibitem[Scholes and Walker(1993)]{Scholes1993african}
R.J. Scholes and B.H. Walker.
\newblock \emph{An African savanna: Synthesis of the Nylsvley study}.
\newblock Cambridge University Press, 1993.

\bibitem[Simioni et~al.(2003)Simioni, Gignoux, and Le~Roux]{Simioni2003tree}
G.~Simioni, J.~Gignoux, and X.~Le~Roux.
\newblock Tree layer spatial structure can affect savanna production and water
  budget: results of a 3-d model.
\newblock \emph{Ecology}, 84\penalty0 (7):\penalty0 1879--1894, 2003.

\bibitem[Staver et~al.(2011{\natexlab{a}})Staver, Archibald, and
  Levin]{Staver2011b}
A.~Carla. Staver, Sally. Archibald, and Simon.~A. Levin.
\newblock The global extent and determinants of savanna and forest as
  alternative biome states.
\newblock \emph{Science}, 334\penalty0 (6053):\penalty0 230--232,
  2011{\natexlab{a}}.

\bibitem[Staver and Levin(2012)]{Staver2012integrating}
A.C. Staver and S.A. Levin.
\newblock Integrating theoretical climate and fire effects on savanna and
  forest systems.
\newblock \emph{Am. Nat.}, 180\penalty0 (2):\penalty0 211--224, 2012.
\newblock \doi{10.1086/666648}.

\bibitem[Staver et~al.(2011{\natexlab{b}})Staver, Archibald, and
  Levin]{Staver2011tree}
A.C. Staver, S.~Archibald, and S.~Levin.
\newblock Tree cover in sub-saharan africa: rainfall and fire constrain forest
  and savanna as alternative stable states.
\newblock \emph{Ecology}, 92\penalty0 (5):\penalty0 1063--1072,
  2011{\natexlab{b}}.
\newblock \doi{10.2307/41151234}.

\bibitem[Tchuint\'e~Tamen(2017)]{PhDTchuinte2017}
A.~Tchuint\'e~Tamen.
\newblock \emph{Study of a Generic Mathematical Model of Forest-Savanna
  Interactions: Case of {C}ameroon}.
\newblock PhD thesis, University of Yaound\'e I, 2017.

\bibitem[Tchuint\'e~Tamen et~al.(2014)Tchuint\'e~Tamen, Tewa, Couteron, Bowong,
  and Dumont]{Tchuinte2014}
A.~Tchuint\'e~Tamen, J.~J. Tewa, P.~Couteron, S.~Bowong, and Y.~Dumont.
\newblock A generic modeling of fire impact in a tree-grass savanna model.
\newblock \emph{BIOMATH}, 3\penalty0 (2):\penalty0 1407191, 2014.

\bibitem[Tchuint\'{e}~Tamen et~al.(2016)Tchuint\'{e}~Tamen, Dumont, Tewa,
  Bowong, and Couteron]{Tchuinte2016}
A.~Tchuint\'{e}~Tamen, Y.~Dumont, J.~J. Tewa, S.~Bowong, and P.~Couteron.
\newblock Tree--grass interaction dynamics and pulsed fires: Mathematical and
  numerical studies.
\newblock \emph{Appl. Math. Model.}, 40:\penalty0 6165--6197, 2016.
\newblock \doi{http://dx.doi.org/10.1016/j.apm.2016.01.019}.

\bibitem[Tchuint\'{e}~Tamen et~al.(2017)Tchuint\'{e}~Tamen, Dumont, Tewa,
  Bowong, and Couteron]{Tchuinte2017}
A.~Tchuint\'{e}~Tamen, Y.~Dumont, J.~J. Tewa, S.~Bowong, and P.~Couteron.
\newblock A minimalistic model of tree--grass interactions using impulsive
  differential equations and non-linear feedback functions of grass biomass
  onto fire-induced tree mortality.
\newblock \emph{Math. Comput. Simulation}, 133:\penalty0 265--297, 2017.
\newblock \doi{http://dx.doi.org/10.1016/j.matcom.2016.03.008}.

\bibitem[Trollope et~al.(2002)Trollope, Trollope, and
  Hartnett]{Trollope2002topkill}
W.~S.~W Trollope, L.~A. Trollope, and D.~C. Hartnett.
\newblock Fire behaviour a key factor in the fire ecology of african grasslands
  and savannas.
\newblock \emph{Forest Fire Research and Wildland Fire Safety, Viegas (ed.),
  Millpress, Rotterdam}, 2002.

\bibitem[UNESCO(1981)]{UNESCO1981}
UNESCO.
\newblock Ecosyst\`{e}mes p\^{a}tur\'{e}s tropicaux. {U}n rapport sur
  l'\'{e}tat des connaissances pr\'{e}par\'{e} par l'{UNESCO}, le {PNUE} et la
  {FAO}.
\newblock In \emph{Recherches sur les Ressources Naturelles, XVI}. Presses
  Univ. de France, Vend\^{o}me 675 p, 144 fig., 169 tabl., 1981.

\bibitem[Van De~Koppel and Rietkerk(2000)]{vanKoppel2000herbivore}
J.~Van De~Koppel and M.~Rietkerk.
\newblock Herbivore regulation and irreversible vegetation change in semi-arid
  grazing systems.
\newblock \emph{Oikos}, 90\penalty0 (2):\penalty0 253--260, 2000.

\bibitem[van~de Koppel et~al.(1997)van~de Koppel, Rietkerk, and
  Weissing]{vandeKoppel1997}
J.~van~de Koppel, M.~Rietkerk, and F.~J. Weissing.
\newblock Catastrophic vegetation shifts and soil degradation in terrestrial
  grazing systems.
\newblock \emph{Trends Ecol. Evol.}, 12\penalty0 (9):\penalty0 352--356, 1997.

\bibitem[Van~Langevelde et~al.(2003)Van~Langevelde, Van De~Vijver, Kumar, Van
  De~Koppel, De~Ridder, Van~Andel, Skidmore, Hearne, Stroosnijder, Bond,
  et~al.]{vanLangevelde2003}
F.~Van~Langevelde, C.A.D.M. Van De~Vijver, L.~Kumar, J.~Van De~Koppel,
  N.~De~Ridder, J.~Van~Andel, A.K. Skidmore, J.W. Hearne, L.~Stroosnijder, W.J.
  Bond, et~al.
\newblock Effects of fire and herbivory on the stability of savanna ecosystems.
\newblock \emph{Ecology}, 84\penalty0 (2):\penalty0 337--350, 2003.

\bibitem[van Leeuwen et~al.(2014)van Leeuwen, Van~der Werf, Hoffmann, Detmers,
  R{\"u}cker, French, Archibald, Carvalho~Jr, Cook, De~Groot, H{\'e}ly,
  Kasischke, Kloster, McCarty, Pettinari, Savadogo, Alvarado, Boschetti,
  Manuri, Meyer, Siegert, Trollope, and Trollope]{vanLeeuwen2014}
T.~T. van Leeuwen, G.~R. Van~der Werf, A.~A. Hoffmann, R.~G. Detmers,
  G.~R{\"u}cker, N.~H.~F. French, S.~Archibald, J.~A. Carvalho~Jr, G.~D. Cook,
  W.~J. De~Groot, C.~H{\'e}ly, E.~C. Kasischke, S.~Kloster, L.~J. McCarty,
  M.~L. Pettinari, P.~Savadogo, E.~C. Alvarado, L.~Boschetti, S.~Manuri, C.~P.
  Meyer, F.~Siegert, L.~A. Trollope, and W.~S.~W. Trollope.
\newblock Biomass burning fuel consumption rates: a field measurement database.
\newblock \emph{Biogeosciences}, 11:\penalty0 7305--7329, 2014.

\bibitem[Van~Nes et~al.(2014)Van~Nes, Hirota, Holmgren, and
  Scheffer]{Nes2014tipping}
E.~H. Van~Nes, Marina Hirota, Milena Holmgren, and Marten Scheffer.
\newblock Tipping points in tropical tree cover: linking theory to data.
\newblock \emph{Global Change Biol.}, 20\penalty0 (3):\penalty0 1016--1021,
  2014.

\bibitem[Van~Wilgen et~al.(2004)Van~Wilgen, Govender, Biggs, Ntsala, and
  Funda]{vanWilgen2004response}
B.W. Van~Wilgen, N.~Govender, H.C. Biggs, D.~Ntsala, and X.N. Funda.
\newblock Response of savanna fire regimes to changing fire-management policies
  in a large african national park.
\newblock \emph{Conserv. Biol.}, 18\penalty0 (6):\penalty0 1533--1540, 2004.
\newblock \doi{10.1111/j.1523-1739.2004.00362.x.}

\bibitem[Walker and Noy-Meir(1982)]{Walker1982aspects}
B.H. Walker and I.~Noy-Meir.
\newblock Aspects of the stability and resilience of savanna ecosystems.
\newblock In \emph{Ecology of tropical savannas}, pages 556--590. Springer,
  1982.
\newblock \doi{10.1007/978-3-642-68786-0}.

\bibitem[Walker et~al.(1981)Walker, Ludwig, Holling, and
  Peterman]{Walker1981stability}
B.H. Walker, D.~Ludwig, C.S. Holling, and R.M. Peterman.
\newblock Stability of semi-arid savanna grazing systems.
\newblock \emph{The Journal of Ecology}, pages 473--498, 1981.

\bibitem[West et~al.(2009)West, Enquist, and Brown]{West2009}
G.~B. West, B.~J. Enquist, and J.~H. Brown.
\newblock A general quantitative theory of forest structure and dynamics.
\newblock \emph{Proc. Natl. Acad. Sci.}, 106\penalty0 (17):\penalty0
  7040--7045, 2009.

\bibitem[Whittaker(1975)]{Whittaker1975}
R.~H. Whittaker.
\newblock \emph{Communities and ecosystems}.
\newblock . Second edition. Macmillan, New York, New York, USA, 1975.

\bibitem[Yatat et~al.(2014)Yatat, Dumont, Tewa, Couteron, and
  Bowong]{Yatat2014}
V.~Yatat, Y.~Dumont, J.J. Tewa, P.~Couteron, and S.~Bowong.
\newblock Mathematical analysis of a size structured tree-grass competition
  model for savanna ecosystems.
\newblock \emph{BIOMATH}, 3\penalty0 (1):\penalty0 1404212, 2014.

\bibitem[Yatat et~al.(2016)Yatat, Couteron, Tewa, Bowong, and
  Dumont]{Yatat2016}
V.~Yatat, P.~Couteron, J.~J. Tewa, S.~Bowong, and Y.~Dumont.
\newblock An impulsive modelling framework of fire occurrence in a
  size-structured model of tree--grass interactions for savanna ecosystems.
\newblock \emph{J. Math. Biol.}, pages 1--58, 2016.
\newblock \doi{10.1007/s00285-016-1060-y}.

\bibitem[Yu and D'Odorico(2014)]{YuDOdoricco2014ecohydrological}
K.~Yu and P.~D'Odorico.
\newblock An ecohydrological framework for grass displacement by woody plants
  in savannas.
\newblock \emph{J. Geophys. Res. Biogeosci.}, 119\penalty0 (3):\penalty0
  192--206, 2014.

\end{thebibliography}

\end{document}